\documentclass[fleqn,usenatbib]{mnras}
\usepackage{newtxtext,newtxmath}
\usepackage[T1]{fontenc}
\usepackage{ae,aecompl}
\usepackage{graphicx}
\usepackage{amsmath,braket,bm,multirow}
\usepackage{ulem}
\usepackage{fancybox}
\numberwithin{equation}{section}
\title[Detectability of the gravitational redshift]%
{Detectability of the gravitational redshift effect from the asymmetric galaxy clustering}
\author[S. Saga et al.]
{Shohei Saga,$^{1}$\thanks{E-mail: shohei.saga@obspm.fr}
Atsushi Taruya,$^{2,3}$
Yann Rasera,$^{4}$
and Michel-Andr{\`e}s Breton,$^{5}$
\\
$^{1}$Laboratoire Univers et Th{\'e}ories, Observatoire de Paris, Universit{\'e} PSL, Universit{\'e} de Paris, CNRS, F-92190 Meudon, France\\
$^{2}$Center for Gravitational Physics, Yukawa Institute for Theoretical Physics, Kyoto University, Kyoto 606-8502, Japan\\
$^{3}$Kavli Institute for the Physics and Mathematics of the Universe (WPI), The University of Tokyo Institutes for Advanced Study, \\
The University of Tokyo, 5-1-5 Kashiwanoha, Kashiwa, Chiba 277-8583, Japan\\
$^{4}$Laboratoire Univers et Th{\'e}ories, Universit{\'e} de Paris, Observatoire de Paris, Universit{\'e} PSL, CNRS, F-92190 Meudon, France\\
$^{5}$Aix Marseille Univ, CNRS, CNES, LAM, Marseille, France\\
}
\thisfancyput(14.5cm,0.5cm){\large{YITP-21-93}}
\date{Accepted XXX. Received YYY; in original form ZZZ}
\pubyear{2021}
\begin{document}
\label{firstpage}
\pagerange{\pageref{firstpage}--\pageref{lastpage}}
\maketitle
\begin{abstract}
It has been recently recognized that the observational relativistic effects, mainly arising from the light propagation in an inhomogeneous universe, induce the dipole asymmetry in the cross-correlation function of galaxies. In particular, the dipole asymmetry at small scales is shown to be dominated by the gravitational redshift effects. 
In this paper, we exploit a simple analytical description for the dipole asymmetry in the cross-correlation function valid at quasi-linear regime. In contrast to the previous model, a new prescription involves only one dimensional integrals, providing a faster way to reproduce the results obtained by Saga et al. (2020).
Using the analytical model, we discuss the detectability of the dipole signal induced by the gravitational redshift effect from upcoming galaxy surveys.
The gravitational redshift effect at small scales enhances the signal-to-noise ratio (S/N) of the dipole, and in most of the cases considered, the S/N is found to reach a maximum at $z\approx0.5$.
We show that current and future surveys such as DESI and SKA provide an idealistic data set, giving a large S/N of $10$--$20$.
Two potential systematics arising from off-centered galaxies are also discussed (transverse Doppler effect and diminution of the gravitational redshift effect), and their impacts are found to be mitigated by a partial cancellation between two competitive effects. Thus, the detection of the dipole signal at small scales is directly linked to the gravitational redshift effect, and should provide an alternative route to test gravity.
\end{abstract}
\begin{keywords}
\textit{(cosmology:)} large-scale structure of Universe -- cosmology: theory -- gravitation
\end{keywords}

\section{Introduction}

Mapping the large-scale structure of the universe with galaxy surveys is currently a major science driver for cosmology. In particular, through its statistical characterizations such as two-point correlation function or power spectrum, the large-scale galaxy distribution enables us to probe the late-time cosmic expansion, growth of structure, and even the primordial fluctuations. However, the observed three-dimensional map of galaxies does not directly reflect the true galaxy distribution because of a number of physical effects.
The most prominent effect is the Doppler effect induced by the peculiar velocities of galaxies, which produces apparent anisotropies along the line-of-sight direction, known as redshift-space distortions (RSD)~\citep{1987MNRAS.227....1K,1992ApJ...385L...5H}.
The RSD has now been recognized as a sensitive probe of the growth of cosmic structure, and the measurement of it provides a unique opportunity for a test of gravity on cosmological scales~\citep[e.g.,][]{2008Natur.451..541G,2008APh....29..336L,2009MNRAS.393..297P,2012MNRAS.426.2719R,2013MNRAS.433.1202S,2017MNRAS.470.2617A}.
The upcoming galaxy surveys will observe an unprecedented number of galaxies and provide us with high-precision measurements of RSD, which can further offer a way to detect small but non-negligible special and general relativistic contributions to RSD~\citep[][]{1987MNRAS.228..653S,2004MNRAS.348..581P,2009PhRvD..80h3514Y,2010PhRvD..82h3508Y,2011PhRvD..84f3505B,2011PhRvD..84d3516C,2014CQGra..31w4001Y}.

Recently, it has been shown that relativistic effects arising from the light propagation in an inhomogeneous universe, e.g., gravitational redshift, integrated Sachs-Wolfe, and weak lensing effects, produce asymmetric distortions to the galaxy distribution along the line-of-sight direction~\citep{2013MNRAS.434.3008C,2012arXiv1206.5809Y,2018JCAP...03..019T}.
This means that with a certain line-of-sight definition, applying the multipole expansion to the cross-correlation function or power spectrum between different biased objects yields non-vanishing odd multipole moments, with the largest signals coming from the dipole moment~\citep[e.g.,][]{2009JCAP...11..026M,2014PhRvD..89h3535B}.
Detection of such relativistic signals would provide a new window to probe gravity on cosmological scales, thus complementary to the measurement of the redshift-space distortions induced by the Doppler effect. Further, it can offer a fundamental or classical test of gravity from a viewpoint of the equivalence principle, helpful to constrain cosmology~\citep[e.g.,][]{2018JCAP...05..061B,2020JCAP...08..004B}.
Recently, \citet{2017MNRAS.470.2822A} have claimed the detection of the asymmetry at the $2.8\sigma$ level using SDSS BOSS DR12 CMASS galaxy sample.
Their results are consistent with the gravitational redshift effect predicted by general relativity \citep[see also][for the detection using clusters of galaxies]{2011Natur.477..567W, 2015PhRvL.114g1103S,2015MNRAS.448.1999J,2021MNRAS.503..669M}.

In our previous studies, toward a solid detection of the non-vanishing relativistic dipole in the cross-correlation function, we have numerically constructed halo catalogues on light cone, taking consistently the observational relativistic effects into account~\citep{2019MNRAS.483.2671B} \citep[see][for recent similar works at lower resolution]{2017MNRAS.471.3899B,2021MNRAS.501.2547G,2020arXiv201112936C}.
At large scales, we found that the standard Doppler effect without taking the distant-observer approximation gives the largest contribution to the dipole~\citep[][]{2020MNRAS.491.4162T}.
On the other hand, at the scales beyond the linear regime, the gravitational redshift effect starts to dominate the dipole, and the linear theory prediction fails to reproduce the simulation results.

In order to quantitatively explain major findings in the numerical simulations, \citet{2020MNRAS.498..981S} developed a quasi-linear model based on the Zel'dovich approximation.
The model considers the standard Doppler and gravitational redshift effects as dominant relativistic contributions, taking also the so-called wide-angle effect of RSD into account in a self-consistent way.
In particular, the model accounts for the non-perturbative contribution to the gravitational redshift effect arising from the halo potential, which is shown to play an important role to describe the small-scale behaviours of the dipole moment, leading to a remarkable agreement with the dipole cross-correlations measured in simulations at quasi-linear scales~($s \gtrsim 5\, {\rm Mpc}/h$).

In this paper, based on the success of our numerical and analytical modelling, we pursue to further investigate the relativistic dipole, focusing specifically on its future detectability.
Several authors have investigated the feasibility to detect the relativistic dipole, but they rely on the linear theory prediction, and consider large scales~\citep{2017PhRvD..95d3530H,2018JCAP...05..043L}.
Contrary to these previous works, our study here is based on a model capable of going beyond linear regime, taking the nonlinear gravitational potential of haloes into account. A similar study focusing on small scales has been recently done by \citet{2020JCAP...07..048B}, using the third-order Eulerian perturbation theory. They considered the power spectrum dipole, i.e., the Fourier counterpart of the dipole cross-correlation function, and dividing a single galaxy population observed by Dark Energy Spectroscopic Instrument\footnote{\url{https://www.desi.lbl.gov/}}~\citep[DESI,][]{2016arXiv161100036D} into more than two subsamples, they found that the signal-to-noise ratio of their cross power spectrum exceeds $10$ if the difference of the (linear) galaxy biases between two subsamples, $\Delta b$, becomes $\Delta b= 1$. In this paper, we estimate the signal-to-noise ratio for the cross-correlation function, and applying the multi-tracer techniques, we discuss systematically the detectability of the relativistic dipole through the combination of various upcoming galaxy surveys. In doing so, we will first present a simple analytical model, which quantitatively reproduces major trends obtained from our previous study~\citep{2020MNRAS.498..981S}. In contrast to our previous model which involves seven dimensional integrals, the prediction of the dipole in the present model needs only the one dimensional integrals, hence providing a faster way to estimate the signal-to-noise ratio. We will then examine the detectability of relativistic dipole in various upcoming surveys:
DESI~\citep[][]{2016arXiv161100036D}, Euclid\footnote{\url{https://www.euclid-ec.org/}}~\citep{2011arXiv1110.3193L}, Subaru Prime Focus Spectrograph\footnote{\url{http://sumire.ipmu.jp/en/}} \citep[PFS,][]{2014PASJ...66R...1T}, and Square Kilometre Array\footnote{\url{https://www.skatelescope.org/}}\citep[SKA,][]{2020PASA...37....7S}.
Moreover, potentially important systematics are also investigated, and incorporating these effects into the analytical model, we quantitatively predict their impacts on the dipole cross-correlation function.

This paper is organized as follows.
In Sec.~\ref{sec: model}, we present a simple analytical model for the relativistic dipole induced by the Doppler and gravitational redshift effects, which involves only one dimensional integrals.
In Sec.~\ref{sec: cov}, we write down the estimator for the dipole moment of the cross-correlation function and compute its covariance matrix following \citet{2016JCAP...08..021B,2017PhRvD..95d3530H}.
This is used in Sec.~\ref{sec: result} to estimate the signal-to-noise ratio of the dipole moment for various upcoming surveys.
In Sec.~\ref{sec: systematics}, we discuss a potential impact of the systematic effects from off-centered galaxies on the dipole moment.
Finally, Sec.~\ref{sec: summary} is devoted to the summary of important findings.

Supplementing with the analysis and results in the main text, Appendices~\ref{app: derivation}, \ref{app: comparison formula}, and \ref{app: other moments} provide respectively key expressions to derive the analytical expression for the dipole cross-correlation function in our simple model, the comparison of its model with an approximate description discussed in our previous paper, and the analytical expressions of the non-vanishing multipoles based on the model. Appendix~\ref{app: magnification} discusses the impact of the effect ignored in our analytical model on the dipole signal, particularly focusing on the Doppler magnification.
In Appendix~\ref{app: future surveys}, we summarize the parameters characterizing upcoming galaxy surveys, which are used to estimate the signal-to-noise ratio of the dipole in Sec.~\ref{sec: result}.
In Appendix~\ref{sec: SN in sim}, we present an alternative way to estimate the signal-to-noise ratio, in which the halo subsamples to cross-correlate are characterized by the minimum halo mass and the width of (logarithmic) halo mass bins.

Throughout this paper, we assume a flat Lambda cold dark matter ($\Lambda$CDM) model. The fiducial values of cosmological parameters are chosen so as to match the numerical simulations \citep{2017MNRAS.471.3899B}, based on the seven-year WMAP results~\citep{2011ApJS..192...18K}:
$\Omega_{\rm m0} = 0.25733$, $\Omega_{\rm b0} = 0.04356$, $\Omega_{\Lambda0} = 0.74259$, and $\Omega_{\rm r0} = 8.076 \times 10^{-5}$ for the density parameters for matter, baryon, dark energy with equation-of-state parameter $w=-1$, and radiation, respectively, at the present time.
The other cosmological parameters are chosen as $h = 0.72$, $n_{\rm s} = 0.963$, and $\sigma_{8} = 0.801$ for the Hubble parameter, scalar spectral index, and the root-mean-square matter density fluctuations with a top-hat filter of radius $8 \; h^{-1}$ Mpc. Throughout the paper, we will work with units of $c=1$.

\section{Model}
\label{sec: model}

The main purpose of this paper is to quantitatively estimate the detectability of the relativistic dipole, arising from the gravitational redshift effects, in upcoming deep and wide surveys. 
In doing so, we first present an analytical model of dipole cross-correlation function in this section. The model presented below involves only one dimensional integrals, and hence it provides a fast way to predict the relativistic dipole as well as to estimate its signal-to-noise ratio based on the covariance matrix calculations. 

In modelling the dipole cross-correlation function, the standard Doppler effect has to be also taken into account, since it gives a dominant contribution to the dipole at large scales through the so-called wide-angle effect~\citep{1994MNRAS.266..219F,1996ApJ...462...25Z,1996MNRAS.278...73H,1998ApJ...498L...1S,2000ApJ...535....1M,2004ApJ...614...51S,2004ApJ...615..573M,2008MNRAS.389..292P}.
Considering both the Doppler and gravitational redshift effects, \citet{2020MNRAS.498..981S} constructed a quasi-linear model based on the Zel'dovich approximation. To account for the non-perturbative contributions at small scales, we combined it with the halo model to predict the relativistic dipole from the halo potential. In Sec.~\ref{sec: model approx}, starting from the expression in our previous work, we derive a simplified expression for the density field by linearizing the displacement fields but still retaining the non-perturbative contribution. 
Then, the expression for the dipole cross-correlation function is simplified, and is presented in Sec.~\ref{sec: correlation function}.

\subsection{Modelling observed density fields}
\label{sec: model approx}

Consider an object at the true position $\bm{x}$ in comoving space. In redshift space, the observed position $\bm{s}$ generally differs from $\bm{x}$, mainly due to the standard Doppler effect. Taking also into account the relativistic corrections, which we denote by $\epsilon$, the relation between the two positions $\bm{x}$ and $\bm{s}$ is given by ~\citep[e.g.,][]{2011PhRvD..84d3516C}:
\begin{align}
\bm{s} = \bm{x} + \frac{1}{aH}\left( \bm{v}\cdot \hat{\bm{x}}\right)\hat{\bm{x}} + \epsilon(\bm{x})\hat{\bm{x}} ~,
\label{eq: s to x}
\end{align}
where $\hat{\bm{x}}$ is the unit vector defined by $\hat{\bm{x}} = \bm{x}/|\bm{x}|$ and $a$, $H$, and $\bm{v}$ are a scale factor, Hubble parameter, and peculiar velocity of the object, respectively.
Note that the expression at Eq.~\eqref{eq: s to x} is valid in the weak-field approximation of metric perturbation, and $|\bm{v}|\ll 1$. In Eq.~\eqref{eq: s to x}, we also ignore the gravitational lensing effect, which has been shown to give a very minor contribution to the asymmetric cross-correlation, i.e., odd multipole anisotropies. The term $\epsilon$ includes the contributions of gravitational redshift, integrated Sachs-Wolfe, transverse Doppler, and Shapiro time-delay effects, among which the gravitational redshift effect gives the most dominant relativistic contribution. Thus, focusing on the major relativistic effect, it is expressed as 
\begin{align}
\epsilon(\bm{x}) = -\frac{1}{aH}\phi(\bm{x}) ~,
\end{align}
where the function $\phi(\bm{x})$ stands for the gravitational potential.
The explicit forms of other relativistic contributions to the observed source position can be found in the literature~\citep[e.g.,][]{2010PhRvD..82h3508Y,2011PhRvD..84d3516C,2011PhRvD..84f3505B}.

To derive a simplified expression for the correlation function, we first follow the analytical treatment given by \citet{2020MNRAS.498..981S}, who applied the Zel'dovich approximation to predict the cross-correlation function beyond linear regime~\citep{1969JETP...30..512N,1970A&A.....5...84Z,1989RvMP...61..185S}. The Zel'dovich approximation, known as the first-order Lagrangian perturbation theory, describes the motion of mass element at the Eulerian position $\bm{x}$, introducing the Lagrangian displacement field, $\bm{\Psi}$, which is given as a function of the Lagrangian position (initial position) $\bm{q}$. Assuming that the objects of our interest follow the velocity flow of mass distributions, the Eulerian position and the velocity of each mass element at $\bm{x}$, $\bm{v}$, at a given time $t$ are generally expressed as
\begin{align}
\bm{x}(\bm{q},t) &= \bm{q} + \bm{\Psi}(\bm{q},t) ~, \label{eq: x to q}\\
\bm{v}(\bm{x}) &= a \frac{{\rm d}\bm{\Psi}}{{\rm d}t} ~. \label{eq: velocity}
\end{align}
The displacement field should satisfy the condition $\bm{\Psi}\to0$ at $t\to0$.
In the Zel'dovich approximation, it is expressed in terms of the (Lagrangian) linear density field $\delta_{\rm L}$ as $\bm{\nabla}_{q}\cdot \bm{\Psi}_{\rm ZA} = -\delta_{\rm L}$, with the operator $\bm{\nabla}_{q}$ being a spatial derivative with respect to the Lagrangian coordinate. Recalling that the linear density field is related to initial density field $\delta_0$ through $\delta_{\rm L}=D_+(t)\delta_0$ with $D_+$ being the linear growth factor, the velocity field is rewritten with
\begin{align}
\bm{v} &= aHf\bm{\Psi}_{\rm ZA} ~,\label{eq: Zel velocity}
\end{align}
where the quantity $f$ is the linear growth rate defined by $f \equiv {\rm d}\ln{D_{+}(a)}/{\rm d}\ln{a}$.

Substituting the expressions at Eqs.(\ref{eq: x to q}) and (\ref{eq: Zel velocity}) into Eq.~(\ref{eq: s to x}), the relation between the redshift-space position $\bm{s}$ and the Lagrangian-space position $\bm{q}$ becomes
\begin{align}
s_{i} &= q_i + \{\delta_{ij}+f\,\hat{x}_i\hat{x}_j\}\Psi_i(\bm{q})+\epsilon(\bm{x})\hat{x}_i
\nonumber
\\
&\simeq q_{i} + R_{ij}(\hat{\bm{q}})\Psi_{j}(\bm{q}) + \epsilon(\bm{q}) \hat{q}_{i}~,
\label{eq: mapping_s_q-space epsilon}
\end{align}
with the matrix $R_{ij}$ defined by $R_{ij}(\hat{\bm{q}}) \equiv \delta_{ij} + f\hat{q}_{i}\hat{q}_{j}$. Here, we used the Einstein summation convention and omit the subscript ZA, simply writing $\bm{\Psi}_{\rm ZA}$ as $\bm{\Psi}$. Note that the second line is valid at first-order displacement field (i.e. Zel’dovich approximation).

Given the relation at Eq.~\eqref{eq: mapping_s_q-space epsilon}, the number density field of the object in redshift space, $n^{\rm(S)}$, is expressed in terms of the quantities defined in Lagrangian space. We have 
\begin{align}
n^{({\rm S})}(\bm{s}){\rm d}^{3}\bm{s} = \overline{n} \left( 1 + b^{\rm L} \delta_{\rm L}(\bm{q})\right){\rm d}^{3}\bm{q} ~,
\end{align}
where the quantity $b^{\rm L}$ is the Lagrangian linear bias parameter, and $\overline{n}$ is the mean number density at a given redshift. The above expression is recast as
\begin{align}
n^{({\rm S})}(\bm{s})
&=
\overline{n} \left( 1 + b^{\rm L} \delta_{\rm L}(\bm{q})\right)\left| \frac{\partial s_{i}}{\partial q_{j}}\right|^{-1}
\notag \\
&= 
\overline{n} \int{\rm d}^{3}\bm{q}\, \left( 1 + b^{\rm L} \delta_{\rm L}(\bm{q})\right) \delta_{\rm D}(s_{i} - q_{i} -R_{ij}\Psi_{j} + \epsilon \hat{q}_{i})
\notag \\
&= \overline{n}\int{\rm d}^{3}\bm{q}\; \int\frac{{\rm d}^{3}\bm{k}}{(2\pi)^{3}}\;
{\rm e}^{{\rm i} k_{i}\left( s_{i}-q_{i}-R_{ij}\Psi_{j} - \epsilon\hat{q}_{i}\right)} 
\left( 1 + b^{\rm L} \delta_{\rm L}(\bm{q})\right) ~. \label{eq: def n^{S}}
\end{align}

Let us now consider the density fluctuation. Denoting it by $\delta^{\rm (S)}$, we define
\begin{align}
\delta^{\rm (S)}(\bm{s}) = \frac{n^{({\rm S})}(\bm{s})}{\Braket{n^{({\rm S})}(\bm{s})}} -1 ~,
\label{eq: def delta}
\end{align}
where the bracket $\Braket{\cdots}$ stands for the ensemble average. Here, it is to be noted that the quantity $\langle n^{\rm(S)}\rangle$ generally differs from $\overline{n}$, due to the directional-dependent matrix $R_{ij}$ and relativistic correction along the line-of-sight direction. In the presence of these terms, a naive substitution of Eq.~\eqref{eq: def n^{S}} into the above yields an intricate expression for the correlation function which involves the multi-dimensional integrals in both numerator and denominator. Indeed, without invoking any approximation, \citet{2020MNRAS.498..981S} derived an exact expression for the cross-correlation function from Eq.~\eqref{eq: def delta} \citep[see also][]{2020MNRAS.491.4162T}, with which the prediction of the dipole moment is made numerically by performing seven dimensional integrals, requiring a time-consuming computation. However, ignoring the relativistic contribution, a detailed comparison of the predictions between the exact expression and the linear theory has shown that the results almost coincide with each other \citep{2020MNRAS.491.4162T}. One can thus linearise the expression at \eqref{eq: def delta} with respect to the displacement field. Further, the relativistic corrections, which are supposed to be small, can be also expanded from the exponent. Then, we obtain
\begin{align}
\delta^{({\rm S})}(\bm{s}) &= \int{\rm d}^{3}\bm{q}\; \int\frac{{\rm d}^{3}\bm{k}}{(2\pi)^{3}}\;
{\rm e}^{{\rm i}\bm{k}\cdot\left( \bm{s}-\bm{q}\right)}
\Biggl[ - \left( \epsilon - \Braket{\epsilon} \right){\rm i}\bm{k}\cdot\hat{\bm{q}}
\notag \\
& \quad
+ \left( 1 - \epsilon ({\rm i}\bm{k}\cdot\hat{\bm{q}}) + 2\frac{\Braket{\epsilon}}{s} \right)
\left( b^{\rm L}\delta_{\rm L} - {\rm i}k_{i} R_{ij}\Psi_{j} \right)
\Biggr] ~ .
\label{eq: delta s pre int}
\end{align}
Here, in computing the density field for galaxies/halos, we have to be careful of dealing with the term $\epsilon$ coming from the gravitational redshift effect. Although the term $\epsilon$ itself should be a small quantity, the gravitational potential at the halo/galaxy position would not be simply characterized by the gravitational potential of the linear density field. since the halos/galaxies are likely to be formed in the presence of a deep potential well through nonlinear processes, it should involve the non-perturbative contribution. Thus, following \citet{2020MNRAS.498..981S}, we decompose the gravitational redshift contribution $\epsilon$ into two pieces:
\begin{align}
\epsilon(\bm{x}) &= \epsilon_{\rm L}(\bm{x}) + \epsilon_{\rm NL}~. \label{eq: epsilon + NL epsilon}
\end{align}
In Eq.~\eqref{eq: epsilon + NL epsilon}, the first term at the right-hand side, $\epsilon_{\rm L}(\bm{x})$, represents the linear-order contribution arising from the gravitational potential of the linear density field, $\phi_{\rm L}$:
\begin{align}
\epsilon_{\rm L}(\bm{x}) &= - \frac{1}{aH}\phi_{\rm L}(\bm{x}) ~. \label{eq: epsilon L}
\end{align}
On the other hand, the second term, $\epsilon_{\rm NL}$ describes the non-perturbative contribution. In this paper, we shall model it with the universal halo density profile called NFW profile by \citet{1996ApJ...462..563N}, as adopted in \citet{2020MNRAS.498..981S}: 
\begin{align}
\epsilon_{\rm NL} = - \frac{1}{aH}\phi_{\rm NFW,0}(z,M) ~ \label{eq: phi NL}
\end{align}
with $\phi_{\rm NFW,0}$ being the halo potential of the NFW profile at the centre (see Appendix~D in \citet{2020MNRAS.498..981S} for the explicit form of the NFW potential $\phi_{\rm NFW,0}$). Here, we assume that the object to cross correlate resides at the halo centre. The potential impact of this assumption will be later discussed in Sec.~\ref{sec: systematics}.
Note that the non-perturbative potential contribution, $\epsilon_{\rm NL}$, is not a random variable but a constant value as a function of the halo mass and redshift through Eq.~\eqref{eq: phi NL}. Thus, we have $\langle\epsilon\rangle=\epsilon_{\rm NL}$.

Keeping the above points in mind, we substitute Eqs.~(\ref{eq: epsilon + NL epsilon}) and (\ref{eq: epsilon L}) into Eq.~(\ref{eq: delta s pre int}). After performing the integration by parts, the density fluctuation $\delta^{\rm(S)}$ is recast in the following form:
\begin{align}
\delta^{({\rm S})}(\bm{s}) &=
\delta^{(\rm std)}(\bm{s}) + \delta^{(\rm pot)}(\bm{s}) + \delta^{(\epsilon_{\rm NL})}(\bm{s}) ~. \label{eq: delta std + delta epsNL}
\end{align}
Here, we classify the density fluctuations into three contributions: the standard Doppler effects without assuming the plane-parallel limit, $\delta^{(\rm std)}(\bm{s})$, the gravitational redshift effect due to the linear density fields, $\delta^{(\rm pot)}(\bm{s})$, and gravitational redshift effect due to the non-linear halo potential, $\delta^{(\epsilon_{\rm NL})}(\bm{s})$. Those contributions are explicitly given by
\begin{align}
\delta^{(\rm std)}(\bm{s}) &\equiv \int\frac{{\rm d}^{3}\bm{k}}{(2\pi)^{3}}{\rm e}^{{\rm i}\bm{k}\cdot\bm{s}}
\Biggl[ b + f\mu_{k}^{2} - {\rm i} f \frac{2}{ks}\mu_{k} \Biggr]\delta_{\rm L}(\bm{k})
\label{eq: def delta std}
~, \\
\delta^{(\rm pot)}(\bm{s}) &\equiv \int\frac{{\rm d}^{3}\bm{k}}{(2\pi)^{3}}{\rm e}^{{\rm i}\bm{k}\cdot\bm{s}}
\Biggl[ \left( {\rm i}\,ks\mu_{k} + 2\right) \frac{\mathcal{M}}{sk^{2}} \Biggr]\delta_{\rm L}(\bm{k})
\label{eq: def delta grav}
~, \\
\delta^{(\epsilon_{\rm NL})}(\bm{s}) & \equiv 
\frac{\epsilon_{\rm NL}}{s} \int\frac{{\rm d}^{3}\bm{k}}{(2\pi)^{3}}{\rm e}^{{\rm i}\bm{k}\cdot\bm{s}}
\Biggl[
-1 + \mu^{2}_{k} - {\rm i}f\frac{2}{ks}\mu_{k}
\notag \\
& \quad
- {\rm i}\, b ks \mu_{k} -2f\mu^{2}_{k}- {\rm i}\frac{2}{ks}\mu_{k} - {\rm i} f ks \mu^{3}_{k}
\Biggr]\delta_{\rm L}(\bm{k}) ~,
\label{eq: def delta epsNL}
\end{align}
with the quantity $\mu$ being the directional cosine defined by $\mu_{k}\equiv \hat{\bm{s}}\cdot\hat{\bm{k}}$. Here, we introduced a new quantity $\mathcal{M} \equiv - 3\Omega_{\rm m0}H^{2}_{0}/(2a^{2}H)$. The quantity $b$ is the Eulerian linear bias parameter, which is related to the Lagrangian linear bias $b^{\rm L}$ through $b = 1 + b^{\rm L}$. Note that in the above, the gravitational potential $\phi_{\rm L}$ is rewritten with the linear density fields through the Poisson equation.
The linear-order contributions given in Eqs.~\eqref{eq: def delta std} and \eqref{eq: def delta grav} reproduce the results obtained previously if one neglects other minor contributions but keep the terms at the $O(aH/k)$ order (see, e.g., Eq.~(A7) in \citet{2014PhRvD..89h3535B} or Eq.~(1) in \citet{2017PhRvD..95d3530H}).

Eq.~\eqref{eq: delta std + delta epsNL} with Eqs.~\eqref{eq: def delta std}--\eqref{eq: def delta epsNL} is the key expression of our analytical model for the dipole cross-correlation function. As we will see in the next subsection, the resultant expression for the dipole moment involves only one dimensional integrals, and the prediction can be made much faster than that of the quasi-linear model by \citet{2020MNRAS.498..981S}, also reproducing the simulation results remarkably well. Hence, the present model can be used to systematically explore the dependence of various parameters characterizing the properties of galaxies as well as the setup of upcoming/ongoing surveys.

\subsection{Cross-correlation function}
\label{sec: correlation function}

\begin{figure}
\centering
\includegraphics[width=0.5\columnwidth]{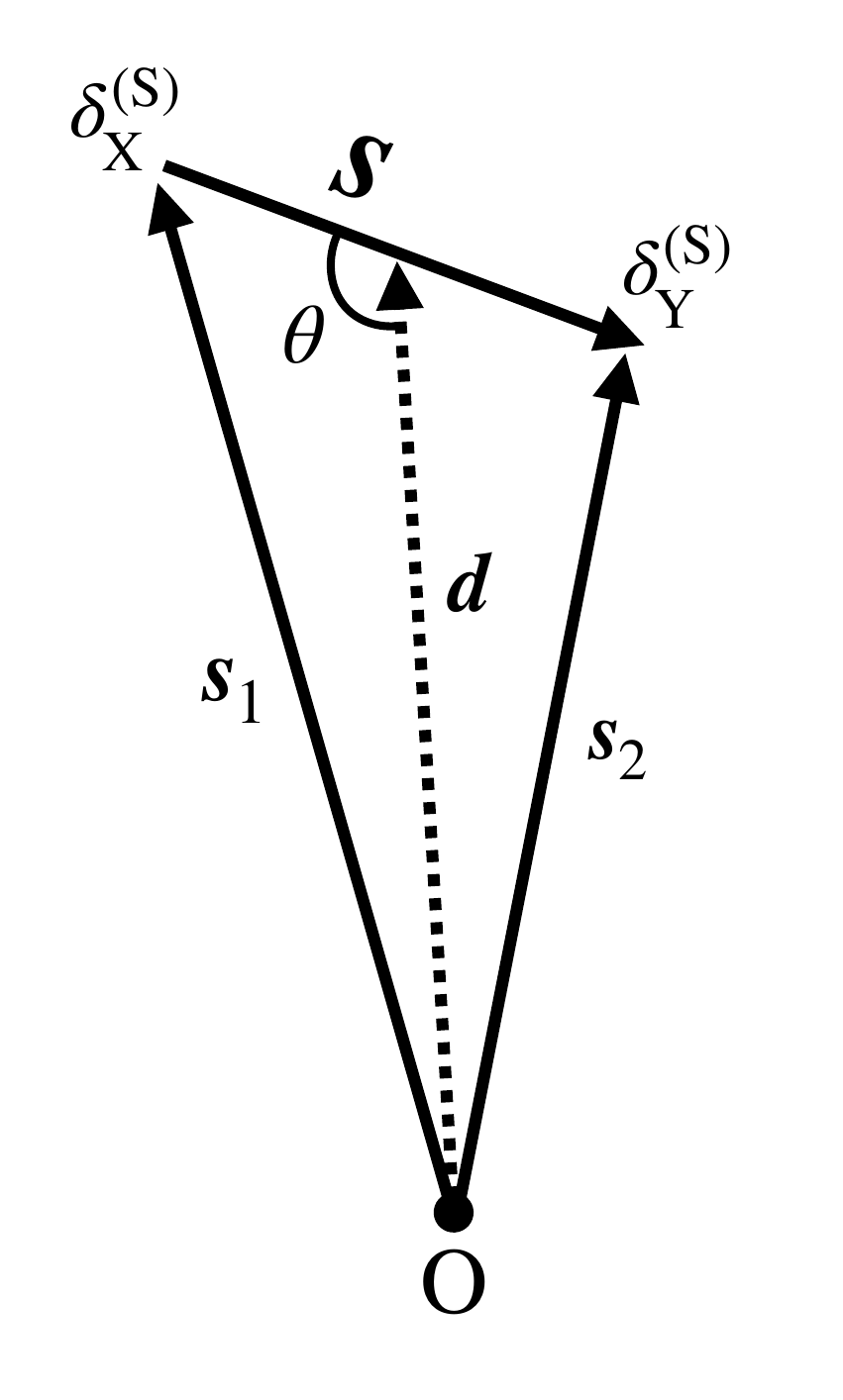}
\caption{The geometric configuration of the cross-correlation function in redshift space.
The biased objects $\delta^{(\rm S)}_{\rm X}$ and $\delta^{(\rm S)}_{\rm Y}$ are, respectively, observed at $\bm{s}_{1}$ and $\bm{s}_{2}$ with respect to the observer (O). Here we assume $b_{\rm X} > b_{\rm Y}$.
The separation vector, line-of-sight vector, and directional cosine are defined by $\bm{s} = \bm{s}_{2} - \bm{s}_{1}$, $\bm{d} = (\bm{s}_{1} + \bm{s}_{2} )/2$, and $\mu \equiv \cos{\theta} = \hat{\bm{s}}\cdot\hat{\bm{d}}$ respectively.
}
\label{fig: configuration}
\end{figure}

We now compute the cross-correlation function and derive an analytical expression for the dipole moment. In doing so, we explicitly write the density field for the objects ${\rm X}$ as $\delta^{\rm(S)}_{\rm X}$. Then, the cross-correlation function between different species ${\rm X}$ and ${\rm Y}$ is given by
\begin{align}
\xi_{\rm XY}(\bm{s}_1,\,\bm{s}_2) \equiv
\Braket{\delta^{({\rm S})}_{\rm X}(\bm{s}_{1}) \delta^{({\rm S})}_{\rm Y}(\bm{s}_{2})} ~, \label{eq: def correlation function}
\end{align}
Taking the directional dependence of the observer's line of sight into account, the statistical homogeneity and isotropy no longer hold, and the cross-correlation function given above cannot be simply characterized as a function of the separation $s=|\bm{s}_2-\bm{s}_1|$. Rather, it also depends on the distances to the objects X and Y, i.e., $|\bm{s}_1|$ and $|\bm{s}_2|$. Equivalently, the function $\xi_{\rm XY}$ is characterized as a function of the separation $s$, the mid-point distance $d=|\bm{s}_1+\bm{s}_2|/2$, and the directional cosine between the separation vector and the mid-point vector, $\mu\equiv\hat{s}\cdot\hat{d}$, with separation vector defined by $\bm{s}\equiv\bm{s}_2-\bm{s}_1$ (see Fig.~\ref{fig: configuration} for the geometric configuration of the cross-correlation function). We shall below write the explicit dependence of $\xi_{\rm XY}$ in its argument as $\xi_{\rm XY}(s,\,d,\,\mu)$.

Substituting Eq.~(\ref{eq: delta std + delta epsNL}) into Eq.~(\ref{eq: def correlation function}), the cross-correlation function $\xi_{\rm XY}$ is given as a collection of several pieces. Since the terms coming from the gravitational redshift effect, i.e., $\delta^{\rm(pot)}$ and $\delta^{\rm(\epsilon_{\rm NL})}$, are supposed to be sub-dominant compared to the standard Doppler term, we can neglect the contributions from their cross talks. We then have 
\begin{align}
\xi_{\rm XY}(s,d,\mu)& \simeq
\Braket{\delta^{(\rm std)}_{\rm X}(\bm{s}_{1})\delta^{(\rm std)}_{\rm Y}(\bm{s}_{2})}
\nonumber
\\
&+ \Bigl\{\Braket{\delta^{(\rm std)}_{\rm X}(\bm{s}_{1})\delta^{(\rm pot)}_{\rm Y}(\bm{s}_{2})}+\Braket{\delta^{(\rm pot)}_{\rm X}(\bm{s}_{1})\delta^{(\rm std)}_{\rm Y}(\bm{s}_{2})}\Bigr\}
\nonumber
\\
& + \Bigl\{
\Braket{\delta^{(\epsilon_{\rm NL})}_{\rm X}(\bm{s}_{1}) \delta^{(\rm std)}_{\rm Y}(\bm{s}_{2})}
+\Braket{\delta^{\rm (std)}_{\rm X}(\bm{s}_{1}) \delta^{(\epsilon_{\rm NL})}_{\rm Y}(\bm{s}_{2})}
\Bigr\}
\nonumber
\\
&\equiv \xi^{(\rm std)}_{\rm XY}(s,d,\mu) + \xi^{(\rm pot)}_{\rm XY}(s,d,\mu) + \xi^{(\epsilon_{\rm NL})}_{\rm XY}(s,d,\mu) ~.
 \label{eq: xi std+rel+eps}
\end{align}
Since we are particularly interested in the dipole moment of the cross-correlation function, we hereafter consider the multipole expansion of the $\xi_{\rm XY}$, taking specifically the mid-point vector, $\bm{d}=(\bm{s}_1+\bm{s}_2)/2$, as the line-of-sight direction: 
\begin{align}
\xi_{{\rm XY}, \ell}(s,d)
& = \frac{2\ell+1}{2}\int^{1}_{-1}{\rm d}\mu\; \xi_{{\rm XY}}(s,d,\mu) \mathcal{L}_{\ell}(\mu)~,
\label{eq: multipole expansion}
\\
&\equiv \xi^{(\rm std)}_{\rm XY,\ell}(s,d) + \xi^{(\rm pot)}_{\rm XY,\ell}(s,d) + \xi^{(\epsilon_{\rm NL})}_{\rm XY,\ell}(s,d) ~, \label{eq: xi1ell std+grav+eps}
\end{align}
with $\mathcal{L}_\ell$ being the Legendre polynomials. Notice that the line-of-sight direction considered here is directional-dependent. Since we do not take the plane-parallel limit, the wide-angle effect comes to play, and the multipole moment of the correlation function, $\xi_{{\rm XY},\ell}$, is not simply given as a function of the separation, but rather given as a bi-variate function of $s$ and $d$. In order to isolate the scale (i.e., separation) dependence of the multipole moment from the line-of-sight dependence, we further expand the multipole moments in powers of $(s/d)$ as follows:
\begin{align}
\xi_{{\rm XY}, \ell}(s,d) &=
\xi^{\rm pp}_{{\rm XY}, \ell}(s) + \left( \frac{s}{d}\right) \xi^{\rm wa}_{{\rm XY}, \ell}(s) + O\left( \left( \frac{s}{d} \right)^{2} \right) ~.\label{eq: wa expansion app}
\end{align}
The first and second terms at the right-hand side, respectively, represent the contributions from the plane-parallel limit $d\to \infty$ and the leading-order wide-angle correction.
In Appendix~\ref{app: derivation}, substituting Eqs.~\eqref{eq: def delta std}--\eqref{eq: def delta epsNL} into Eq.~(\ref{eq: xi std+rel+eps}), the multipole expansion is applied up to the plane-parallel limit and wide-angle correction, and the terms defined above are derived in each contribution. The resultant expressions for the dipole moment $(\ell=1)$, including only the non-vanishing contributions, are summarized as follows (see Appendix~\ref{app: other moments} for other multipoles):
\begin{align}
 \xi^{(\rm std)}_{\rm XY,1}(s,d) &= \Bigl(\frac{s}{d}\Bigr)\, 2f (b_{\rm X}-b_{\rm Y}) \left( \Xi^{(1)}_{1}(s) - \frac{1}{5} \Xi^{(0)}_{2}(s) \right) + \mathcal{O}\left( \left( \frac{s}{d} \right)^{2} \right) ,
 \label{eq: xi1 std}
 \\
 \xi^{(\rm pot)}_{\rm XY,1}(s,d) &=- (b_{\rm X}-b_{\rm Y}) \mathcal{M}\,s \,\Xi^{(1)}_{1}(s) 
 +\mathcal{O}\left( \left( \frac{s}{d} \right)^{2} \right) ~, \label{eq: xi1 grav}
 \\
 \xi^{(\epsilon_{\rm NL})}_{\rm XY,1}(s,d) & = - \frac{1}{s} (\epsilon_{\rm NL,X}-\epsilon_{\rm NL,Y})
\notag 
\\
& \qquad\times \left( b_{\rm X}b_{\rm Y} + \frac{3}{5}(b_{\rm X}+b_{\rm Y})f + \frac{3}{7}f^{2}\right) \Xi^{(-1)}_{1}(s)
\notag \\
& \quad +\mathcal{O}\left( \left( \frac{s}{d} \right)^{2} \right) 
~, 
\label{eq: xi1 eps}
\end{align}
with the function $\Xi^{(n)}_{\ell}$ defined by
\begin{align}
\Xi^{(n)}_{\ell}(s) \equiv \int \frac{k^{2}\, {\rm d}k}{2\pi^{2}}\, \frac{j_{\ell}(ks)}{(ks)^{n}} P_{\rm L}(k) ~,
\end{align}
where the functions $j_{\ell}$ and $P_{\rm L}(k)$ are, respectively, the spherical Bessel function and the linear power spectrum defined in Eq.~(\ref{eq: def PL(k)}).

The analytical expressions at Eqs.~(\ref{eq: xi1 std})--(\ref{eq: xi1 eps}) are one of the main result in the present paper. As we see, the expressions of the dipole moment involve only one dimensional integrals, and for a given redshift $z$, they are characterized by the (Eulerian) bias parameters $b_{\rm X/Y}$ and the non-perturbative halo potentials $\epsilon_{\rm NL,X/Y}$, the latter of which are predicted with the NFW profile for given halo masses. 
We note that, in the derivations above, the magnification bias caused by the fact that the galaxy samples are flux limited is ignored~\citep[see e.g.,][]{2014PhRvD..89h3535B,2017PhRvD..95d3530H}. In Appendix~\ref{app: magnification}, the impact of the magnification bias, particularly induced by the Doppler effect (potentially the most dominant contribution), is discussed in detail, showing that such an effect is sub-dominant, and becomes negligibly small at higher redshifts ($z\gtrsim 0.1$).

\begin{figure}
\centering
\includegraphics[width=0.85\columnwidth]{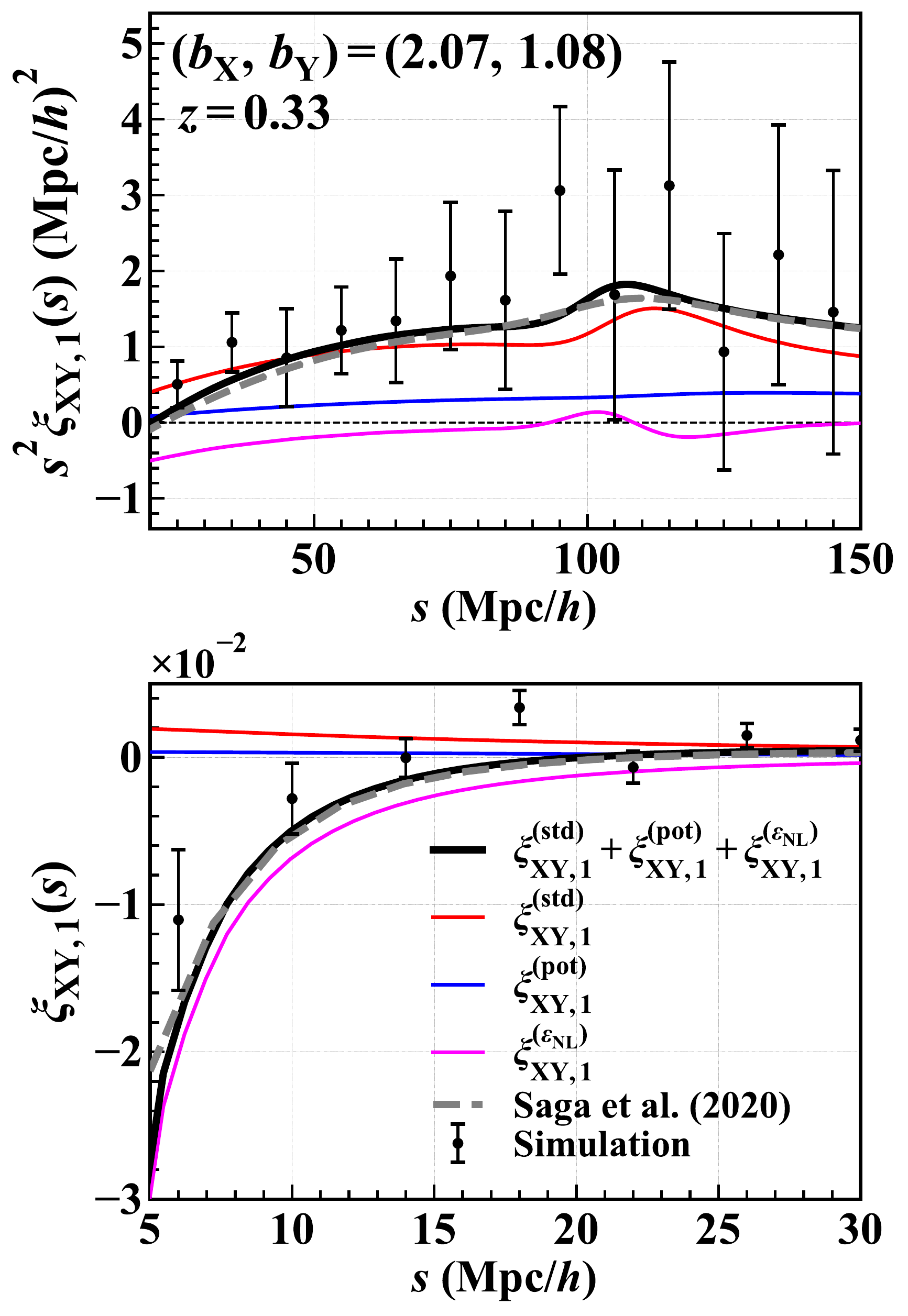}
\caption{Dipole moment of the cross-correlation function between halos having different bias parameters on large ({\it top}) and small ({\it bottom}) scales. The results of analytical model predictions presented in this paper are particularly shown at $z=0.33$, together with the measured results from the halo catalogues, RayGalGroupSims, in which all possible special and general relativistic effects arising from the light propagation in an inhomogeneous universe are consistently taken into account (filled circles with errorbars). Note that in the upper panel, to clarify the large-scale behaviour, the dipole moment multiplied by the square of separation, i.e., $s^2\xi_{\rm XY,1}$, is plotted. In each panel, 
black solid lines are the predictions of the analytical model (see Eq.~(\ref{eq: xi1ell std+grav+eps}) with Eqs.~(\ref{eq: xi1 std})--(\ref{eq: xi1 eps})). The 
coloured solid lines show the breakdown of these predictions, and the red, blue, and magenta respectively represent the contributions from the standard 
Doppler ($\xi_{\rm XY,1}^{\rm (std)}$, Eq.~(\ref{eq: xi1 std})), the gravitational redshift from linear-order potential ($\xi_{\rm XY,1}^{\rm (pot)}$, Eq.~(\ref{eq: xi1 grav})), and the gravitational redshift from the non-perturbative halo potential ($\xi_{\rm XY,1}^{(\epsilon_{\rm NL})}$, Eq.~(\ref{eq: xi1 eps})). For reference, we also plot the predictions based on \citet{2020MNRAS.498..981S} (gray dashed), in which the dipole cross correlation is computed based on the Zel'dovich approximation by performing numerically seven dimensional integrals. In all predictions, we adopt the bias parameters and halo masses of the data {\tt data$\_$H$_{1600}$} and {\tt data$\_$H$_{100}$}, listed Table 1 of \citet{2020MNRAS.498..981S}, and the potentials at the halo centre are predicted to be $\overline{\phi}_{\rm NFW,0, X} = -1.63\times10^{-5}$ and $\overline{\phi}_{\rm NFW,0, Y} = -0.285\times10^{-5}$ (bias parameters are also indicated in the upper panel).
In the top panel, the horizontal black dotted line represents $\xi_{{\rm XY},1} = 0$.
}
\label{fig: dipole}
\end{figure}

To see the quantitative behaviour of our model presented here, in Fig.~\ref{fig: dipole}, 
the predictions of the dipole moment of the cross-correlation function, $\xi_{\rm XY,1}$, are plotted. The results at $z=0.33$ are particularly shown, and for comparison, we also plot the measured results from the simulated halo catalogue, RayGalGroupSims\footnote{\url{https://cosmo.obspm.fr/public-datasets/}}, which consistently take into account all the relativistic corrections by solving the geodesic equation in the presence of matter inhomogeneities. Here, the plotted results show the cross-correlation between the halos of {\tt data$\_$H$_{1600}$} and {\tt data$\_$H$_{100}$}, whose bias parameters are respectively given by $b_{\rm X}=2.07$ and $b_{\rm Y}=1.08$.
In each halo sample, the potentials at the halo centre are predicted to be $\overline{\phi}_{\rm NFW,0, X} = -1.63\times10^{-5}$ and $\overline{\phi}_{\rm NFW,0, Y} = -0.285\times10^{-5}$. These values are taken from Table 1 of \citet{2020MNRAS.498..981S}, assuming the NFW profile. We use them to estimate the size of the gravitational redshift effect at each halo, $\epsilon_{\rm NL,X/Y}$, and obtain $\epsilon_{\rm NL,X}>\epsilon_{\rm NL,Y}>0$. 

In Fig.~\ref{fig: dipole}, the black solid lines are the predictions of our analytical model. Also, their building blocks, i.e., $\xi_{{\rm XY},1}^{\rm (std)}$, $\xi_{{\rm XY},1}^{\rm (pot)}$, and $\xi_{\rm XY,1}^{\rm (\epsilon_{\rm NL})}$, are separately plotted as red, blue, and magenta lines. The predicted behaviours of the dipole moment reproduce the simulation result including all the relativistic corrections well at both large and small scales. Also, it is rather close to the predictions based on the quasi-linear model of \citet{2020MNRAS.498..981S}, depicted as grey dashed lines. Thus, our present model not only successfully explain the overall trend, but also quantitatively describe the halo cross-correlation both at small and large scales. Hence, we can use it for a quantitative study on the detectability of the gravitational redshift effect. Finally, we note that the dipole moment of the cross-correlation function is dominated by the standard Doppler effect at large scales, while the gravitational redshift effect turns to be dominant at small scales, leading to the sign flip of the amplitude of $\xi_{\rm XY,1}$ at $s\approx 20$--$30\,h^{-1}$\,Mpc. Thus, these behaviours play a crucial role to detect the gravitational redshift effect, and in this respect, the predictions beyond linear scales would be indispensable.

\section{Covariance matrix}
\label{sec: cov}

In estimating the signal-to-noise ratio of the relativistic dipole in the upcoming surveys, the covariance matrix between different scales plays a crucial role. This is in particular the case for the statistics defined in the configuration space as we consider. In this paper, to compute the covariance matrix, we adopt the formalism developed by \citet{2016JCAP...08..021B,2017PhRvD..95d3530H}. This is a generalization of the previous formulae for the Gaussian covariance \citep[e.g.,][]{2009MNRAS.400..851S,2016MNRAS.457.1577G,2006NewA...11..226C} to include the anisotropies in the correlation function and multi-tracer technique, taking also the orientation-dependent weight function into account. 
In Sec.~\ref{sec: formula}, we present their analytical formulae for the covariance matrix. We then estimate the covariance matrix, specifically focusing on the dipole cross-correlation, in Sec.~\ref{sec: covariance numerical}.

\subsection{Covariance matrix of dipole cross-correlation function}
\label{sec: formula}

To give the analytical formulae for the Gaussian covariance, let us first define
the estimator for the dipole moment of the cross-correlation function. Here, we assume that the cross-correlation function can be written as a function of the separation between two objects, $\bm{s}$. This assumption is validated if we take the plane-parallel limit:
\begin{align}
\hat{\xi}_{{\rm XY}, 1}(s) = \frac{3}{2} \int^{1}_{-1}{\rm d}\mu\, \mu\, \int\frac{{\rm d}^{3}\bm{r}}{V}
\delta_{\rm X}(\bm{r}-\bm{s}/2) \delta_{\rm Y}(\bm{r}+\bm{s}/2)
~, \label{eq: estimator xi1}
\end{align}
where the quantities $V$ and $\delta_{\rm X/Y}$ are respectively the survey volume and the measured density fluctuation of the objects X/Y. The quantity $\mu$ is the directional cosine between the (fixed) line-of-sight $\hat{\bm{z}}$ and separation vectors defined by $\mu = \hat{\bm{s}}\cdot\hat{\bm{z}}$. It is to be noted that while the wide-angle effect indeed comes to play an important role in the signal part, 
its impact on the covariance matrix has been shown to be negligible at the scales below $190\,{\rm Mpc}/h$ \citep[][]{2018JCAP...05..043L}.

Taking the contribution arising from the discreteness of the galaxy samples into consideration, the ensemble average of the quadrature, $\delta_{\rm X}(\bm{r}_1)\delta_{\rm Y}(\bm{r}_2)$, becomes 
\begin{align}
\Braket{\delta_{\rm X}(\bm{r}_{1}) \delta_{\rm Y}(\bm{r}_{2})} = \xi_{\rm XY}(\bm{r}_{2}-\bm{r}_{1}) + \frac{\delta^{\rm K}_{\rm X,Y}}{n_{\rm X}}\delta_{\rm D}(\bm{r}_{2}-\bm{r}_{1}) ~,
\label{eq:def_correlation}
\end{align}
where the quantity $\delta_{\rm X,Y}^{\rm K}$ is the Kronecker's delta and the function $\delta_{\rm D}$ is the Dirac's delta function. 
The first term, $\xi_{\rm XY}$, represents the cross-correlation function arising purely from the intrinsic clustering properties. The second term characterizes the contribution from the Poisson sampling process, which becomes non-vanishing only in the self-correlation case (i.e., ${\rm X}={\rm Y}$ and $\bm{r}_1=\bm{r}_2$). Using the expression at Eq.~(\ref{eq:def_correlation}), the estimator given at Eq.~(\ref{eq: estimator xi1}) is shown to be an unbiased estimator of the dipole cross-correlation, i.e., $\Braket{\hat{\xi}_{{\rm XY}, 1}(s) }=\xi_{{\rm XY}, 1}(s)$ unless $X=Y$ and $s=0$. 

We then define the covariance of the dipole moment as follows:
\begin{align}
{\rm COV}(s,s')
\equiv
\Braket{\hat{\xi}_{\rm XY, 1}(s)\hat{\xi}_{\rm XY, 1}(s')} - \Braket{\hat{\xi}_{\rm XY, 1}(s)} \Braket{\hat{\xi}_{\rm XY, 1}(s')} ~ .
\end{align}
With the definition given above, \citet{2017PhRvD..95d3530H} derived the analytical formula for the covariance, which only involves one dimensional integrals:
\begin{align}
{\rm COV}(s,s')
& = \frac{9}{V}\int\frac{k^{2}{\rm d}k}{2\pi^{2}}\; j_{1}(ks)j_{1}(ks')
\notag \\
&\quad \times 
\sum_{\ell_{1},\ell_{2}}G^{\ell_{2}\ell_{1}}_{11}\Biggl( P_{\rm XX, \ell_{1}}P_{\rm YY, \ell_{2}} - P_{\rm XY, \ell_{1}}P_{\rm XY, \ell_{2}}\Biggr)
\notag \\
& 
+ \frac{3}{V}\int\frac{k^{2}{\rm d}k}{2\pi^{2}}\; j_{1}(ks)j_{1}(ks')
\notag \\
& \quad
\times \Biggl[
\left( P_{\rm XX, 0} + \frac{2}{5}P_{\rm XX, 2} \right)\frac{1}{n_{\rm Y}} 
+ \left( P_{\rm YY, 0} + \frac{2}{5}P_{\rm YY, 2} \right) \frac{1}{n_{\rm X}}
\Biggr]
\notag \\
&
+ \frac{\delta^{\rm K}_{s,s'}}{4\pi s^{2} L_{\rm p}}\frac{3}{n_{\rm X}n_{\rm Y}V} ~,
\label{eq: dipole covariance}
\end{align}
where we define the square pixels of the side-length $L_{\rm p}$.
The coefficient $G^{\ell_{2}\ell_{1}}_{11}$ is defined by
\begin{align}
G^{\ell_{2}\ell_{1}}_{\ell'\ell}
= \sum_{\ell_{3}}(2\ell_{3}+1)
\left(
\begin{array}{c c c}
\ell_{1} & \ell_{2} & \ell_{3}\\
0 & 0 & 0
\end{array}
\right)^{2}
\left(
\begin{array}{c c c}
\ell & \ell' & \ell_{3}\\
0 & 0 & 0
\end{array}
\right)^{2}
~ . \label{eq: Gell}
\end{align}
The functions $P_{\rm XY, \ell}$ are the Fourier counterparts of the multipole correlation function in the plane-parallel limit:
\begin{align}
\xi_{\rm XY, \ell}(s) &= (-{\rm i})^{\ell}\int \frac{k^{2}\, {\rm d}k}{2\pi^{2}}\,
P_{\rm XY, \ell}(k,z)j_{\ell}(ks)
~. \label{eq: P ell pp}
\end{align}

In Eq.~(\ref{eq: dipole covariance}), the covariance matrix consists of the three contributions. The first term at the right-hand side represents the contributions arising purely from the cosmic variance, which we call the CV$\times$CV term. On the other hand, the second term describes the cross-talk between the cosmic variance and Poisson noise, and the third term originates from the Poisson noise. We respectively call these two terms the CV$\times$P and the P$\times$P terms. It is to be noted that for the CV$\times$CV term, the summation over the non-zero even multipoles $\ell_1$ and $\ell_2$ leads to \citep[][]{2016JCAP...08..021B}
\begin{align}
\sum_{\ell_{1},\ell_{2}={\rm even}}G^{\ell_{2}\ell_{1}}_{11}\Bigl( P^{({\rm std})}_{{\rm XX},\ell_{1}}P^{({\rm std})}_{{\rm YY},\ell_{2}} - P^{({\rm std})}_{{\rm XY},\ell_{1}}P^{({\rm std})}_{{\rm XY},\ell_{2}}\Bigr) = 0 ~.
\label{eq: cancel CVCV}
\end{align}
This cancellation shows that the even multipoles of the standard Doppler terms do not contribute to the CV$\times$CV term. On the other hand, the CV$\times$P term contains the non-vanishing even multipoles coming from the standard Doppler terms. These suggest that the CV$\times$CV term is a sub-dominant contribution to the covariance matrix. Indeed, as we will see later, the covariance matrix is mostly dominated by the two terms, CV$\times$P and P$\times$P, with a negligible contribution of the CV$\times$CV term.

To sum up, Eq.~(\ref{eq: dipole covariance}) is the covariance matrix of the dipole cross-correlation function used in the subsequent analysis. Given the multipole power spectra $P_{{\rm XX},\ell}$, $P_{{\rm YY},\ell}$ and $P_{{\rm XY},\ell}$, the covariance matrix ${\rm COV}(s,s')$ is characterized by the number densities of the objects X and Y (i.e., $n_{\rm X}$ and $n_{\rm Y}$), the side-length of the square pixel $L_{\rm p}$, and the survey volume $V$. In what follows, we follow \citet{2018JCAP...05..043L}, and set the pixel size $L_{\rm p}$ to $2\, {\rm Mpc}/h$. Note that the choice of this parameter does not change the results significantly as long as we consider the scales above $L_{\rm p}$. 
Ignoring the survey masks and window functions, the survey volume of a hypothetical galaxy survey with the fractional sky coverage $f_{\rm sky}$ and redshift width $\Delta z$ is expressed as $V = (4\pi/3)f_{\rm sky}\left\{ r^{3}(z + \Delta z/2) - r^{3}(z - \Delta z/2)\right\}$, with $z$ being the mean redshift. Here, the function $r(z)$ represents the comoving distance at redshift $z$. Thus, provided the survey specification parameters (i.e., $n_{\rm X/Y}$, $z$, $\Delta z$), the remaining pieces in estimating the covariance matrix are the multipole auto- and cross-power spectra, which are characterized in our model of cross-correlation function by the linear bias parameters $b_{\rm X/Y}$ and the non-perturbative potentials $\phi_{\rm NL,X/Y}$ for a given cosmological model. In Appendix \ref{app: other moments}, we present the explicit expressions for the multipole power spectra. Since we ignored the wide-angle effect to derive the covariance matrix above, it is sufficient to consider the contributions from the plane-parallel limit, summarized in Appendix \ref{app:plane-parallel_limit}.

\subsection{Numerical results of the dipole covariance}
\label{sec: covariance numerical}

In this subsection, before computing the signal-to-noise ratio for upcoming surveys, we shall elucidate the basic properties of the covariance matrix. As we saw in the previous section, the covariance matrix ${\rm COV}(s,s')$ includes several parameters characterizing both the galaxy survey and intrinsic clustering properties. In order to relate these parameters, we adopt the halo model, and compute the covariance of the halo cross-correlation function. For halos in the mass range $[M-\Delta M/2,M+\Delta M/2]$, the model predicts the number density $n$ and the bias parameter $b$ from the halo mass function, for which we use the fitting form given by \citet{1999MNRAS.308..119S}. Further, through the NFW profile, the non-perturbative potential at the halo centre $\phi_{\rm NFW,0}$ is also predicted. In other words, given the halo bias and number density, the mass of halos and the width of mass range are determined uniquely, from which one can estimate the central halo potential\footnote{In the actual computation, the width of the halo mass $\Delta M$ turns out to be narrow enough so that the bias parameter and halo potential averaged over the halo mass range $[M-\Delta M/2,\,M+\Delta M/2]$ are simply replaced with those evaluated at the central halo mass, $M$, i.e., $\Braket{b}\simeq b(M)$ and $\Braket{\phi_{\rm NFW,0}}\simeq \phi_{\rm NFW,0}(M)$. Also, the number density of halos can be approximately estimated by the halo mass function $dn/dM$ multiplied by the width of halo mass, i.e., $n\simeq (dn/dM)\Delta M$. }.

With the halo model prescription mentioned above, we set the bias parameters and number densities for the halo populations X and Y to $(b_{\rm X},\,n_{\rm X})=(2.5,\,3\times 10^{-4}\,({\rm Mpc}/h)^{-3})$ and $(b_{\rm Y},\,n_{\rm Y})=(1.5,\, 10^{-3}\,({\rm Mpc}/h)^{-3})$. These are representative values among various upcoming surveys summarized in Appendix~\ref{app: future surveys}. Then, in Fig.~\ref{fig: cov fixz}, the covariance matrix of the dipole cross-correlation function is plotted as a function of the separation, focusing specifically on the diagonal component, i.e., $s=s'$. Here, we consider a hypothetical full-sky survey $(f_{\rm sky}=1)$ having the redshift width $\Delta z=0.1$, varying the central redshift from $0.1$ (purple) to $1.7$ (yellow). Dividing the diagonal covariance into the three contributions, the results normalized by the dipole moment squared, i.e., ${\rm COV}/(\xi_1)^2$, are separately shown: CV$\times$CV (left), CV$\times$P (middle), and P$\times$P (right). That is, ignoring the off-diagonal components of the covariance matrix, Fig.~\ref{fig: cov fixz} effectively represents the inverse of the square of the signal-to-noise ratio for a fixed separation. Indeed, the off-diagonal components of the covariance matrix are shown to play a minor role, and the estimated signal-to-noise mostly come from the diagonal components, as we will see later in Sec.~\ref{sec: SN smin zred}. 

\begin{figure*}\centering
\includegraphics[width=0.8\textwidth]{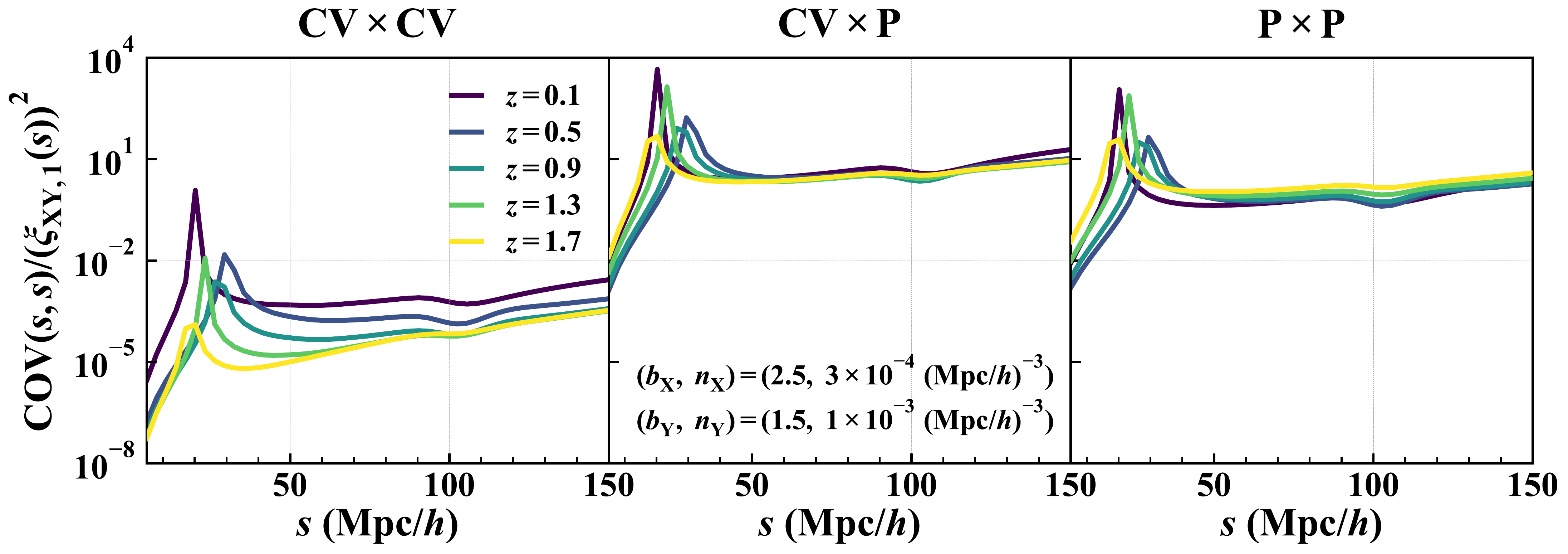}
\caption{
Diagonal components of the covariance matrix divided by the square of the dipole cross-correlation at various redshifts, plotted as a function of the separation $s$.
From {\it left} to {\it right}, we present the contributions of the CV$\times$CV term, the CV$\times$P term, and the P$\times$P term, respectively.
The depth of redshift and fractional sky coverage are set to $\Delta z = 0.1$ and $f_{\rm sky} = 1$, respectively. We choose the bias parameter and number density indicated in the middle panel, which are the typical values of upcoming surveys.
Note that the sharp feature near $s\approx 20$--$30\, {\rm Mpc}/h$ arises from the zero-crossing of the dipole moment.
}
\label{fig: cov fixz}
\end{figure*}

In Fig.~\ref{fig: cov fixz}, in all three cases, 
the normalized covariance stays almost constant at large scales, $s\gtrsim40$\,Mpc$/h$, where no clear redshift dependence is seen. On the other hand, at the scales of $s=20$--$40$\,Mpc$/h$, we see a sharp peak. This characteristic feature merely comes from the denominator, $(\xi_1)^2$, which exhibits the zero crossing, as shown in Fig.~\ref{fig: dipole}. In \citet{2020MNRAS.498..981S}, the zero-crossing point where the amplitude of the dipole moment eventually flips the sign is shown to scale as $b_{\rm X}b_{\rm Y}/(b_{\rm X}-b_{\rm Y})|\Delta\phi_{\rm NL}|\{H_0(1+z)/H(z)\}$, with $\Delta\phi_{\rm NL}$ defined by $\Delta\phi_{\rm NL}\equiv \phi_{\rm NFW,0,X}-\phi_{\rm NFW,0,Y}$. For halos considered here, the zero-crossing point typically appears at $s\approx 20$--$40$\,Mpc$/h$ for the redshifts $0.1\leq z\leq 1.7$. Below this scale, the normalized covariance starts to fall off, and a rather clear redshift dependence becomes manifest, compared to the one at large scales. This implies that the signal-to-noise ratio of the dipole moment would be dominated by the behaviour below the zero-crossing point. Although these features are common in all three panels, the amplitude of the ratio for the CV$\times$CV (left) is substantially smaller than the other two contributions, meaning that the contribution coming from the cosmic variance is sub-dominant in the covariance matrix of the dipole moment. This is consistent with what was discussed in the previous section (see Eq.~(\ref{eq: cancel CVCV}) below). The results of Fig.~\ref{fig: cov fixz} thus show that the detectability of the relativistic dipole is mostly governed by the covariance structure of the CV$\times$P and P$\times$P terms below the zero-crossing point. 

\begin{figure}
\centering
\includegraphics[width=0.8\columnwidth]{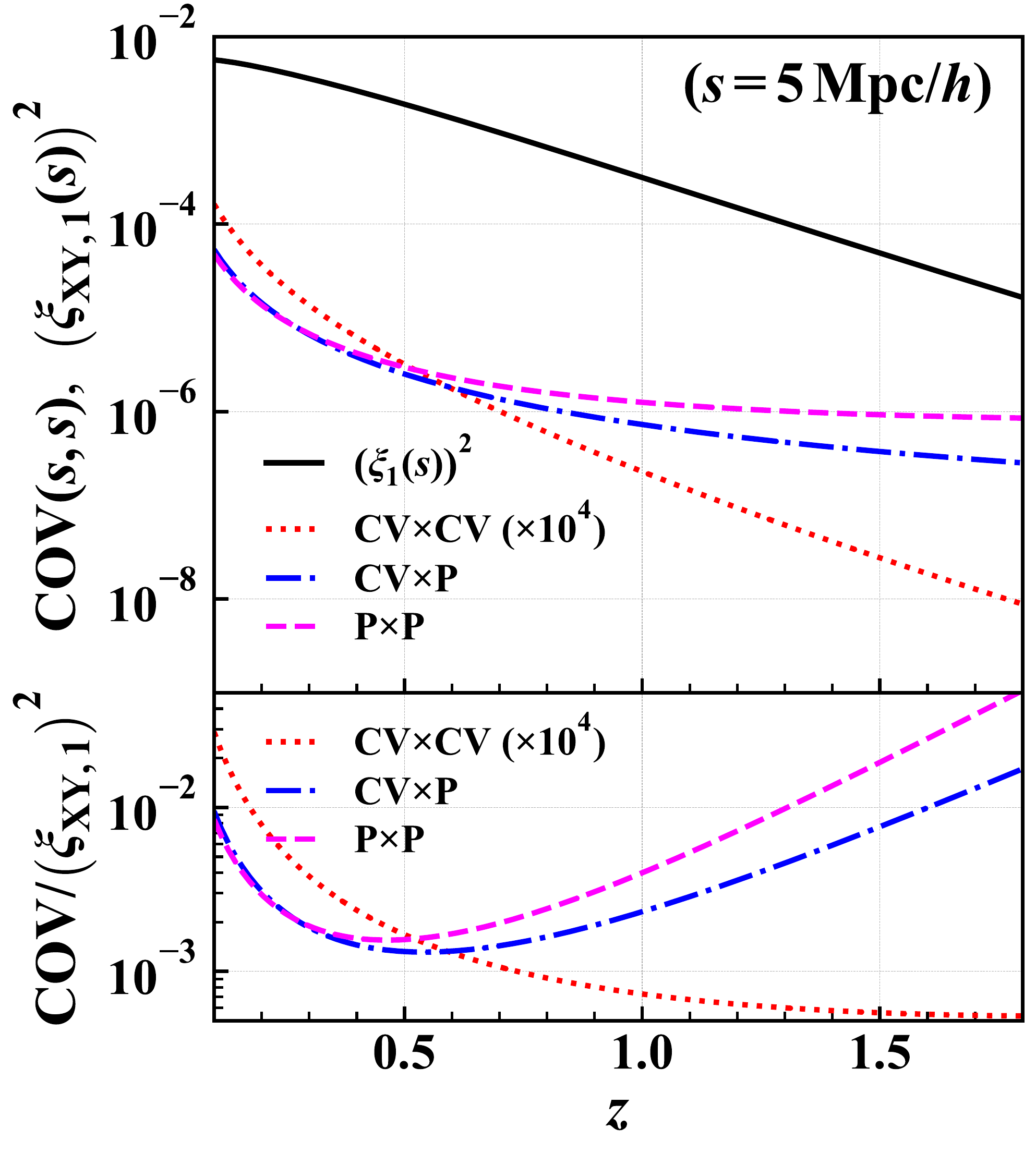}
\caption{
({\it Top}) Redshift dependence of the diagonal components of the covariance matrix, fixing the separations to $s=s'=5\, {\rm Mpc}/h$.
Contributions from CV$\times$CV~(red dotted), CV$\times$P~(blue dot-dashed), and P$\times$P~(magenta dashed) terms are separately plotted. For comparison, the square of the dipole moment, $\left( \xi_{1}(s) \right)^{2}$, is also shown (black solid).
({\it Bottom}) Redshift dependence of the ratio, ${\rm COV}(s,s)/(\xi_{\rm XY,1}(s))^{2}$ at $s=5\, {\rm Mpc}/h$, with contributions from CV$\times$CV, CV$\times$P and P$\times$P separately plotted.
In both panels, the contributions from CV$\times$CV are multiplied by $10^{5}$ for clarity.
The depth of redshift, fractional sky coverage, bias, and number density are chosen to be the same as in Fig.~\ref{fig: cov fixz}.
}
\label{fig: cov fixs}
\end{figure}

In Fig.~\ref{fig: cov fixs}, to see more clearly the redshift dependence of the normalized covariance at small scales, we fix the separation $s$ to $5\,{\rm Mpc}/h$, and plot the three contributions as a function of the redshift, again focusing on the diagonal components of the covariance matrix. The upper panel of Fig.~\ref{fig: cov fixs} shows the diagonal components of the covariance matrix and the square of the dipole moment, while the lower panel plots their ratios. It is to be noted that the ratio ${\rm COV}/(\xi_1)^2$ exhibit a non-monotonic behaviour. That is, the result of each contribution first decreases with the redshift, and then turns to increase at $z\gtrsim0.5$. These behaviours come from the competition of the redshift dependence between the numerator and denominator, as is explicitly shown in the upper panel. Due to the survey volume dependence of the covariance matrix dominated by the P$\times$P term, the numerator rapidly decreases at $z\lesssim0.5-1$, but beyond that, it asymptotically approaches a constant value. On the other hand, the denominator, $(\xi_1)^2$, monotonically decreases its amplitude through the redshift evolution of the linear growth factor and the halo potential at the centre. Thus, taking the ratio, ${\rm COV}/(\xi_1)^2$, yields a non-trivial behaviour which takes a minimum value around $z\approx 0.5$. Although Fig.~\ref{fig: cov fixs} shows a part of the covariance matrix, the trends seen in the diagonal component generically appear in the signal-to-noise ratio for various survey setup, and these indeed dominate the behaviours of the signal-to-noise ratio, as we will see later.

\section{Results: Estimating signal-to-noise ratio in upcoming surveys}
\label{sec: result}

Provided the analytical model describing the relativistic dipole and the covariance matrix in the previous section, we are in a position to estimate the signal-to-noise ratio of the relativistic dipole. We define the signal-to-noise ratio, $({\rm S/N})$: 
\begin{align}
\left( \frac{\rm S}{\rm N}\right)^{2}
& \equiv
\sum^{s_{\rm max}}_{s,s' = s_{\rm min}}
\xi_{\rm XY, 1}(s)\, {\rm COV}^{-1}(s,s')\, \xi_{\rm XY, 1}(s')
~, \label{eq: def SN}
\end{align}
Here, the minimum and maximum separation, $s_{\rm min}$ and $s_{\rm max}$, have to be specified in computing the signal-to-noise ratio.
In what follows, we fix the maximum separation $s_{\rm max}$ to $150$\,Mpc$/h$. As long as we set it to a scale larger than the zero-crossing point of the dipole signal (typically at $20$--$40$\,Mpc$/h$), the change of $s_{\rm max}$ hardly affects the signal-to-noise ratio. On the other hand, we see that our analytical prediction of the dipole quantitatively reproduces the simulation results even at $s\sim 5\,{\rm Mpc}/h$, below which the dipole amplitude seems to be further increased with a negative sign. However, the baryonic effects ignored in our analytical model and simulations potentially affect the dipole, and their impacts may have to be taken into account as a possible systematic effect, which needs further study. For this reason, we restrict the signal-to-noise estimation to the scales where such an effect is neglected, and set the minimum separation $s_{\rm min}$ to $5$\,Mpc$/h$.

Then, in Sec.~\ref{sec: SN smin zred}, varying the minimum separation and redshift, we study the basic behaviours of the signal-to-noise ratio, and discuss its key properties. 
In Sec.~\ref{sec: samples SN}, we change parameters for galaxy surveys and galaxy/halo clustering properties to investigate the general trend of the signal-to-noise ratio. Finally, Sec.~\ref{sec: future observations} estimates the signal-to-noise ratio for upcoming surveys.

\subsection{Scale and redshift dependence}
\label{sec: SN smin zred}

\begin{figure}
\centering
\includegraphics[width=0.8\columnwidth]{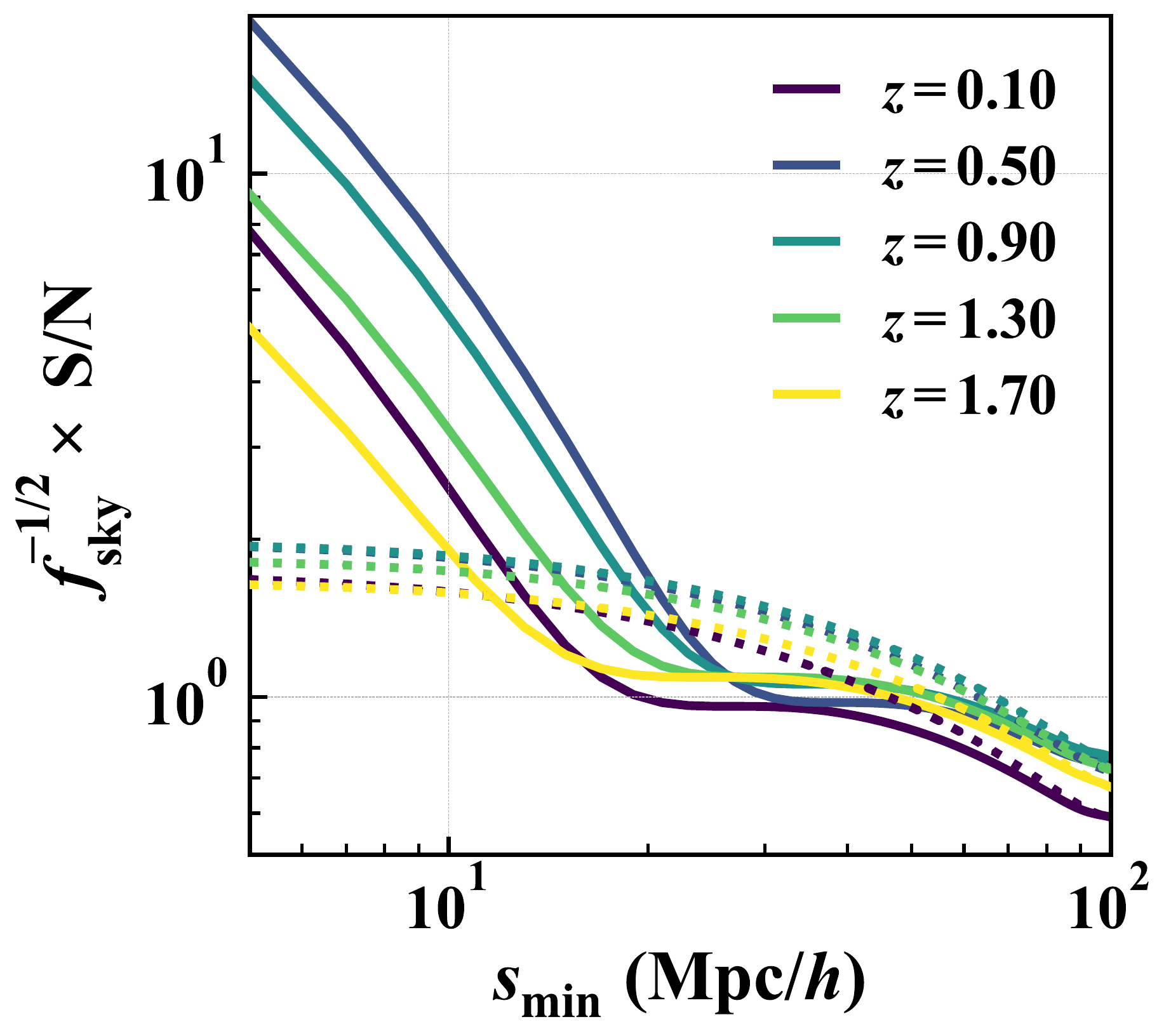}
\caption{
Signal-to-noise ratio normalized by the square root of the fractional sky coverage, $f^{-1/2}_{\rm sky}({\rm S/N})$, plotted as a function of the minimum separation $s_{\rm min}$ fixing the maximum separation to $s_{\rm max} = 150\, {\rm Mpc}/h$, results at various redshifts are shown in different colours.
The solid and dotted lines represent the results based on our model with and without the non-perturbative correction $\xi^{(\epsilon_{\rm NL})}_{{\rm XY}, 1}$, respectively.
The redshift depth, bias, and number density are chosen to be the same as in Fig.~\ref{fig: cov fixz}.
}
\label{fig: SN smin}
\end{figure}

Let us look at the basic behaviour of the signal-to-noise ratio. First consider the dependence of the signal-to-noise ratio on the minimum separation $s_{\rm min}$. In Fig.~\ref{fig: SN smin}, assuming the same halo populations as considered in Figs.~\ref{fig: cov fixz} and \ref{fig: cov fixs}, we plot the signal-to-noise ratio with (solid) and without (dotted) the halo potential contributions, $\xi_1^{\epsilon_{\rm NL}}$. Here, the results at different redshifts are shown as a function of $s_{\rm min}$, keeping the redshift depth fixed to $\Delta z=0.1$. Since the signal-to-noise ratio generally scales as $({\rm S}/{\rm N})\,\propto\,f_{\rm sky}^{1/2}$, the plotted results are normalized by $f_{\rm sky}^{1/2}$. 

Overall, the signal-to-noise ratio generally gets increased as decreasing $s_{\rm min}$. 
A notable point is that in the presence of the halo potential term, the signal-to-noise ratio deviates from the one ignoring the halo potential at $s\lesssim 40\, {\rm Mpc}/h$. As decreasing the minimum separation, it first tends to stay constant, but eventually turns to increase, finally exceeding the signal-to-noise ratio without the halo potential contribution. These behaviours are indeed expected from the behaviour of the signal part, $\xi_{\rm XY,1}$. That is, the plateau and amplification of the signal-to-noise ratio are respectively linked to the sign flip and the sharp drop with negative amplitude of the dipole cross-correlation function, as shown in Fig.~\ref{fig: dipole}. Thus, the signal-to-noise ratio at the small minimum separation can be dominated by the gravitational redshift effect from the halo potential, and because of this, the dipole signal would be detectable at a statistically significant level.

\begin{figure}
\centering
\includegraphics[width=0.8\columnwidth]{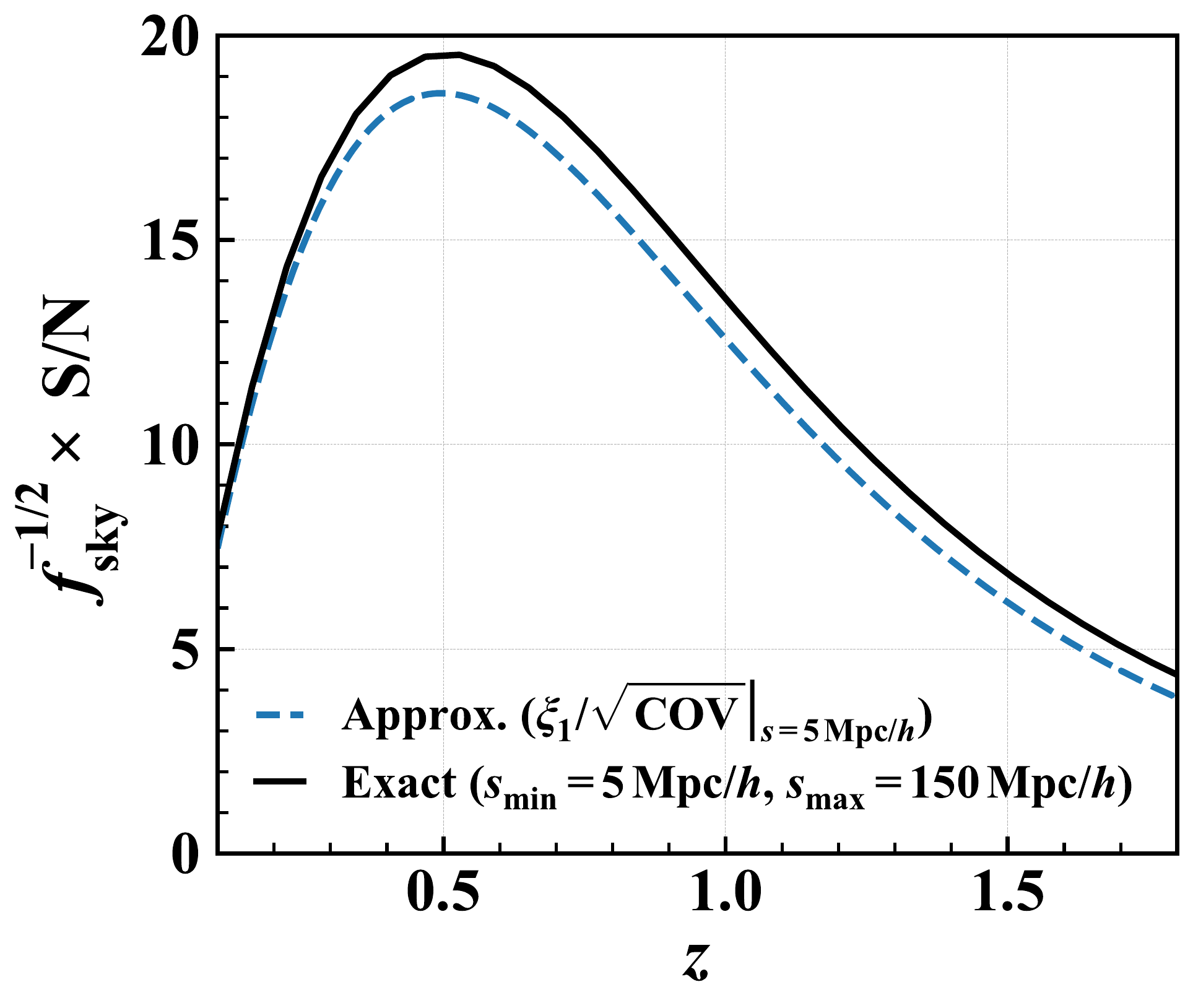}
\caption{
Redshift dependence of the signal-to-noise ratio normalized by the fractional sky coverage, $f^{-1/2}_{\rm sky}({\rm S/N})$, fixing the minimum and maximum separations to $s_{\rm min} = 5\, {\rm Mpc}/h$ and $s_{\rm max} = 150\, {\rm Mpc}/h$, respectively (black solid).
The redshift depth, bias, and number density are chosen to be the same as in Fig.~\ref{fig: cov fixz}.
Blue-dashed line represents the ratio, $\xi_{{\rm XY}, 1}(s)/\sqrt{\rm COV}(s,s)$, at $s=5\, {\rm Mpc}/h$, which approximately describes the black-solid line.
}
\label{fig: SN zred}
\end{figure}

In Fig.~\ref{fig: SN smin}, another notable point is that the signal-to-noise ratio in the presence of halo potential contribution shows a non-trivial redshift dependence on its amplitude at $s_{\rm min}\lesssim 10$\,Mpc$/h$. To look closely at the redshift dependence, we next plot in Fig.~\ref{fig: SN zred} the signal-to-noise ratio as a function of the redshift, fixing the minimum separation to $s_{\rm min} = 5\, {\rm Mpc}/h$. The result depicted as a black solid line has a peak at $z\approx 0.5$. Ignoring the contribution of the off-diagonal covariance, this non-monotonic behaviour is indeed inferred from the lower panel of Fig.~\ref{fig: cov fixs}, where we see the diagonal covariance normalized by $(\xi_1)^2$ has a minimum at $z\approx 0.5$. This indicates that the estimated signal-to-noise ratio is dominated by the contribution from the diagonal part of the covariance matrix, which is mainly determined by the terms CV$\times$P and P$\times$P . To prove this, in Fig.~\ref{fig: SN zred}, we plot the ratio, $\xi_{\rm XY,1}(s)/{\rm COV}(s,s)$, evaluated at $s=5$\,Mpc$/h$ (blue dashed). We then find that the resultant ratio nicely explains the redshift dependence of the signal-to-noise ratio. Thus, the non-monotonic redshift dependence of the signal-to-noise ratio, having a maximum at $z\approx 0.5$, is shown to be originated from the two competitive behaviours of the cross-correlation function and diagonal covariance, as shown in Fig.~\ref{fig: cov fixs}. We will see below that based on the halo model prescription, these are rather generic features, irrespective of the survey parameters.

\subsection{Dependence of target samples}
\label{sec: samples SN}

\begin{figure*}
\centering
\includegraphics[width=0.7\textwidth]{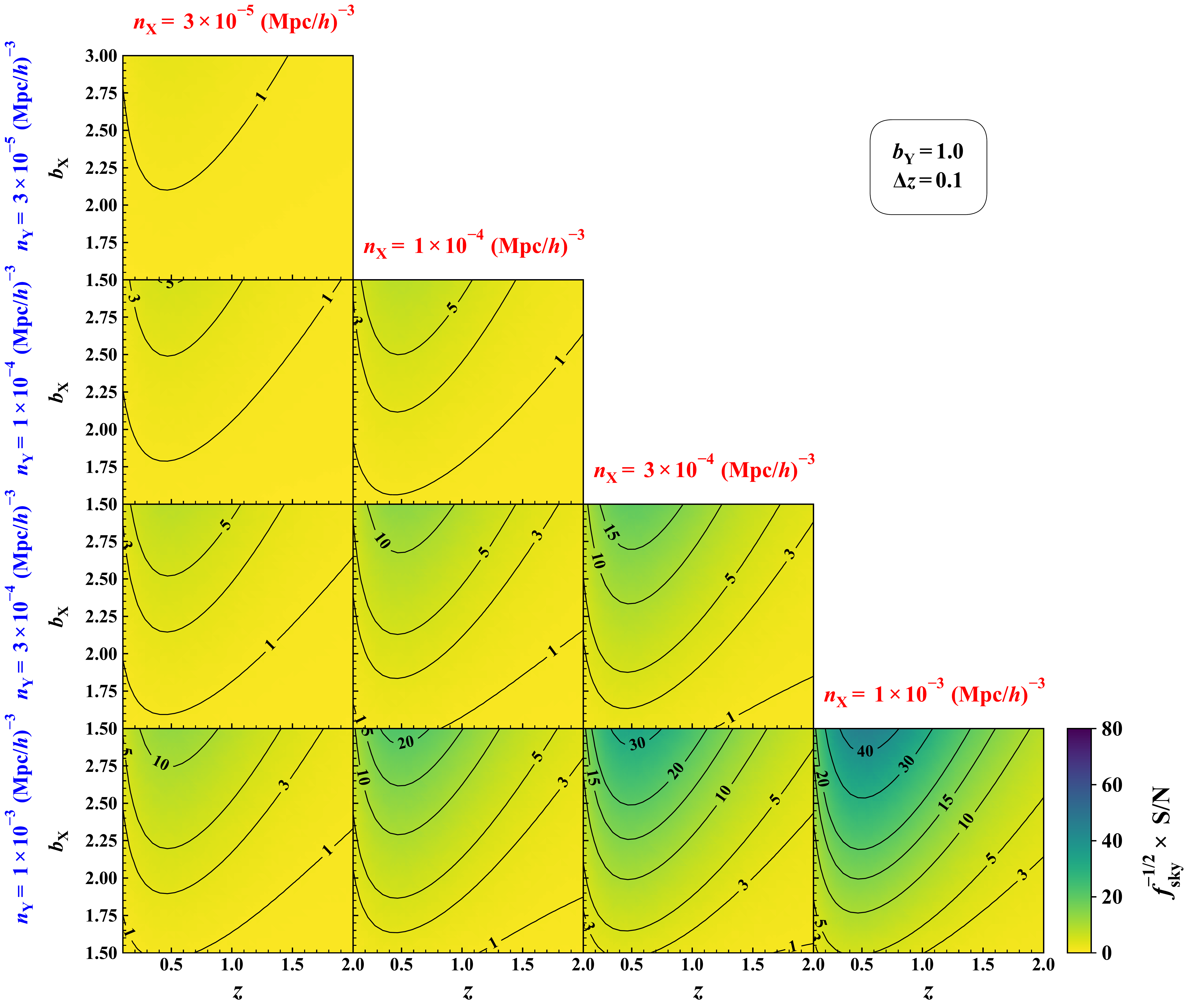}
\caption{
Two-dimensional plot of the signal-to-noise ratio as a function of $b_{\rm X}$ and $z$, where $b_{\rm X}$ is the bias of massive halo populations and $z$ is the redshift of the survey assuming the range $[z-0.05, z+0.05]$.
The bias of less massive halo population is fixed to $b_{\rm Y} = 1.0$.
In each panel, the colour scale and black contours indicate the signal-to-noise ratio normalized by the square of the fractional sky coverage, $f^{-1/2}_{\rm sky}({\rm S/N})$ (see the rightmost colour bar).
Panels show the results adopting various number densities of halo populations, $n_{\rm X}$ and $n_{\rm Y}$, ranging from $3\times 10^{-5}\, ({\rm Mpc}/h)^{-3}$ to $10^{-3}\, ({\rm Mpc}/h)^{-3}$, as indicated in the blue and red texts.
}
\label{fig: SN 2D bY1}
\end{figure*}

\begin{figure*}
\centering
\includegraphics[width=0.7\textwidth]{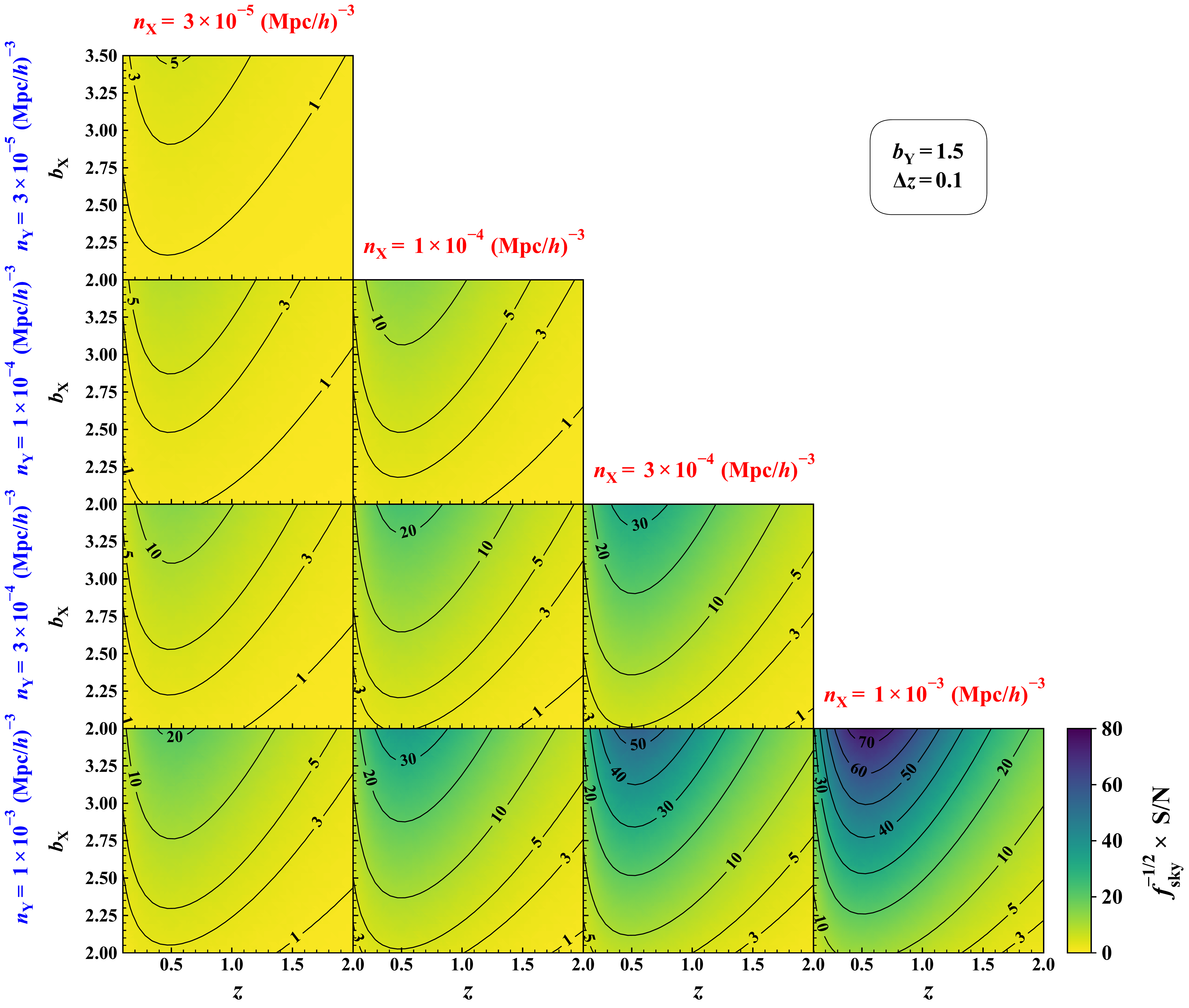}
\caption{
Same as Fig.~\ref{fig: SN 2D bY1} but for the bias of less massive halo population, $b_{\rm Y} = 1.5$.
}
\label{fig: SN 2D bY2}
\end{figure*}

So far, we have studied the behaviours of the covariance matrix and signal-to-noise ratio for specific halo samples, fixing the halo bias and halo number density, $(b_{\rm X/Y}, \, n_{\rm X/Y})$. Here, we investigate the dependence of the halo samples on the signal-to-noise ratio. To do this, we vary the parameters $b_{\rm X}$, $n_{\rm X}$, and $n_{\rm Y}$.
To be precise, we first set the bias for the halo sample Y to $b_{\rm Y}=1$ (or $1.5$). We then compute the signal-to-noise ratio for various set of parameters $b_{\rm X}$, $n_{\rm X}$, and $n_{\rm Y}$, with $b_{\rm X}$ being larger than $b_{\rm Y}$.
Note that we ignore the contributions from the magnification bias, among which the most dominant contribution coming from the Doppler effect is discussed in Appendix~\ref{app: magnification}, showing it to be negligible.
The results normalized by $f_{\rm sky}^{1/2}$ are plotted as a function of the halo bias $b_{\rm X}$ and the central redshift of the surveys, shown in Figs.~\ref{fig: SN 2D bY1} and \ref{fig: SN 2D bY2}. Here, the redshift depth of the survey is fixed to $\Delta z=0.1$. Note that given the halo bias and number density, one can uniquely determine the halo mass range, from which the halo potential is predicted through the NFW profile, as we did in Sec.~\ref{sec: covariance numerical}.

In Figs.~\ref{fig: SN 2D bY1} and \ref{fig: SN 2D bY2}, the estimated results of $f_{\rm sky}^{-1/2}\,({\rm S}/{\rm N})$ are shown for the halos with the number density of $n_{\rm X/Y}=3\times 10^{-5}$, $10^{-4}$, $3\times10^{-4}$, and $10^{-3}$, restricting the cases to $n_{\rm X}\leq n_{\rm Y}$. In all cases, we see that the signal-to-noise ratio has a peak at $z\approx 0.5$. In particular, for the halo samples having the large number density $n_{\rm X}=n_{\rm Y}=10^{-3}$\, Mpc$/h$ (bottom right), the signal-to-noise ratio reaches $f^{-1/2}_{\rm sky}\left({\rm S}/{\rm N}\right) = 45.8$ and $75.5$, respectively in Figs.~\ref{fig: SN 2D bY1} and \ref{fig: SN 2D bY2}, which correspond to the halo samples with the biases of $(b_{\rm X},\,b_{\rm Y}) =(3,\,1)$ and $(b_{\rm X},\,b_{\rm Y}) =(3.5,\,1.5)$. Comparing between the results in both figures, while the width of the plot range in the vertical axis are the same, i.e., $\Delta b=b_{\rm X}-b_{\rm Y}=2$, the resultant signal-to-noise ratios are overall enhanced in the cases with $b_{\rm Y}=1.5$ (Fig.~\ref{fig: SN 2D bY2}). Ignoring the halo potential contribution, the dipole moment of the cross-correlation function scales as $\xi_{\rm XY,1}\propto(b_{\rm X}-b_{\rm Y})$ (see Eqs.~(\ref{eq: xi1 grav}) and (\ref{eq: xi1 std})). That is, in the absence of the halo potential, the resultant signal-to-noise ratio should be the same in both Figs.~\ref{fig: SN 2D bY1} and \ref{fig: SN 2D bY2}. This implies that the difference between the two figures is attributed to the contribution from the halo potential in the dipole moment. Since the halos with a larger bias tend to have larger halo masses, the halo potential also becomes deeper as increasing the bias. The important point is that the depth of the potential is not linearly proportional to the halo mass. As a result, the difference of the potential $\Delta\phi_{\rm NL}=\phi_{\rm NFW,0,X}-\phi_{\rm NFW,0,Y}$ gets large as increasing the bias or halo mass, leading to an additional enhancement of the signal-to-noise ratio for halos with large biases. 

The behaviours shown in Figs.~\ref{fig: SN 2D bY1} and \ref{fig: SN 2D bY2} provide a useful guideline to discuss the feasibility to detect the relativistic dipole. In the next subsection, based on these results, we will estimate the detectability of the dipole moment.

\subsection{Future observations}
\label{sec: future observations}

\begin{table}\centering
\caption{
The upcoming surveys considered in this paper. In Appendix \ref{app: future surveys}, we summarize each survey parameters in Tables \ref{table: DESI1}--\ref{table: SKA2}.
}
\begin{tabular}{c c c c c}
survey
& target samples
& $f_{\rm sky}\, ({\rm deg}^{2})$
& redshift range
\\
\hline\hline
 \multirow{3}{*}{DESI} & BGS & \multirow{3}{*}{14,000} & $[0.05, 0.45]$ \\
 & LRG & & $[0.65, 1.15]$ \\
 & ELG & & $[0.65, 1.65]$ \\
\hline
 Euclid & H$\alpha$ emitter & 15,000 & $[0.9, 1.8]$ \\
\hline
 PFS & (OII) ELG & 1,464 & $[0.6, 2.4]$ \\
\hline
 SKA1 & \multirow{2}{*}{HI galaxies} & 1,500 & $[0.05, 0.45]$ \\
 SKA2 & & 30,000 & $[0.23, 1.81]$ \\
\end{tabular}
\label{table: surveys}
\end{table}

\begin{figure}
\centering
\includegraphics[width=0.85\columnwidth]{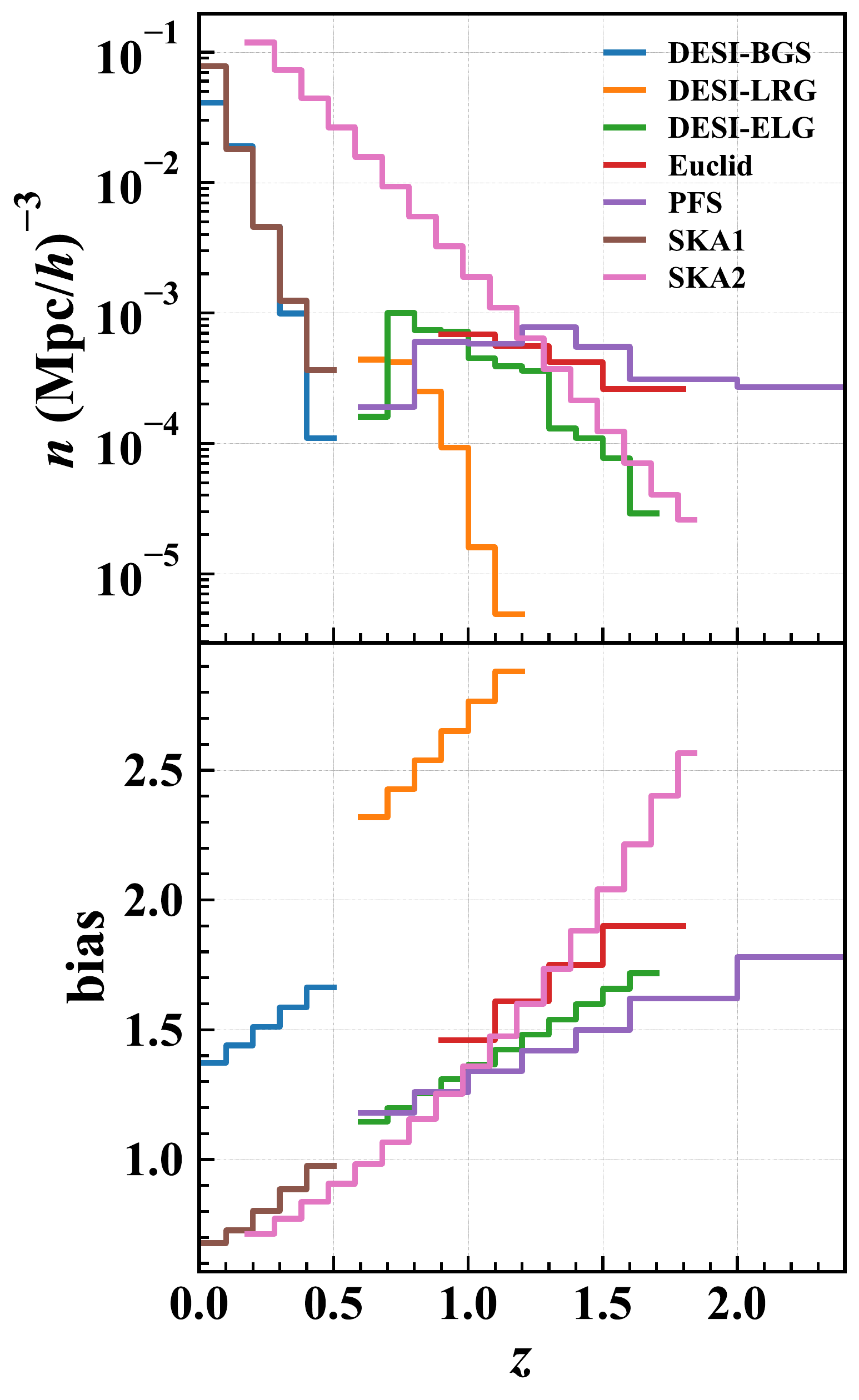}
\caption{Expected number density of galaxies ({\it top}) and bias parameter ({\it bottom}) for the surveys listed in Table~\ref{table: surveys}. The plotted data are taken from the tables summarized in Appendix~\ref{app: future surveys}.}
\label{fig: observations}
\end{figure}

Having studied the general behaviours of the signal-to-noise ratio, let us now focus on the upcoming galaxy surveys, and estimate the signal-to-noise ratio of the dipole moment. The surveys considered here are listed in Table \ref{table: surveys}. In Fig.~\ref{fig: observations}, we summarize the redshift dependence of the bias and number density for the target galaxies in each survey, which are based on Tables \ref{table: DESI1}--\ref{table: SKA2}, summarized in Appendix \ref{app: future surveys}.

In detecting the relativistic dipole, we need two galaxy samples having different values of the bias parameters. There are in general two strategies to measure the dipole cross-correlation functions. One is to divide a single galaxy population in a given survey into two subsamples. Another is to cross correlate two different samples obtained from multiple surveys (or single survey). In what follows, we set 
$s_{\rm min}=5$\,Mpc$/h$ and $s_{\rm max}=150$\,Mpc$/h$, and 
separately consider the two cases in estimating the signal-to-noise ratios.

\subsubsection{Cross-correlating two divided populations from the single target}
\label{subsubsec: cross single}

We first focus on a single galaxy population, and dividing the sample into two subsamples, we take a cross-correlation between them. Depending on how we divide the sample into two, the number densities and the bias parameters of the two subsamples differ from each other as well as those of the original sample. Thus, the signal-to-noise ratio of the relativistic dipole varies on how we divide the sample into two. Here, we shall estimate the {\it best} signal-to-noise ratio based on the halo model prescription, assuming that the galaxies of our interest follow the halo distribution whose halo masses are larger than $M_{\rm min}$. We then divide the galaxies into two subsamples Y and X hosted respectively by the halos with the mass ranges $[M_{\rm min}, M_*]$ and $[M_*, \infty]$.

Denoting the number density of the galaxies before division by $n_{\rm obs}$, their bias parameters $b_{\rm X/Y}$ and number densities $n_{\rm X/Y}$ are given by
\begin{align}
n_{\rm X}(M_{*}) &= n_{\rm obs} \frac{\int^{\infty}_{\ln{M_{*}}}\, \frac{{\rm d}n}{{\rm d}\ln{M}}\, {\rm d}\ln{M}}
{\int^{\infty}_{\ln{M_{\rm min}}}\, \frac{{\rm d}n}{{\rm d}\ln{M}}\, {\rm d}\ln{M}} ~,\\
b_{\rm X}(M_{*}) &=
\frac{\int^{\infty}_{\ln{M_{*}}}\, b_{\rm ST}(M)\frac{{\rm d}n}{{\rm d}\ln{M}}\, {\rm d}\ln{M}}
{\int^{\infty}_{\ln{M_{*}}}\, \frac{{\rm d}n}{{\rm d}\ln{M}}\, {\rm d}\ln{M}}
~, \label{eq: bias X Mstar}
\end{align}
for the massive population, and 
\begin{align}
n_{\rm Y}(M_{*}) &=
n_{\rm obs} \frac{\int^{\ln{M_{*}}}_{\ln{M_{\rm min}}}\, \frac{{\rm d}n}{{\rm d}\ln{M}}\, {\rm d}\ln{M}}
{\int^{\infty}_{\ln{M_{\rm min}}}\, \frac{{\rm d}n}{{\rm d}\ln{M}}\, {\rm d}\ln{M}}
~, \\
b_{\rm Y}(M_{*}) &=
\frac{\int^{\ln{M_{*}}}_{\ln{M_{\rm min}}}\, b_{\rm ST}(M)\frac{{\rm d}n}{{\rm d}\ln{M}}\, {\rm d}\ln{M}}
{\int^{\ln{M_{*}}}_{\ln{M_{\rm min}}}\, \frac{{\rm d}n}{{\rm d}\ln{M}}\, {\rm d}\ln{M}}
~. \label{eq: bias Y Mstar}
\end{align}
for the less massive population. Here, the functions $b_{\rm ST}$ and ${\rm d}n/{\rm d}\ln M$ are the halo bias and mass function, for which we use the expressions given by \citet{1999MNRAS.308..119S}. Note that we also examined the prescription given by \citet{2008ApJ...688..709T,2010ApJ...724..878T}, and found that the estimated halo potential changes at most by a few percent, and thus the results are insensitive to the choice of the model.
With this prescription, we have $b_{\rm X}>b_{\rm Y}$, and $n_{\rm obs} = n_{\rm X}(M_{*}) + n_{\rm Y}(M_{*})$. 
Note that, because of the idealistic treatment in the above, i.e., two subsamples having the mass ranges $[M_{\rm min}, M_*]$ and $[M_*, \infty]$, the value of the parameter $M_{*}$ tends to be large when we obtain the {\it best} signal-to-noise ratio.
In Appendix~\ref{app: future surveys}, we summarize the ratio of the number densities $n_{\rm X}(M_{*})/n_{\rm obs}$ when the signal-to-noise ratio reaches its maximum. This will give us a guideline for future observations when we divide the sample into two subsamples.

In the expressions given above, the minimum halo mass $M_{\rm min}$ and the threshold mass $M_*$ are the parameters, but the former is determined by the bias of the original sample, $b_{\rm obs}$:
\begin{align}
b_{\rm obs} = \frac{\int^{\infty}_{\ln{M_{\rm min}}}\, b_{\rm ST}(M)\frac{{\rm d}n}{{\rm d}\ln{M}}\, {\rm d}\ln{M}}{\int^{\infty}_{\ln{M_{\rm min}}}\, \frac{{\rm d}n}{{\rm d}\ln{M}}\, {\rm d}\ln{M}} ~.
\label{eq: b obs to Mmin}
\end{align}
That is, provided the value of $b_{\rm obs}$ for a given survey, the minimum mass $M_{\rm min}$ is obtained by solving Eq.~(\ref{eq: b obs to Mmin}). Thus, the threshold mass is the only free parameter that controls the signal-to-noise ratio, and we determine it by maximizing the signal-to-noise ratio. Note that in evaluating $({\rm S}/{\rm N})$, the halo potential contribution to the relativistic dipole, $\phi_{\rm NFW,0,X}$ and $\phi_{\rm NFW,0,Y}$, are averaged over the mass ranges $[M_*,\,\infty]$ and $[M_{\rm min},\,M_*]$, respectively, as similarly to the biases given in Eqs.(\ref{eq: bias X Mstar}) and (\ref{eq: bias Y Mstar}). 
We note that, in the lowest redshift bin of SKA1 ($z=0.05$), the bias parameter given in \citet{2015aska.confE..24B} does not fulfill the condition given at Eq.~(\ref{eq: b obs to Mmin}), and we cannot obtain the solution for $M_{\rm min}$. Hence, only for this case, we do not use Eq.~(\ref{eq: b obs to Mmin}), but instead fix the minimum mass to $M_{\rm min} = 10^{8}\, M_{\odot}/h$, based on \citet{2015MNRAS.450.2251Y}.

\begin{figure}
\centering
\includegraphics[width=0.9\columnwidth]{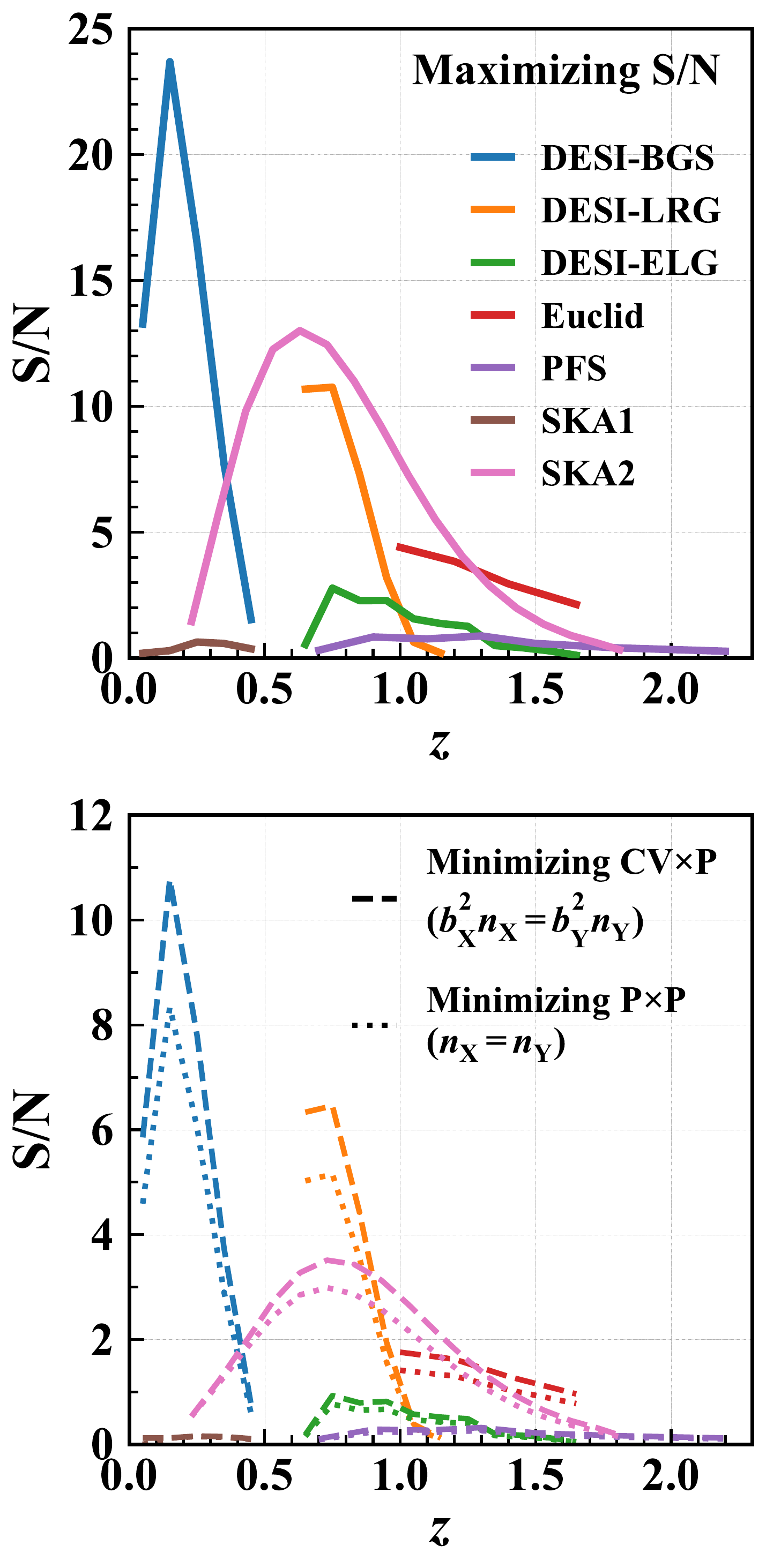}
\caption{
Expected signal-to-noise ratio for the surveys listed in Table~\ref{table: surveys}, using the single galaxy population.
({\it Top}) Dividing the sample into two subsamples to cross-correlate, we choose the threshold halo mass $M_*$ so that the signal-to-noise ratio is maximized at each redshift bin (see text in detail in Sec.~\ref{subsubsec: cross single}).
({\it Bottom}) Same as the top panel, but the threshold halo mass $M_*$ is chosen so that the CV$\times$P (dashed lines) and P$\times$P (dotted lines) contributions are minimized by imposing the conditions, $b^{2}_{\rm X}n_{\rm X} = b^{2}_{\rm Y}n_{\rm Y}$ and $n_{\rm X} = n_{\rm Y}$, respectively.
Note that accounting for the halo occupation number, the signal-to-noise ratio for DESI-BGS would be optimistic (see the main text, fourth paragraph in Sec.~\ref{subsubsec: cross single} for details).
}
\label{fig: SN obs split}
\end{figure}

Top panel of Fig.~\ref{fig: SN obs split} shows the results of the optimal signal-to-noise ratio for each galaxy population of upcoming surveys. We find that among those considered, the DESI-BGS sample gives the largest ${\rm S}/{\rm N}$. Since the cosmic variance is not the main source for the statistical error, surveys with a larger number density can give a higher signal-to-noise ratio, irrespective of the survey volume. Further increasing the difference of the biases $b_{\rm X}$--$b_{\rm Y}$, the signal-to-noise ratio for the DESI-BGS sample eventually reaches the maximum value ${\rm S}/{\rm N}=23$ at $0.1\leq z\leq0.2$, above which the signal-to-noise ratio sharply falls off due to a rapid decrease of the number density. 
Note cautiously that with the minimum mass $M_{\rm min}$ determined by the bias $b_{\rm obs}$, the number density of the DESI-BGS sample $n_{\rm obs}$ exceeds the one inferred from the halo mass function. This implies that the host halo generally contains multiple DESI-BGS samples. Since these galaxies do not necessarily reside at the halo centre, the non-perturbative potential contribution to the relativistic dipole would be suppressed. In this respect, the resultant ${\rm S}/{\rm N}$ for the DESI-BGS samples should be considered as a theoretical upper bound. A more realistic estimation of the signal-to-noise ratio needs a model based on the halo occupation distribution approach. We leave specific modelling for the DESI-BGS samples to our future work. This issue is a priori less severe in other surveys where the halo occupation number is less than unity.

Apart from the low-$z$ galaxy survey, other notable results having large signal-to-noise ratios ($1 \lesssim {\rm S/N}$) are found from the Euclid, DESI-ELG, SKA2 and DESI-LRG samples, among which the last two exceed ${\rm S}/{\rm N}=10$ around $z\approx 0.7$.
Interestingly, looking at Fig.~\ref{fig: observations}, the number density of the DESI-LRG sample is substantially smaller than that of the SKA2 by more than one order of magnitude. However, the bias of DESI-LRG sample is larger than that of the SKA2 sample, and the difference amounts to $\Delta b\approx 1.5$. As a result, at $z\approx 0.7$--$0.8$, their signal-to-noise ratios are comparable and reach maximum values. This implies that for a solid detection of the relativistic dipole, samples having a large bias are preferable. In other words, samples with a small bias $b\approx 1$--$1.5$ tend to have small signal-to-noise ratios, as indeed shown for other surveys in Fig.~\ref{fig: SN obs split}.
It is to be noted that even though the bias and number density of the samples considered are not constant over the redshifts, the overall trends seen in Fig.~\ref{fig: SN obs split} resemble those shown in Figs.~\ref{fig: SN 2D bY1} and \ref{fig: SN 2D bY2}.

Finally, to illustrate how the ${\rm S}/{\rm N}$ shown in the left panel of Fig.~\ref{fig: SN obs split} is robust and optimal against the strategies to create two subsamples, we consider alternative ways to divide the sample into two, and estimate their signal-to-noise ratios. The bottom panel of Fig.~\ref{fig: SN obs split} plots the results derived from the two strategies. One is to minimize the CV$\times$P term in the covariance matrix (dashed), and the other is to minimize the P$\times$P term (dotted). Recalling from Eq.~(\ref{eq: dipole covariance}) that the CV$\times$P and P$\times$P terms are roughly proportional to ${\rm COV}_{\rm XY}\,\propto\,b^{2}_{\rm X}/n_{\rm Y} + b^{2}_{\rm Y}/n_{\rm X}$ and $1/(n_{\rm X}n_{\rm Y})$, the conditions that minimize these two contributions are found to be $b^{2}_{\rm X}n_{\rm X}=b^{2}_{\rm Y}n_{\rm Y}$ and $n_{\rm X}=n_{\rm Y}$ (a popular choice), respectively. In our treatment, these conditions are satisfied by choosing an appropriate mass threshold $M_*$.
Note that these strategies are considered from a perspective of the error minimization, ignoring the role of the signal part itself. In this respect, they do not necessarily provide an optimal signal-to-noise ratio.
Accordingly, the signal-to-noise ratio is changed, and one finds that in all surveys considered, the resultant value of ${\rm S}/{\rm N}$ almost halves the optimal signal-to-noise ratio. The results imply that both the CV$\times$P and P$\times$P contributions play an equal role in estimating the signal-to-noise ratio, suggesting that a careful sample cut needs to be considered in practical observations in optimizing the ${\rm S}/{\rm N}$.

\subsubsection{Cross-correlating two different targets}
\label{subsubsec: cross multi}

The signal-to-noise ratio of the relativistic dipole considered in Sec.~\ref{subsubsec: cross single} depends on how we divide the sample into two subsamples, and thus it would be sensitive to the internal properties of the galaxy populations. Now, let us next consider the cross-correlation between two different samples, obtained either from different surveys or single survey, without creating subsamples. This is achieved with the samples whose observed regions are overlapped with each other. In order to maximize the detectability of the relativistic dipole, we here consider an idealistic setup where the observed areas of galaxy surveys considered are perfectly overlapped with each other without survey masks. To be precise, based on Tables \ref{table: DESI1}--\ref{table: PFS} in Appendix \ref{app: future surveys}, we follow the halo model prescription in Sec.~\ref{subsubsec: cross single} and first determine the minimum halo mass $M_{\rm min}$ in each sample from Eq.~(\ref{eq: b obs to Mmin}). Then, we estimate the non-perturbative contribution to the halo potential, $\phi_{\rm NFW,0}$, which we take an average over the mass range $[M_{\rm min},\infty]$. Plugging this potential into the dipole cross-correlation function, the signal-to-noise ratio is computed, and we examine all possible combinations of overlapping surveys in redshift. In practice, one may encounter the case that redshift slices of the two samples do not coincide with each other. In such a case, we adopt the redshift bin for the sample having a larger value of the bias as our fiducial redshift slice, and compute the signal-to-noise ratio for this redshift bin, with the bias and number density of the less biased galaxies redefined, as described in Appendix \ref{app: cross sample}.
This treatment would lead to an optimistic ${\rm S}/{\rm N}$, particularly for the cases including the DESI-BGS sample.

\begin{figure*}
\centering
\includegraphics[width=\textwidth]{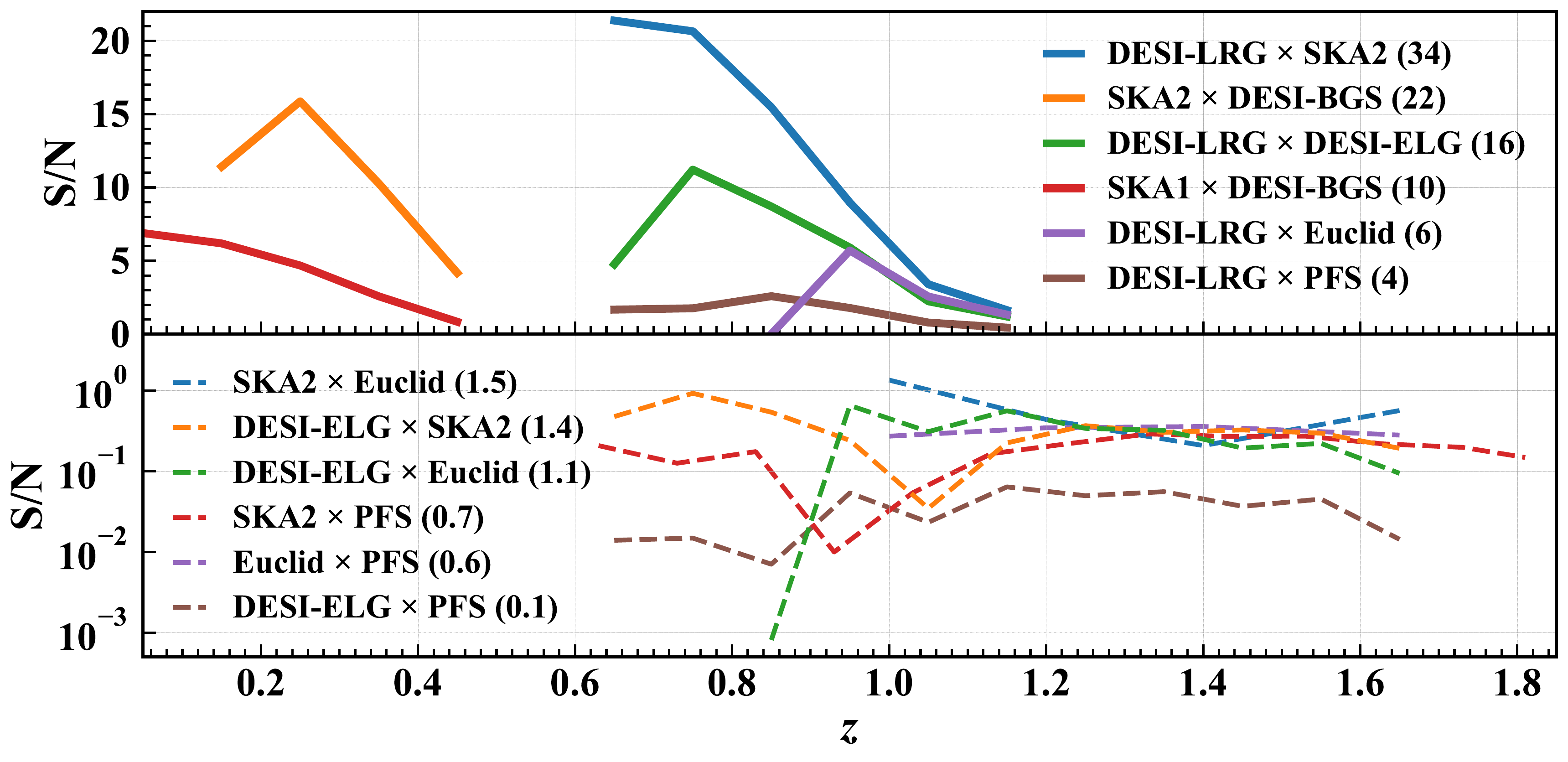}
\caption{
Expected signal-to-noise ratio for the cross-correlation between two different samples without creating subsamples.
The target samples are obtained either from different surveys or single survey listed in Table~\ref{table: surveys}. 
The top (bottom) panel summarizes the results for which the cumulative signal-to-noise ratio combining multiple redshift slices, given by $\sqrt{\sum_{z}({\rm S/N})^{2}}$, is greater (less) than 2. The estimated values of the cumulative signal-to-noise ratio are summarized in the legend (see parentheses).
Note that the signal-to-noise ratio may be optimistic for the cases including the DESI-BGS sample (see the fourth paragraph in Sec.~\ref{subsubsec: cross single} for details).
}
\label{fig: SN obs cross}
\end{figure*}

Fig.~\ref{fig: SN obs cross} summarizes the results of the signal-to-noise ratio for various cross-correlated galaxy samples. The top (bottom) panels show the results in which the cumulative signal-to-noise ratio combining all redshift bins, $\sqrt{\sum_{z}({\rm S/N})^2}$, is larger (smaller) than 2, for presentation purpose. We find that the cross-correlation between DESI and SKA2 surveys gives a large value of ${\rm S}/{\rm N}$, and a statistically significant detection of the relativistic dipole is expected particularly for DESI-BGS and SKA2 (purple), DESI-LRG and SKA2 (blue). Also, the cross-correlation between the DESI samples, i.e., LRG and ELG (orange), gives a large signal-to-noise ratio ${\rm S}/{\rm N}\approx 10$ around $z=0.7$.
The detection of the dipole signal from these surveys would provide a new way to probe gravity at cosmological scales. Furthermore, making use of the cross-correlation technique, the signal-to-noise ratio becomes improved, and SKA1 and Euclid surveys are capable of detecting the relativistic dipole at high statistical significance (${\rm S}/{\rm N}\gtrsim5$) if we combine them with the DESI-LRG and Euclid galaxy samples, respectively.
The results having a small signal-to-noise ratio, shown in the bottom panel, mainly come from the cross-correlation between emission-line galaxies which typically have small bias parameters.
Compared to the single-tracer cases in Sec.~\ref{subsubsec: cross single}, the advantage of the present method is that the impact of the shot noise contribution is mitigated, also helping to reduce unknown systematics inherent in each survey. In this respect, combining multiple tracers would be rather suited for detecting the dipole moment induced by the gravitational redshift effects.

\section{Systematic effects from off-centered galaxies}
\label{sec: systematics}

So far, we have considered the detectability of the relativistic dipole, taking only the gravitational redshift and Doppler effects into account. In this section, we discuss a potential impact of the systematics ignored so far. 

In our analytical treatment, one crucial assumption is that each of the galaxies to cross correlate strictly reside at the halo centre, and thus no virialized random motion is invoked. This is an idealistic situation, and there are galaxies whose positions are away from the halo center~\citep[e.g.,][]{2013MNRAS.435.2345H}. The off-centered galaxy positions lead to two possible systematics in the dipole signal. One is the diminution of the non-perturbative halo potential contribution to the gravitational redshift effect. Another is to introduce the virialized random motion to the off-centered galaxies. This can give a non-negligible amount of the transverse Doppler effect as the second-order special relativistic effect, which is known to produce the dipole cross-correlation signal~\citep{2013PhRvD..88d3013Z,2013MNRAS.435.1278K,2017MNRAS.468.1981C,2017MNRAS.471.2345Z,2019MNRAS.483.2671B}. Note that there are other relativistic effects that induce the dipole asymmetry in the cross-correlation function, and their impacts on the detection of gravitational redshift effect have been studied in both numerical and analytical treatments~\citep{2017MNRAS.471.2345Z,2019JCAP...04..050D,2019MNRAS.483.2671B,2020JCAP...07..048B}. Below, we analytically estimate the impacts of these two effects on the dipole signal.

Let us first discuss the suppressed gravitational potential. Following \citet{2013MNRAS.435.2345H}, we introduce the probability distribution function of the galaxy position inside each halo, $p_{\rm off}$, normalized as follows: 
\begin{align}
\int^{r_{\rm vir}}_{0}\, 4\pi r^{2}p_{\rm off}(r; R_{\rm off})\, {\rm d}r = 1 ~.
\label{eq: norm poff}
\end{align}
We model it to be Gaussian distribution, i.e., $p_{\rm off}(r; R_{\rm off}) \propto \exp{\left( -(r/R_{\rm off})^{2}/2\right)}$ with $R_{\rm off}$ being the offset parameter. Using the distribution function $p_{\rm off}$, the halo potential at the off-centered galaxy position can be estimated to be 
\begin{align}
\overline{\phi}_{\rm NFW}(z,M,R_{\rm off}) &= \int^{r_{\rm vir}}_{0}\, 4\pi r^{2}\phi_{\rm NFW}(r,z,M)p_{\rm off}(r;R_{\rm off})\, {\rm d}r ~, \label{eq: average OC}
\end{align}
where the explicit form of the NFW potential $\phi_{\rm NFW}(r,z,M)$ can be found in Appendix~D of \citet{2020MNRAS.498..981S}. Note that in the limit of $R_{\rm off}\to0$, the distribution function becomes $p_{\rm off}(r) = \delta_{\rm D}(r)/(4\pi r^{2})$, and we consistently reproduce $\overline{\phi}_{\rm NFW}(z,M,R_{\rm off}) = \phi_{\rm NFW,0}(z,M)$ . Adopting Eq.~(\ref{eq: average OC}), we substitute $\bar{\phi}_{\rm NFW}$ into the expression of $\epsilon_{\rm NL}$ in Eq.~(\ref{eq: phi NL}), instead of the central potential $\phi_{\rm NFW,0}$. Then the dipole cross-correlation with the suppressed halo potential contribution is estimated through the analytical formulas in Sec.~\ref{sec: correlation function}.

Next consider the transverse Doppler effect from the off-centered galaxies. To estimate its qualitative impact, we compute the velocity dispersion of galaxies, $\sigma_v^2$, which is expressed as a sum of the two contributions~\citep[e.g.,][]{2001MNRAS.322..901S}:
\begin{align}
\sigma^{2}_{v}(r,z,M) = \sigma^{2}_{\rm vir}(r,z,M) + \sigma^{2}_{\rm halo}(z,M)
~. \label{eq: v2 sum}
\end{align}
Here, the first and second terms at the right-hand side are originated respectively from the virial motion within a halo and the large-scale coherent motion of the host haloes.
Note that the second term is non-vanishing even if the galaxies reside at the centre of the haloes.
Although we include it for self-consistency, we confirmed that the transverse Doppler effect is dominated by the virial motion.

To compute the velocity dispersion of the virial motion, $\sigma^{2}_{\rm vir}$, we adopt the halo model prescription and use the analytical formula for the velocity dispersion of the NFW density profile~\citep[see Eq.~(14) of ][]{2001MNRAS.321..155L}:
\begin{align}
\sigma^{2}_{\rm vir}(r,z,M) = \alpha(r,z,M)\frac{GM}{r_{\rm vir}}
~, \label{eq: v2 vir}
\end{align}
with the function $\alpha(r,z,M)$ given by
\begin{align}
\alpha(r,z,M) &= \frac{3}{2}c^{2} g(c)x(1+cx)^{2}\Biggl[
6{\rm Li}_{2}(-cx) + \pi^{2} -\ln{(cx)} - \frac{1}{cx}
\notag \\
& \qquad 
- \frac{1}{(1+cx)^{2}} - \frac{6}{1+cx} +3\ln^{2}(1+cx)
\notag \\
& \qquad 
+ \ln{(1+cx)}\left( 1 + \frac{1}{(cx)^{2}} - \frac{4}{cx} - \frac{2}{1+ cx}\right)
\Biggr]
~, \label{eq: alpha}
\end{align}
where the quantities $c$, $x$, and function ${\rm Li}_{2}(x)$ respectively stand for the concentration parameter~\citep{2001MNRAS.321..559B,2002PhR...372....1C}, the radius normalized by the virial radius, $x \equiv r/r_{\rm vir}$, and the dilogarithm.
The function $g(c)$ is defined as $g(c) \equiv \left[ \ln(1+c)-c/(1+c)\right]^{-1}$.

For the velocity dispersion, $\sigma^{2}_{\rm halo}$, we estimate it using the prediction of the peak theory based on the linear Gaussian density fields~\citep[][]{1986ApJ...304...15B,2001MNRAS.322..901S}:
\begin{align}
\sigma^{2}_{\rm halo}(z,M) = (aHfD_{+})^{2}\sigma^{2}_{-1}(M)\left( 1-\frac{\sigma^{4}_{0}(M)}{\sigma^{2}_{1}(M)\sigma^{2}_{-1}(M)} \right)
~, \label{eq: v2 halo}
\end{align}
where we define the function $\sigma_{n}$ by
\begin{align}
\sigma^{2}_{n}(M) = \int\frac{k^{2}{\rm d}k}{2\pi^{2}}\, k^{2n}P_{\rm L}(k)W^{2}(kR) ~.
\end{align}
Here the function $W(x) = 3j_{1}(x)/x$ is the Fourier transform of the real space top-hat window function, and the radius $R$ is related to the mass of the halo $M$ through $M = 4\pi \bar{\rho} R^{3}/3$, where the quantity $\bar{\rho}$ is the background matter density.

Given the velocity dispersion from the above analytical formulae, the total impact of the off-centering effects, including the transverse Doppler effect, is estimated by replacing the $\epsilon_{\rm NL}$ in Eq.~(\ref{eq: phi NL}) with 
\begin{align}
\epsilon_{\rm NL} \to \overline{\epsilon}_{\rm NL} = -\frac{1}{aH}\overline{\phi}_{\rm NFW}(z,M, R_{\rm off}) + \frac{1}{aH}\frac{1}{2}\overline{\sigma}^{2}_{v}(z,M,R_{\rm off}) ~. \label{eq: phi NFW0 + TD}
\end{align}
Here, the second term at the right-hand side represents the transverse Doppler effect, and the velocity dispersion, $\overline{\sigma}_v^2$, is obtained by averaging $\sigma_v^2$ over the radius with the probability distribution function, $p_{\rm off}$, similarly to the first term (see Eq.~(\ref{eq: average OC})). Eq.~(\ref{eq: phi NFW0 + TD}) provides an analytical way to estimate the impact of the off-centering effects on the dipole signal, but we note that there are several assumptions and simplifications in deriving Eq.~(\ref{eq: phi NFW0 + TD}). For instance, the velocity dispersion $\sigma_{\rm vir}^2$ at Eq.~(\ref{eq: alpha}) has been derived under the assumption of the isotropic velocity distribution, which is known to be inaccurate for the haloes in $N$-body simulations. Further, the bulk velocity dispersion $\sigma_{\rm halo}^2$ at Eq.~(\ref{eq: v2 halo}) is based on the linear theory, and it under-predicts the actual velocity dispersion for simulated haloes. Our primary focus here is to study the qualitative impacts of the off-centering effects, and a more accurate estimation will have to be addressed based on numerical simulations. This is left for our future work.

\begin{figure}
\centering
\includegraphics[width=\columnwidth]{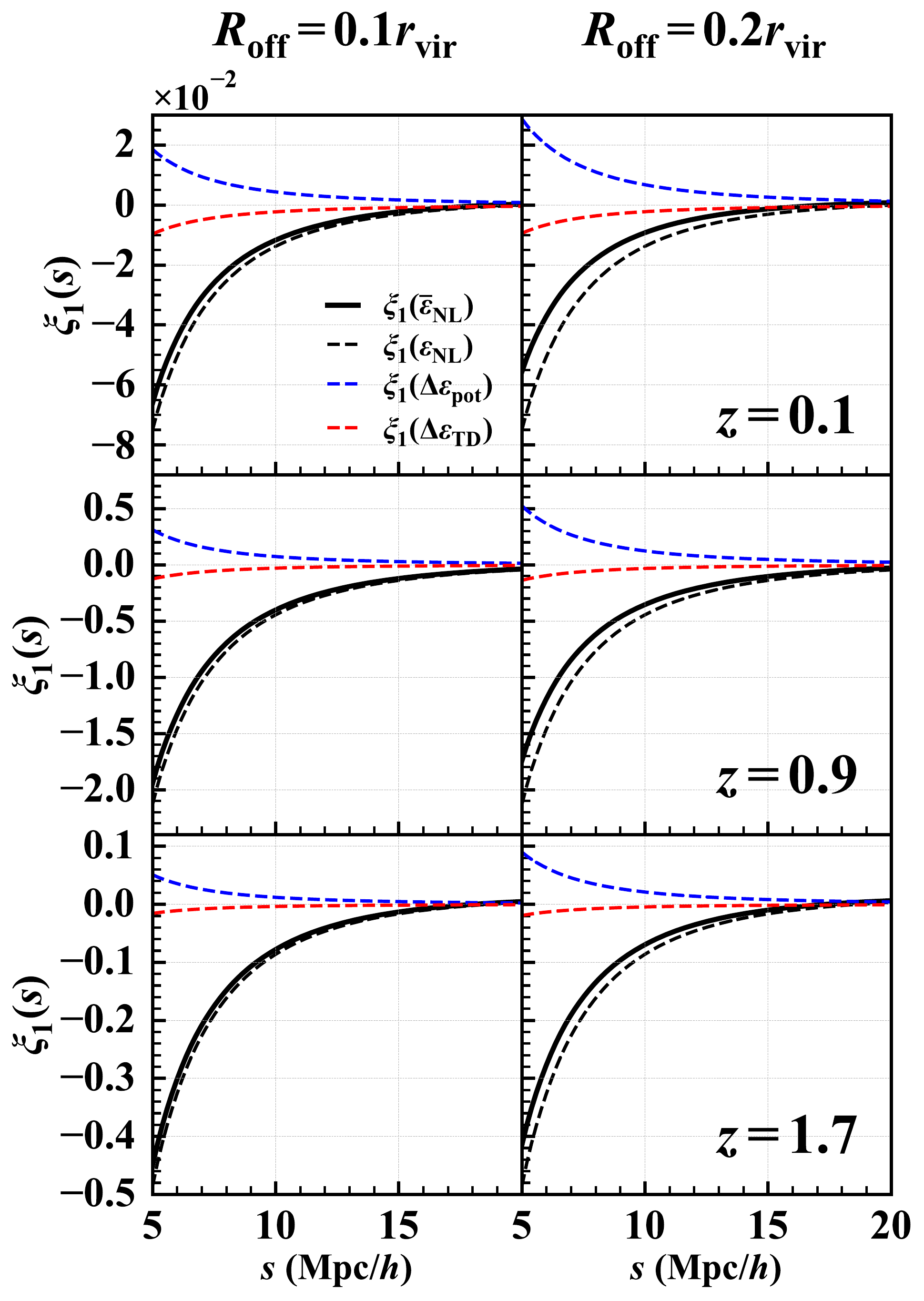}
\caption{
Impacts of the off-centered galaxies on the dipole cross-correlation function at $z=0.1$, $0.9$, and $1.7$ (from {\it top} to {\it bottom}).
The black-solid and black-dashed lines, respectively, represent the results including and neglecting the off-centering effects.
The off-centered galaxies induce two effects: lowering the halo potential and introducing the virial motion which gives rise to the transverse Doppler effect.
Contributions of these two effects are, respectively, shown in the blue ($\xi_{1}(\Delta\epsilon_{\rm pot})$, Eq.~(\ref{eq: eps A})) and red ($\xi_{1}(\Delta\epsilon_{\rm TD})$, Eq.~(\ref{eq: eps B})) dashed lines.
The effects of the off-centered galaxies are characterized by the parameter $R_{\rm off}$ (see below Eq.~(\ref{eq: norm poff})).
In the {\it left} and {\it right} panels, we set it to $R_{\rm off} = 0.2r_{\rm vir}$ and $R_{\rm off} = 0.1r_{\rm vir}$, respectively.
The bias parameters are fixed to be $b_{\rm X} = 2.5$ and $b_{\rm Y} = 1.5$.
}
\label{fig: xi eps NL}
\end{figure}

Fig.~\ref{fig: xi eps NL} shows the impacts of the off-centering effects on the dipole moment obtained from the analytical treatment at redshifts, $z=0.1$ (top), $0.9$ (middle), and $1.7$ (bottom). Here, we particularly focus on the dipole cross-correlation function at $s=5$--$20$\,Mpc$/h$, where the gravitational redshift effect dominates the standard Doppler effect, and it dominantly contributes to the signal-to-noise ratio. To elucidate how their impacts are changed with the off-centering parameter, we examine the two cases: $R_{\rm off}=0.1\,r_{\rm vir}$ (left) and $0.2\,r_{\rm vir}$ (right), as typical values considered in \citet{2013MNRAS.435.2345H}. In each panel, black solid and dashed lines are the dipole cross-correlation functions with and without the systematics, respectively (labelled by $\xi_1(\overline{\epsilon}_{\rm NL})$ and $\xi_{1}(\epsilon_{\rm NL})$ in Fig.~\ref{fig: xi eps NL}). Overall, the systematics arising from the off-centered galaxies lower the dipole signals. 
The fractional changes in dipole amplitude are typically $7$--$25$ \% at $s\lesssim10\,$Mpc$/h$. That is, the gravitational redshift effect still dominates the dipole signal at small scales.

To better understand the impact of the off-centering effects, we divide the expression of $\overline{\epsilon}_{\rm NL}$ into the three pieces as 
$\overline{\epsilon}_{\rm NL}=\epsilon_{\rm NL}+\Delta\epsilon_{\rm pot}+\Delta\epsilon_{\rm TD}$, where the last two terms represent respectively the diminution of the halo potential and the contribution from the transverse Doppler effect, defined by 
\begin{align}
\Delta\epsilon_{\rm pot} &= -\frac{1}{aH}\Bigl\{\overline{\phi}_{\rm NFW}(z,M,R_{\rm off})
- \phi_{\rm NFW,0}(z,M)\Bigr\},
\label{eq: eps A}
\\
\Delta\epsilon_{\rm TD} &= \frac{1}{aH}\frac{1}{2}\overline{\sigma}^{2}_{v}(z,M,R_{\rm off}).
\label{eq: eps B}
\end{align}
Since the model considered here involves the terms that is linearly proportional to $\epsilon_{\rm NL}$, the dipole signal taking the off-centering effects into account, $\xi_1(\overline{\epsilon}_{\rm NL})$, is decomposed into the three pieces: 
\begin{align}
\xi_{1}(\overline{\epsilon}_{\rm NL})= \xi_{1}(\epsilon_{\rm NL})+
\xi_{1}(\Delta\epsilon_{\rm pot})+\xi_{1}(\Delta\epsilon_{\rm TD}). 
\end{align}
In Fig.~\ref{fig: xi eps NL}, the two contributions $\xi_{1}(\Delta\epsilon_{\rm pot})$ and $\xi_{1}(\Delta\epsilon_{\rm TD})$ are respectively plotted in blue and red dashed lines. We find that these two contributions are competitive, and have different signs. 
That is, a small impact of the off-centering effects is partly ascribed to the cancellation between the two competitive effects. Note that the negative amplitude of the term $\xi_1(\Delta\epsilon_{\rm TD})$ comes from the fact that the velocity dispersion of galaxies, $\sigma_v^2$, is dominated by the virial motion inside the halo, and the dispersion $\sigma_{\rm vir}^2$ monotonically increases with the halo mass\footnote{If one considers the situation that the virial motion is ignorable, the sign of $\xi_1(\Delta\epsilon_{\rm TD})$ becomes positive. This is because the velocity dispersion $\sigma_v^2\simeq \sigma_{\rm halo}^2$ now becomes a decreasing function of the halo mass. Such a situation has been considered in \citet{2019MNRAS.483.2671B,2013MNRAS.435.1278K,2013PhRvD..88d3013Z}.}. 
These trends would hold even if we consider a more elaborate estimation of the transverse Doppler effect, the cancellation of the off-centering effects is expected to still happen for more accurate modelling, and thus their impact on the dipole signal would be small.

\section{Summary and perspectives}
\label{sec: summary}

It has been recognized that the observational relativistic effects, mainly arising from the light propagation in an inhomogeneous universe, induce the dipole asymmetry in the cross-correlation function between the haloes or galaxies having different clustering biases. In particular, the dipole asymmetry at small scales has been recently found to be dominated by the gravitational redshift effects~\citep{2019MNRAS.483.2671B,2020MNRAS.498..981S}. Thus, the detection of the dipole signal at small scales would provide an interesting opportunity for an alternative test of gravity. In this paper, we have studied analytically the future detectability of the dipole signal induced by the gravitational redshift effect. 

In doing so, we have exploited a simple analytical description for the dipole cross-correlation function valid at quasi-linear regime. Previously, \citet{2020MNRAS.498..981S} presented a quasi-linear model of the cross-correlation function. Taking the two major relativistic effects, i.e., the standard Doppler and gravitational redshift effects into account (but ignoring other minor contributions including magnification bias), we adopted the Zel'dovich approximation and halo model prescription to describe the dipole signals beyond the linear scales. While the quantitative model predictions successfully explain the dipole cross-correlation functions measured from the halo catalogues into which all possible relativistic effects arising from the light propagation are fully incorporated \citep[]{2019MNRAS.483.2671B}, the analytical model involves seven dimensional integrals, and the time-consuming numerical integration needs to be performed. To remedy this, in this paper, we derive new approximate expressions for the galaxy/halo density field based on the Lagrangian perturbative treatment, including also the halo model prediction to account for the non-perturbative potential contributions (see Eqs.~(\ref{eq: delta std + delta epsNL})--(\ref{eq: def delta epsNL})). These results enable us to obtain rather simplified analytical expression for the dipole cross-correlation function, and we found it to quantitatively reproduce the previous result of \citet{2020MNRAS.498..981S} as well as the measured dipole signals in numerical simulations. The new analytical model of dipole cross-correlation function, presented in Eqs.~(\ref{eq: xi1 std})--(\ref{eq: xi1 eps}), involves only one dimensional integrals, and thus one can quickly predict the dipole signal, making the practical application of it to the Bayesian parameter estimation with Markov chain Monte Carlo technique possible.

Based on the new analytical model, we have computed analytically the covariance matrix of the dipole cross-correlation function, and investigated its behaviours. We found that the Gaussian covariance is mostly dominated by the two contributions, i.e., the term characterizing the cross-talk between the cosmic variance and Poisson noise, and the term purely originating from the Poisson shot noises, as similarly found by \citet{2017PhRvD..95d3530H}. As a result, the covariance matrix is shown to sensitively depend on not only the survey parameters (redshift depth and survey area of the galaxy surveys) but also the bias and number density of the galaxies/haloes to cross correlate. 

Plugging further the analytical predictions of both the dipole signal and covariance matrix into the definition of signal-to-noise ratio, we have quantitatively explored, in various setup for upcoming surveys, the feasibility to detect the dipole cross-correlation function, especially focusing on the scales where the gravitational redshift effect starts to be dominated and changes the sign of the dipole amplitudes. Our main findings are summarized as follows:

\begin{itemize}
 \item In most of the cases we examined, the signal-to-noise ratio of the dipole cross-correlation functions becomes maximum around $z\approx 0.5$ (see Figs.~\ref{fig: SN zred}, \ref{fig: SN 2D bY1}, and \ref{fig: SN 2D bY2}). For the non-perturbative halo potential described by the NFW profile, the trend would generically appear true if one considers surveys with a fixed redshift interval in the universe close to the $\Lambda$CDM model. 

 \item Generally, cross-correlating between galaxies having large number densities with a larger difference of the clustering biases enhances the signal-to-noise ratio. Also, the signal-to-noise ratio becomes further increasing if the bias parameters for both of the galaxies gets large. For an idealistic situation with the galaxies of the number density $n_{\rm X,Y}\approx 10^{-3}\,({\rm Mpc}/h)^{-3}$ and the biases $(b_{\rm X},b_{\rm Y})=(3.5,1.5)$, it reaches $f^{-1/2}_{\rm sky}\,{\rm S/N}=75.5$ for a survey at $z=0.5$ with the interval of $\Delta z=0.1$ (see Fig.~\ref{fig: SN 2D bY2})). 

 \item For planned future galaxy surveys considered, if one divides the galaxy samples in each survey into two subsamples, a statistically significant detection of the dipole signal is expected from DESI-BGS, DESI-LRG, and SKA2 samples, and the signal-to-noise ratios of these samples reach $23$, $11$, and $13$, respectively (see Fig.~\ref{fig: SN obs split}).

 \item On the other hand, if the survey regions of the two different samples are overlapped, one can take a cross-correlation between them without dividing the samples into two. In this case, the dipole cross-correlation between DESI-LRG and SKA2 samples gives the largest signal-to-noise ratio, ${\rm S/N}\approx 21$. A solid detection of the dipole signal is also expected from the cross-correlations between DESI-LRG and DESI-ELG samples, and SKA2 and DESI-BGS samples, leading respectively to the signal-to-noise ratios, ${\rm S/N}= 11$ and $16$ (see Fig.~\ref{fig: SN obs cross}).

\item As possible systematic effects arising from the off-centered galaxies, the diminution of the gravitational redshift effect from the halo potential and the non-vanishing transverse Doppler effect can change the dipole signal at small scales. However, these two effects are found to be competitive, leading to different signs of the dipole cross-correlations (blue and red dashed lines in Fig.~\ref{fig: xi eps NL}). As a result of the partial cancellation, the net result of their contributions becomes small, and the dipole signal at $s\lesssim10$\,${\rm Mpc}/h$ is shown to be still dominated by the gravitational redshift effect.
\end{itemize}

Our forecast study suggests that upcoming surveys enable us to detect dipole signals at a statistically significant level, and this would offer a unique probe of the depth of the halo gravitational potential. Exploiting the dipole to test the fundamental physics would be also an interesting subject through a precision measurement of the gravitational redshift effect, and this is left to our future work.

Note that the major findings summarized above rely on several assumptions and simplification based on the halo model. In particular, our analysis assumes the one-to-one correspondence between halo and galaxy distributions. For more realistic estimations, a proper account of the halo-galaxy connection would be crucial, using e.g., the halo-occupation distribution approach, in which the contribution of the so-called satellites would play a substantial role to detect the dipole signal. Furthermore, in this paper, the gravitational redshift effect from the halo potential is computed from the NFW profile, whose potential depth is solely determined by the halo mass and redshift for a given cosmological model. However, even for a fixed halo mass, halo clustering features have been known to depend on secondary halo properties that correlate with halo assembly history, referred to as the halo assembly bias \citep[see e.g.,][]{2005MNRAS.363L..66G,2005ApJ...624..505Z}. This effect would give a systematic impact on the estimation of the halo potential, and proper modelling of it needs further study.

Finally, we have investigated the detectability of the dipole signal, restricting the scales to $s\geq 5\,{\rm Mpc}/h$, where our analytical prediction of the dipole cross-correlation is shown to reproduce quantitatively the simulation results well. Nevertheless, below this scale, the amplitude of the dipole cross-correlation is expected to become further large (with a negative sign), and thus the signal-to-noise ratio would be improved if one uses the cross-correlation data at small scales. In doing so, however, the analytical treatment based on perturbation theory may not be adequate, and one has to exploit a method to quantitatively predict the dipole cross-correlation function, taking consistently not only the nonlinear gravitational clustering but also the baryonic effects on the galaxy distribution into account. This is a challenging task, but is worth for further investigation toward a decisive detection of the gravitational redshift effect.

\section*{Acknowledgments}
This work was initiated during the invitation program of JSPS Grant No. L16519.
Numerical simulation was granted access to HPC resources of TGCC through allocations made by GENCI (Grand Equipement National de Calcul Intensif) under the allocations A0030402287, A0050402287, A0070402287 and A0090402287.
Numerical computation was also carried out partly at the Yukawa Institute Computer Facility.
This work was supported by Grant-in-Aid for JSPS Fellows No.~17J10553 (SS) and in part by MEXT/JSPS KAKENHI Grant Numbers Nos. JP17H06359, JP20H05861, and 21H01081 (AT). AT also acknowledges the support from JST AIP Acceleration Research Grant No. JP20317829, Japan.
The authors thank the Yukawa Institute for Theoretical Physics at Kyoto University. Discussions during the YITP workshop YITP-T-21-06 on ``Galaxy shape statistics and cosmology'' were useful to complete this work.

\section*{Data availability}
The data underlying this article are available in RayGalGroupSims Relativistic Halo Catalogs at \url{https://cosmo.obspm.fr/public-datasets/}.

\bibliographystyle{mnras}
\bibliography{ref}

\begin{thebibliography}{}
\makeatletter
\relax
\def\mn@urlcharsother{\let\do\@makeother \do\$\do\&\do\#\do\^\do\_\do\%\do\~}
\def\mn@doi{\begingroup\mn@urlcharsother \@ifnextchar [ {\mn@doi@}
  {\mn@doi@[]}}
\def\mn@doi@[#1]#2{\def\@tempa{#1}\ifx\@tempa\@empty \href
  {http://dx.doi.org/#2} {doi:#2}\else \href {http://dx.doi.org/#2} {#1}\fi
  \endgroup}
\def\mn@eprint#1#2{\mn@eprint@#1:#2::\@nil}
\def\mn@eprint@arXiv#1{\href {http://arxiv.org/abs/#1} {{\tt arXiv:#1}}}
\def\mn@eprint@dblp#1{\href {http://dblp.uni-trier.de/rec/bibtex/#1.xml}
  {dblp:#1}}
\def\mn@eprint@#1:#2:#3:#4\@nil{\def\@tempa {#1}\def\@tempb {#2}\def\@tempc
  {#3}\ifx \@tempc \@empty \let \@tempc \@tempb \let \@tempb \@tempa \fi \ifx
  \@tempb \@empty \def\@tempb {arXiv}\fi \@ifundefined
  {mn@eprint@\@tempb}{\@tempb:\@tempc}{\expandafter \expandafter \csname
  mn@eprint@\@tempb\endcsname \expandafter{\@tempc}}}

\bibitem[\protect\citeauthoryear{{Alam} et~al.,}{{Alam}
  et~al.}{2017a}]{2017MNRAS.470.2617A}
{Alam} S.,  et~al., 2017a, \mn@doi [\mnras] {10.1093/mnras/stx721}, \href
  {https://ui.adsabs.harvard.edu/abs/2017MNRAS.470.2617A} {470, 2617}

\bibitem[\protect\citeauthoryear{{Alam}, {Zhu}, {Croft}, {Ho}, {Giusarma}  \&
  {Schneider}}{{Alam} et~al.}{2017b}]{2017MNRAS.470.2822A}
{Alam} S.,  {Zhu} H.,  {Croft} R. A.~C.,  {Ho} S.,  {Giusarma} E.,
  {Schneider} D.~P.,  2017b, \mn@doi [\mnras] {10.1093/mnras/stx1421}, \href
  {https://ui.adsabs.harvard.edu/abs/2017MNRAS.470.2822A} {470, 2822}

\bibitem[\protect\citeauthoryear{{Bardeen}, {Bond}, {Kaiser}  \&
  {Szalay}}{{Bardeen} et~al.}{1986}]{1986ApJ...304...15B}
{Bardeen} J.~M.,  {Bond} J.~R.,  {Kaiser} N.,   {Szalay} A.~S.,  1986, \mn@doi
  [\apj] {10.1086/164143}, \href
  {https://ui.adsabs.harvard.edu/abs/1986ApJ...304...15B} {304, 15}

\bibitem[\protect\citeauthoryear{{Beutler} \& {Di Dio}}{{Beutler} \& {Di
  Dio}}{2020}]{2020JCAP...07..048B}
{Beutler} F.,  {Di Dio} E.,  2020, \mn@doi [\jcap]
  {10.1088/1475-7516/2020/07/048}, \href
  {https://ui.adsabs.harvard.edu/abs/2020JCAP...07..048B} {2020, 048}

\bibitem[\protect\citeauthoryear{{Bonvin} \& {Durrer}}{{Bonvin} \&
  {Durrer}}{2011}]{2011PhRvD..84f3505B}
{Bonvin} C.,  {Durrer} R.,  2011, \mn@doi [\prd] {10.1103/PhysRevD.84.063505},
  \href {https://ui.adsabs.harvard.edu/abs/2011PhRvD..84f3505B} {84, 063505}

\bibitem[\protect\citeauthoryear{{Bonvin} \& {Fleury}}{{Bonvin} \&
  {Fleury}}{2018}]{2018JCAP...05..061B}
{Bonvin} C.,  {Fleury} P.,  2018, \mn@doi [\jcap]
  {10.1088/1475-7516/2018/05/061}, \href
  {https://ui.adsabs.harvard.edu/abs/2018JCAP...05..061B} {2018, 061}

\bibitem[\protect\citeauthoryear{{Bonvin}, {Hui}  \& {Gazta{\~n}aga}}{{Bonvin}
  et~al.}{2014}]{2014PhRvD..89h3535B}
{Bonvin} C.,  {Hui} L.,   {Gazta{\~n}aga} E.,  2014, \mn@doi [\prd]
  {10.1103/PhysRevD.89.083535}, \href
  {https://ui.adsabs.harvard.edu/abs/2014PhRvD..89h3535B} {89, 083535}

\bibitem[\protect\citeauthoryear{{Bonvin}, {Hui}  \& {Gaztanaga}}{{Bonvin}
  et~al.}{2016}]{2016JCAP...08..021B}
{Bonvin} C.,  {Hui} L.,   {Gaztanaga} E.,  2016, \mn@doi [\jcap]
  {10.1088/1475-7516/2016/08/021}, \href
  {https://ui.adsabs.harvard.edu/abs/2016JCAP...08..021B} {2016, 021}

\bibitem[\protect\citeauthoryear{{Bonvin}, {Oliveira Franco}  \&
  {Fleury}}{{Bonvin} et~al.}{2020}]{2020JCAP...08..004B}
{Bonvin} C.,  {Oliveira Franco} F.,   {Fleury} P.,  2020, \mn@doi [\jcap]
  {10.1088/1475-7516/2020/08/004}, \href
  {https://ui.adsabs.harvard.edu/abs/2020JCAP...08..004B} {2020, 004}

\bibitem[\protect\citeauthoryear{{Borzyszkowski}, {Bertacca}  \&
  {Porciani}}{{Borzyszkowski} et~al.}{2017}]{2017MNRAS.471.3899B}
{Borzyszkowski} M.,  {Bertacca} D.,   {Porciani} C.,  2017, \mn@doi [\mnras]
  {10.1093/mnras/stx1423}, \href
  {https://ui.adsabs.harvard.edu/abs/2017MNRAS.471.3899B} {471, 3899}

\bibitem[\protect\citeauthoryear{{Breton}, {Rasera}, {Taruya}, {Lacombe}  \&
  {Saga}}{{Breton} et~al.}{2019}]{2019MNRAS.483.2671B}
{Breton} M.-A.,  {Rasera} Y.,  {Taruya} A.,  {Lacombe} O.,   {Saga} S.,  2019,
  \mn@doi [\mnras] {10.1093/mnras/sty3206}, \href
  {https://ui.adsabs.harvard.edu/\#abs/2019MNRAS.483.2671B} {483, 2671}

\bibitem[\protect\citeauthoryear{{Bull}, {Camera}, {Raccanelli}, {Blake},
  {Ferreira}, {Santos}  \& {Schwarz}}{{Bull}
  et~al.}{2015}]{2015aska.confE..24B}
{Bull} P.,  {Camera} S.,  {Raccanelli} A.,  {Blake} C.,  {Ferreira} P.,
  {Santos} M.,   {Schwarz} D.~J.,  2015, in Advancing Astrophysics with the
  Square Kilometre Array (AASKA14). p.~24 (\mn@eprint {arXiv} {1501.04088})

\bibitem[\protect\citeauthoryear{{Bullock}, {Kolatt}, {Sigad}, {Somerville},
  {Kravtsov}, {Klypin}, {Primack}  \& {Dekel}}{{Bullock}
  et~al.}{2001}]{2001MNRAS.321..559B}
{Bullock} J.~S.,  {Kolatt} T.~S.,  {Sigad} Y.,  {Somerville} R.~S.,  {Kravtsov}
  A.~V.,  {Klypin} A.~A.,  {Primack} J.~R.,   {Dekel} A.,  2001, \mn@doi
  [¥mnras] {10.1046/j.1365-8711.2001.04068.x}, \href
  {https://ui.adsabs.harvard.edu/abs/2001MNRAS.321..559B} {321, 559}

\bibitem[\protect\citeauthoryear{{Cai}, {Kaiser}, {Cole}  \& {Frenk}}{{Cai}
  et~al.}{2017}]{2017MNRAS.468.1981C}
{Cai} Y.-C.,  {Kaiser} N.,  {Cole} S.,   {Frenk} C.,  2017, \mn@doi [\mnras]
  {10.1093/mnras/stx469}, \href
  {https://ui.adsabs.harvard.edu/abs/2017MNRAS.468.1981C} {468, 1981}

\bibitem[\protect\citeauthoryear{{Challinor} \& {Lewis}}{{Challinor} \&
  {Lewis}}{2011}]{2011PhRvD..84d3516C}
{Challinor} A.,  {Lewis} A.,  2011, \mn@doi [\prd]
  {10.1103/PhysRevD.84.043516}, \href
  {https://ui.adsabs.harvard.edu/abs/2011PhRvD..84d3516C} {84, 043516}

\bibitem[\protect\citeauthoryear{{Coates}, {Adamek}, {Bull}, {Guandalin}  \&
  {Clarkson}}{{Coates} et~al.}{2020}]{2020arXiv201112936C}
{Coates} L.,  {Adamek} J.,  {Bull} P.,  {Guandalin} C.,   {Clarkson} C.,  2020,
  arXiv e-prints, \href {https://ui.adsabs.harvard.edu/abs/2020arXiv201112936C}
  {p. arXiv:2011.12936}

\bibitem[\protect\citeauthoryear{{Cohn}}{{Cohn}}{2006}]{2006NewA...11..226C}
{Cohn} J.~D.,  2006, \mn@doi [\na] {10.1016/j.newast.2005.08.002}, \href
  {https://ui.adsabs.harvard.edu/abs/2006NewA...11..226C} {11, 226}

\bibitem[\protect\citeauthoryear{{Cooray} \& {Sheth}}{{Cooray} \&
  {Sheth}}{2002}]{2002PhR...372....1C}
{Cooray} A.,  {Sheth} R.,  2002, \mn@doi [¥physrep]
  {10.1016/S0370-1573(02)00276-4}, \href
  {https://ui.adsabs.harvard.edu/abs/2002PhR...372....1C} {372, 1}

\bibitem[\protect\citeauthoryear{{Croft}}{{Croft}}{2013}]{2013MNRAS.434.3008C}
{Croft} R. A.~C.,  2013, \mn@doi [\mnras] {10.1093/mnras/stt1223}, \href
  {https://ui.adsabs.harvard.edu/abs/2013MNRAS.434.3008C} {434, 3008}

\bibitem[\protect\citeauthoryear{{DESI Collaboration} et~al.,}{{DESI
  Collaboration} et~al.}{2016}]{2016arXiv161100036D}
{DESI Collaboration} et~al., 2016, arXiv e-prints, \href
  {https://ui.adsabs.harvard.edu/abs/2016arXiv161100036D} {p. arXiv:1611.00036}

\bibitem[\protect\citeauthoryear{{Di Dio} \& {Seljak}}{{Di Dio} \&
  {Seljak}}{2019}]{2019JCAP...04..050D}
{Di Dio} E.,  {Seljak} U.,  2019, \mn@doi [\jcap]
  {10.1088/1475-7516/2019/04/050}, \href
  {https://ui.adsabs.harvard.edu/abs/2019JCAP...04..050D} {2019, 050}

\bibitem[\protect\citeauthoryear{{Euclid Collaboration} et~al.,}{{Euclid
  Collaboration} et~al.}{2019}]{2019arXiv191009273E}
{Euclid Collaboration} et~al., 2019, arXiv e-prints, \href
  {https://ui.adsabs.harvard.edu/abs/2019arXiv191009273E} {p. arXiv:1910.09273}

\bibitem[\protect\citeauthoryear{{Fisher}, {Scharf}  \& {Lahav}}{{Fisher}
  et~al.}{1994}]{1994MNRAS.266..219F}
{Fisher} K.~B.,  {Scharf} C.~A.,   {Lahav} O.,  1994, \mn@doi [\mnras]
  {10.1093/mnras/266.1.219}, \href
  {https://ui.adsabs.harvard.edu/abs/1994MNRAS.266..219F} {266, 219}

\bibitem[\protect\citeauthoryear{{Gao}, {Springel}  \& {White}}{{Gao}
  et~al.}{2005}]{2005MNRAS.363L..66G}
{Gao} L.,  {Springel} V.,   {White} S. D.~M.,  2005, \mn@doi [\mnras]
  {10.1111/j.1745-3933.2005.00084.x}, \href
  {https://ui.adsabs.harvard.edu/abs/2005MNRAS.363L..66G} {363, L66}

\bibitem[\protect\citeauthoryear{{Grieb}, {S{\'a}nchez}, {Salazar-Albornoz}  \&
  {Dalla Vecchia}}{{Grieb} et~al.}{2016}]{2016MNRAS.457.1577G}
{Grieb} J.~N.,  {S{\'a}nchez} A.~G.,  {Salazar-Albornoz} S.,   {Dalla Vecchia}
  C.,  2016, \mn@doi [\mnras] {10.1093/mnras/stw065}, \href
  {https://ui.adsabs.harvard.edu/abs/2016MNRAS.457.1577G} {457, 1577}

\bibitem[\protect\citeauthoryear{{Guandalin}, {Adamek}, {Bull}, {Clarkson},
  {Abramo}  \& {Coates}}{{Guandalin} et~al.}{2021}]{2021MNRAS.501.2547G}
{Guandalin} C.,  {Adamek} J.,  {Bull} P.,  {Clarkson} C.,  {Abramo} L.~R.,
  {Coates} L.,  2021, \mn@doi [\mnras] {10.1093/mnras/staa3890}, \href
  {https://ui.adsabs.harvard.edu/abs/2021MNRAS.501.2547G} {501, 2547}

\bibitem[\protect\citeauthoryear{{Guzzo} et~al.,}{{Guzzo}
  et~al.}{2008}]{2008Natur.451..541G}
{Guzzo} L.,  et~al., 2008, \mn@doi [\nat] {10.1038/nature06555}, \href
  {https://ui.adsabs.harvard.edu/abs/2008Natur.451..541G} {451, 541}

\bibitem[\protect\citeauthoryear{{Hall} \& {Bonvin}}{{Hall} \&
  {Bonvin}}{2017}]{2017PhRvD..95d3530H}
{Hall} A.,  {Bonvin} C.,  2017, \mn@doi [\prd] {10.1103/PhysRevD.95.043530},
  \href {https://ui.adsabs.harvard.edu/abs/2017PhRvD..95d3530H} {95, 043530}

\bibitem[\protect\citeauthoryear{{Hamilton}}{{Hamilton}}{1992}]{1992ApJ...385L...5H}
{Hamilton} A.~J.~S.,  1992, \mn@doi [\apjl] {10.1086/186264}, \href
  {https://ui.adsabs.harvard.edu/abs/1992ApJ...385L...5H} {385, L5}

\bibitem[\protect\citeauthoryear{{Hamilton} \& {Culhane}}{{Hamilton} \&
  {Culhane}}{1996}]{1996MNRAS.278...73H}
{Hamilton} A.~J.~S.,  {Culhane} M.,  1996, \mn@doi [\mnras]
  {10.1093/mnras/278.1.73}, \href
  {https://ui.adsabs.harvard.edu/abs/1996MNRAS.278...73H} {278, 73}

\bibitem[\protect\citeauthoryear{{Hikage}, {Mandelbaum}, {Takada}  \&
  {Spergel}}{{Hikage} et~al.}{2013}]{2013MNRAS.435.2345H}
{Hikage} C.,  {Mandelbaum} R.,  {Takada} M.,   {Spergel} D.~N.,  2013, \mn@doi
  [\mnras] {10.1093/mnras/stt1446}, \href
  {https://ui.adsabs.harvard.edu/abs/2013MNRAS.435.2345H} {435, 2345}

\bibitem[\protect\citeauthoryear{{Jimeno}, {Broadhurst}, {Coupon}, {Umetsu}  \&
  {Lazkoz}}{{Jimeno} et~al.}{2015}]{2015MNRAS.448.1999J}
{Jimeno} P.,  {Broadhurst} T.,  {Coupon} J.,  {Umetsu} K.,   {Lazkoz} R.,
  2015, \mn@doi [\mnras] {10.1093/mnras/stv117}, \href
  {https://ui.adsabs.harvard.edu/abs/2015MNRAS.448.1999J} {448, 1999}

\bibitem[\protect\citeauthoryear{{Kaiser}}{{Kaiser}}{1987}]{1987MNRAS.227....1K}
{Kaiser} N.,  1987, \mn@doi [\mnras] {10.1093/mnras/227.1.1}, \href
  {https://ui.adsabs.harvard.edu/abs/1987MNRAS.227....1K} {227, 1}

\bibitem[\protect\citeauthoryear{{Kaiser}}{{Kaiser}}{2013}]{2013MNRAS.435.1278K}
{Kaiser} N.,  2013, \mn@doi [\mnras] {10.1093/mnras/stt1370}, \href
  {https://ui.adsabs.harvard.edu/abs/2013MNRAS.435.1278K} {435, 1278}

\bibitem[\protect\citeauthoryear{{Komatsu} et~al.,}{{Komatsu}
  et~al.}{2011}]{2011ApJS..192...18K}
{Komatsu} E.,  et~al., 2011, \mn@doi [¥apjs] {10.1088/0067-0049/192/2/18},
  \href {https://ui.adsabs.harvard.edu/abs/2011ApJS..192...18K} {192, 18}

\bibitem[\protect\citeauthoryear{{Laureijs} et~al.,}{{Laureijs}
  et~al.}{2011}]{2011arXiv1110.3193L}
{Laureijs} R.,  et~al., 2011, arXiv e-prints, \href
  {https://ui.adsabs.harvard.edu/abs/2011arXiv1110.3193L} {p. arXiv:1110.3193}

\bibitem[\protect\citeauthoryear{{Lepori}, {Di Dio}, {Villa}  \&
  {Viel}}{{Lepori} et~al.}{2018}]{2018JCAP...05..043L}
{Lepori} F.,  {Di Dio} E.,  {Villa} E.,   {Viel} M.,  2018, \mn@doi [\jcap]
  {10.1088/1475-7516/2018/05/043}, \href
  {https://ui.adsabs.harvard.edu/abs/2018JCAP...05..043L} {2018, 043}

\bibitem[\protect\citeauthoryear{{Linder}}{{Linder}}{2008}]{2008APh....29..336L}
{Linder} E.~V.,  2008, \mn@doi [Astroparticle Physics]
  {10.1016/j.astropartphys.2008.03.002}, \href
  {https://ui.adsabs.harvard.edu/abs/2008APh....29..336L} {29, 336}

\bibitem[\protect\citeauthoryear{{{\L}okas} \& {Mamon}}{{{\L}okas} \&
  {Mamon}}{2001}]{2001MNRAS.321..155L}
{{\L}okas} E.~L.,  {Mamon} G.~A.,  2001, \mn@doi [\mnras]
  {10.1046/j.1365-8711.2001.04007.x}, \href
  {https://ui.adsabs.harvard.edu/abs/2001MNRAS.321..155L} {321, 155}

\bibitem[\protect\citeauthoryear{{Matsubara}}{{Matsubara}}{2000}]{2000ApJ...535....1M}
{Matsubara} T.,  2000, \mn@doi [\apj] {10.1086/308827}, \href
  {https://ui.adsabs.harvard.edu/abs/2000ApJ...535....1M} {535, 1}

\bibitem[\protect\citeauthoryear{{Matsubara}}{{Matsubara}}{2004}]{2004ApJ...615..573M}
{Matsubara} T.,  2004, \mn@doi [\apj] {10.1086/424561}, \href
  {https://ui.adsabs.harvard.edu/abs/2004ApJ...615..573M} {615, 573}

\bibitem[\protect\citeauthoryear{{McDonald}}{{McDonald}}{2009}]{2009JCAP...11..026M}
{McDonald} P.,  2009, \mn@doi [Journal of Cosmology and Astro-Particle Physics]
  {10.1088/1475-7516/2009/11/026}, \href
  {https://ui.adsabs.harvard.edu/\#abs/2009JCAP...11..026M} {2009, 026}

\bibitem[\protect\citeauthoryear{{Mpetha} et~al.,}{{Mpetha}
  et~al.}{2021}]{2021MNRAS.503..669M}
{Mpetha} C.~T.,  et~al., 2021, \mn@doi [\mnras] {10.1093/mnras/stab544}, \href
  {https://ui.adsabs.harvard.edu/abs/2021MNRAS.503..669M} {503, 669}

\bibitem[\protect\citeauthoryear{{Navarro}, {Frenk}  \& {White}}{{Navarro}
  et~al.}{1996}]{1996ApJ...462..563N}
{Navarro} J.~F.,  {Frenk} C.~S.,   {White} S. D.~M.,  1996, \mn@doi [\apj]
  {10.1086/177173}, \href
  {https://ui.adsabs.harvard.edu/abs/1996ApJ...462..563N} {462, 563}

\bibitem[\protect\citeauthoryear{{Novikov}}{{Novikov}}{1969}]{1969JETP...30..512N}
{Novikov} E.~A.,  1969, Soviet Journal of Experimental and Theoretical Physics,
  \href {https://ui.adsabs.harvard.edu/abs/1969JETP...30..512N} {30, 512}

\bibitem[\protect\citeauthoryear{{P{\'a}pai} \& {Szapudi}}{{P{\'a}pai} \&
  {Szapudi}}{2008}]{2008MNRAS.389..292P}
{P{\'a}pai} P.,  {Szapudi} I.,  2008, \mn@doi [\mnras]
  {10.1111/j.1365-2966.2008.13572.x}, \href
  {https://ui.adsabs.harvard.edu/\#abs/2008MNRAS.389..292P} {389, 292}

\bibitem[\protect\citeauthoryear{{Percival} \& {White}}{{Percival} \&
  {White}}{2009}]{2009MNRAS.393..297P}
{Percival} W.~J.,  {White} M.,  2009, \mn@doi [\mnras]
  {10.1111/j.1365-2966.2008.14211.x}, \href
  {https://ui.adsabs.harvard.edu/abs/2009MNRAS.393..297P} {393, 297}

\bibitem[\protect\citeauthoryear{{Pyne} \& {Birkinshaw}}{{Pyne} \&
  {Birkinshaw}}{2004}]{2004MNRAS.348..581P}
{Pyne} T.,  {Birkinshaw} M.,  2004, \mn@doi [\mnras]
  {10.1111/j.1365-2966.2004.07362.x}, \href
  {https://ui.adsabs.harvard.edu/abs/2004MNRAS.348..581P} {348, 581}

\bibitem[\protect\citeauthoryear{{Reid} et~al.,}{{Reid}
  et~al.}{2012}]{2012MNRAS.426.2719R}
{Reid} B.~A.,  et~al., 2012, \mn@doi [\mnras]
  {10.1111/j.1365-2966.2012.21779.x}, \href
  {https://ui.adsabs.harvard.edu/abs/2012MNRAS.426.2719R} {426, 2719}

\bibitem[\protect\citeauthoryear{{Sadeh}, {Feng}  \& {Lahav}}{{Sadeh}
  et~al.}{2015}]{2015PhRvL.114g1103S}
{Sadeh} I.,  {Feng} L.~L.,   {Lahav} O.,  2015, \mn@doi [\prl]
  {10.1103/PhysRevLett.114.071103}, \href
  {https://ui.adsabs.harvard.edu/abs/2015PhRvL.114g1103S} {114, 071103}

\bibitem[\protect\citeauthoryear{{Saga}, {Taruya}, {Breton}  \&
  {Rasera}}{{Saga} et~al.}{2020}]{2020MNRAS.498..981S}
{Saga} S.,  {Taruya} A.,  {Breton} M.-A.,   {Rasera} Y.,  2020, \mn@doi
  [\mnras] {10.1093/mnras/staa2232}, \href
  {https://ui.adsabs.harvard.edu/abs/2020MNRAS.498..981S} {498, 981}

\bibitem[\protect\citeauthoryear{{S{\'a}nchez} et~al.,}{{S{\'a}nchez}
  et~al.}{2013}]{2013MNRAS.433.1202S}
{S{\'a}nchez} A.~G.,  et~al., 2013, \mn@doi [\mnras] {10.1093/mnras/stt799},
  \href {https://ui.adsabs.harvard.edu/abs/2013MNRAS.433.1202S} {433, 1202}

\bibitem[\protect\citeauthoryear{{Sasaki}}{{Sasaki}}{1987}]{1987MNRAS.228..653S}
{Sasaki} M.,  1987, \mn@doi [\mnras] {10.1093/mnras/228.3.653}, \href
  {https://ui.adsabs.harvard.edu/abs/1987MNRAS.228..653S} {228, 653}

\bibitem[\protect\citeauthoryear{{Shandarin} \& {Zeldovich}}{{Shandarin} \&
  {Zeldovich}}{1989}]{1989RvMP...61..185S}
{Shandarin} S.~F.,  {Zeldovich} Y.~B.,  1989, \mn@doi [Reviews of Modern
  Physics] {10.1103/RevModPhys.61.185}, \href
  {https://ui.adsabs.harvard.edu/abs/1989RvMP...61..185S} {61, 185}

\bibitem[\protect\citeauthoryear{{Sheth} \& {Diaferio}}{{Sheth} \&
  {Diaferio}}{2001}]{2001MNRAS.322..901S}
{Sheth} R.~K.,  {Diaferio} A.,  2001, \mn@doi [\mnras]
  {10.1046/j.1365-8711.2001.04202.x}, \href
  {https://ui.adsabs.harvard.edu/abs/2001MNRAS.322..901S} {322, 901}

\bibitem[\protect\citeauthoryear{{Sheth} \& {Tormen}}{{Sheth} \&
  {Tormen}}{1999}]{1999MNRAS.308..119S}
{Sheth} R.~K.,  {Tormen} G.,  1999, \mn@doi [\mnras]
  {10.1046/j.1365-8711.1999.02692.x}, \href
  {https://ui.adsabs.harvard.edu/abs/1999MNRAS.308..119S} {308, 119}

\bibitem[\protect\citeauthoryear{{Smith}}{{Smith}}{2009}]{2009MNRAS.400..851S}
{Smith} R.~E.,  2009, \mn@doi [\mnras] {10.1111/j.1365-2966.2009.15490.x},
  \href {https://ui.adsabs.harvard.edu/abs/2009MNRAS.400..851S} {400, 851}

\bibitem[\protect\citeauthoryear{{Square Kilometre Array Cosmology Science
  Working Group} et~al.,}{{Square Kilometre Array Cosmology Science Working
  Group} et~al.}{2020}]{2020PASA...37....7S}
{Square Kilometre Array Cosmology Science Working Group} et~al., 2020, \mn@doi
  [\pasa] {10.1017/pasa.2019.51}, \href
  {https://ui.adsabs.harvard.edu/abs/2020PASA...37....7S} {37, e007}

\bibitem[\protect\citeauthoryear{{Szalay}, {Matsubara}  \& {Landy}}{{Szalay}
  et~al.}{1998}]{1998ApJ...498L...1S}
{Szalay} A.~S.,  {Matsubara} T.,   {Landy} S.~D.,  1998, \mn@doi [\apjl]
  {10.1086/311293}, \href
  {https://ui.adsabs.harvard.edu/abs/1998ApJ...498L...1S} {498, L1}

\bibitem[\protect\citeauthoryear{{Szapudi}}{{Szapudi}}{2004}]{2004ApJ...614...51S}
{Szapudi} I.,  2004, \mn@doi [\apj] {10.1086/423168}, \href
  {https://ui.adsabs.harvard.edu/abs/2004ApJ...614...51S} {614, 51}

\bibitem[\protect\citeauthoryear{{Takada} et~al.,}{{Takada}
  et~al.}{2014}]{2014PASJ...66R...1T}
{Takada} M.,  et~al., 2014, \mn@doi [\pasj] {10.1093/pasj/pst019}, \href
  {https://ui.adsabs.harvard.edu/abs/2014PASJ...66R...1T} {66, R1}

\bibitem[\protect\citeauthoryear{{Tansella}, {Bonvin}, {Durrer}, {Ghosh}  \&
  {Sellentin}}{{Tansella} et~al.}{2018}]{2018JCAP...03..019T}
{Tansella} V.,  {Bonvin} C.,  {Durrer} R.,  {Ghosh} B.,   {Sellentin} E.,
  2018, \mn@doi [\jcap] {10.1088/1475-7516/2018/03/019}, \href
  {https://ui.adsabs.harvard.edu/abs/2018JCAP...03..019T} {2018, 019}

\bibitem[\protect\citeauthoryear{{Taruya}, {Saga}, {Breton}, {Rasera}  \&
  {Fujita}}{{Taruya} et~al.}{2020}]{2020MNRAS.491.4162T}
{Taruya} A.,  {Saga} S.,  {Breton} M.-A.,  {Rasera} Y.,   {Fujita} T.,  2020,
  \mn@doi [\mnras] {10.1093/mnras/stz3272}, \href
  {https://ui.adsabs.harvard.edu/abs/2020MNRAS.491.4162T} {491, 4162}

\bibitem[\protect\citeauthoryear{{Tinker}, {Kravtsov}, {Klypin}, {Abazajian},
  {Warren}, {Yepes}, {Gottl{\"o}ber}  \& {Holz}}{{Tinker}
  et~al.}{2008}]{2008ApJ...688..709T}
{Tinker} J.,  {Kravtsov} A.~V.,  {Klypin} A.,  {Abazajian} K.,  {Warren} M.,
  {Yepes} G.,  {Gottl{\"o}ber} S.,   {Holz} D.~E.,  2008, \mn@doi [\apj]
  {10.1086/591439}, \href
  {https://ui.adsabs.harvard.edu/abs/2008ApJ...688..709T} {688, 709}

\bibitem[\protect\citeauthoryear{{Tinker}, {Robertson}, {Kravtsov}, {Klypin},
  {Warren}, {Yepes}  \& {Gottl{\"o}ber}}{{Tinker}
  et~al.}{2010}]{2010ApJ...724..878T}
{Tinker} J.~L.,  {Robertson} B.~E.,  {Kravtsov} A.~V.,  {Klypin} A.,  {Warren}
  M.~S.,  {Yepes} G.,   {Gottl{\"o}ber} S.,  2010, \mn@doi [\apj]
  {10.1088/0004-637X/724/2/878}, \href
  {https://ui.adsabs.harvard.edu/abs/2010ApJ...724..878T} {724, 878}

\bibitem[\protect\citeauthoryear{{Wojtak}, {Hansen}  \& {Hjorth}}{{Wojtak}
  et~al.}{2011}]{2011Natur.477..567W}
{Wojtak} R.,  {Hansen} S.~H.,   {Hjorth} J.,  2011, \mn@doi [\nat]
  {10.1038/nature10445}, \href
  {https://ui.adsabs.harvard.edu/abs/2011Natur.477..567W} {477, 567}

\bibitem[\protect\citeauthoryear{{Yahya}, {Bull}, {Santos}, {Silva},
  {Maartens}, {Okouma}  \& {Bassett}}{{Yahya}
  et~al.}{2015}]{2015MNRAS.450.2251Y}
{Yahya} S.,  {Bull} P.,  {Santos} M.~G.,  {Silva} M.,  {Maartens} R.,  {Okouma}
  P.,   {Bassett} B.,  2015, \mn@doi [\mnras] {10.1093/mnras/stv695}, \href
  {https://ui.adsabs.harvard.edu/abs/2015MNRAS.450.2251Y} {450, 2251}

\bibitem[\protect\citeauthoryear{{Yoo}}{{Yoo}}{2010}]{2010PhRvD..82h3508Y}
{Yoo} J.,  2010, \mn@doi [\prd] {10.1103/PhysRevD.82.083508}, \href
  {https://ui.adsabs.harvard.edu/abs/2010PhRvD..82h3508Y} {82, 083508}

\bibitem[\protect\citeauthoryear{{Yoo}}{{Yoo}}{2014}]{2014CQGra..31w4001Y}
{Yoo} J.,  2014, \mn@doi [Classical and Quantum Gravity]
  {10.1088/0264-9381/31/23/234001}, \href
  {https://ui.adsabs.harvard.edu/abs/2014CQGra..31w4001Y} {31, 234001}

\bibitem[\protect\citeauthoryear{{Yoo}, {Fitzpatrick}  \& {Zaldarriaga}}{{Yoo}
  et~al.}{2009}]{2009PhRvD..80h3514Y}
{Yoo} J.,  {Fitzpatrick} A.~L.,   {Zaldarriaga} M.,  2009, \mn@doi [\prd]
  {10.1103/PhysRevD.80.083514}, \href
  {https://ui.adsabs.harvard.edu/abs/2009PhRvD..80h3514Y} {80, 083514}

\bibitem[\protect\citeauthoryear{{Yoo}, {Hamaus}, {Seljak}  \&
  {Zaldarriaga}}{{Yoo} et~al.}{2012}]{2012arXiv1206.5809Y}
{Yoo} J.,  {Hamaus} N.,  {Seljak} U.,   {Zaldarriaga} M.,  2012, arXiv
  e-prints, \href {https://ui.adsabs.harvard.edu/abs/2012arXiv1206.5809Y} {p.
  arXiv:1206.5809}

\bibitem[\protect\citeauthoryear{{Zaroubi} \& {Hoffman}}{{Zaroubi} \&
  {Hoffman}}{1996}]{1996ApJ...462...25Z}
{Zaroubi} S.,  {Hoffman} Y.,  1996, \mn@doi [\apj] {10.1086/177124}, \href
  {https://ui.adsabs.harvard.edu/abs/1996ApJ...462...25Z} {462, 25}

\bibitem[\protect\citeauthoryear{{Zel'dovich}}{{Zel'dovich}}{1970}]{1970A&A.....5...84Z}
{Zel'dovich} Y.~B.,  1970, \aap, \href
  {https://ui.adsabs.harvard.edu/abs/1970A%26A.....5...84Z} {5, 84}

\bibitem[\protect\citeauthoryear{{Zentner}, {Berlind}, {Bullock}, {Kravtsov}
  \& {Wechsler}}{{Zentner} et~al.}{2005}]{2005ApJ...624..505Z}
{Zentner} A.~R.,  {Berlind} A.~A.,  {Bullock} J.~S.,  {Kravtsov} A.~V.,
  {Wechsler} R.~H.,  2005, \mn@doi [\apj] {10.1086/428898}, \href
  {https://ui.adsabs.harvard.edu/abs/2005ApJ...624..505Z} {624, 505}

\bibitem[\protect\citeauthoryear{{Zhao}, {Peacock}  \& {Li}}{{Zhao}
  et~al.}{2013}]{2013PhRvD..88d3013Z}
{Zhao} H.,  {Peacock} J.~A.,   {Li} B.,  2013, \mn@doi [\prd]
  {10.1103/PhysRevD.88.043013}, \href
  {https://ui.adsabs.harvard.edu/abs/2013PhRvD..88d3013Z} {88, 043013}

\bibitem[\protect\citeauthoryear{{Zhu}, {Alam}, {Croft}, {Ho}  \&
  {Giusarma}}{{Zhu} et~al.}{2017}]{2017MNRAS.471.2345Z}
{Zhu} H.,  {Alam} S.,  {Croft} R. A.~C.,  {Ho} S.,   {Giusarma} E.,  2017,
  \mn@doi [\mnras] {10.1093/mnras/stx1644}, \href
  {https://ui.adsabs.harvard.edu/abs/2017MNRAS.471.2345Z} {471, 2345}

\makeatother
\end{thebibliography}
\appendix

\section{Derivations of the multipole moments}
\label{app: derivation}

In this appendix, we summarize key expressions to derive the dipole cross-correlation function presented in Sec.~\ref{sec: correlation function}.

Based on the density fields given at Eq.~(\ref{eq: delta std + delta epsNL}) together with Eqs.~(\ref{eq: def delta std})--(\ref{eq: def delta epsNL}), let us first compute cross-correlation function. Substituting these equations into Eq.~(\ref{eq: xi std+rel+eps}), we obtain
\begin{align}
\xi^{({\rm std})}_{\rm XY} & =
\int\frac{{\rm d}^{3}k}{(2\pi)^{3}}{\rm e}^{{\rm i}\bm{k}\cdot\bm{s}}
\left( b^{\rm E}_{\rm X} + f\mu^{2}_{k1} + {\rm i} f \frac{2}{ks_{1}}\mu_{k1}\right)
\notag \\ 
& \qquad \times 
\left( b^{\rm E}_{\rm Y} + f\mu^{2}_{k2} - {\rm i} f \frac{2}{ks_{2}}\mu_{k2} \right)
P_{\rm L}(k)
\label{eq: xi std full appendix}
~,\\
\xi^{({\rm pot})}_{\rm XY} & =
\int\frac{{\rm d}^{3}k}{(2\pi)^{3}}{\rm e}^{i\bm{k}\cdot\bm{s}}
\Biggl[
\left( b^{\rm E}_{\rm X} + f\mu^{2}_{k1} + {\rm i} f \frac{2}{ks_{1}}\mu_{k1}\right)
\left( {\rm i}k \mu_{k2} + \frac{2}{s_{2}}\right) 
\notag \\
& \qquad
+ 
\left( b^{\rm E}_{\rm Y} + f\mu^{2}_{k2} - {\rm i} f \frac{2}{ks_{2}}\mu_{k2} \right)
\left( - {\rm i}k \mu_{k1} + \frac{2}{s_{1}}\right) 
\Biggr]
\frac{\mathcal{M}}{k^{2}}
P_{\rm L}(k) 
\label{eq: xi grav full appendix}
~,\\
\xi^{(\epsilon_{\rm NL})}_{\rm XY} & = 
\int\frac{{\rm d}^{3}k}{(2\pi)^{3}}{\rm e}^{{\rm i}\bm{k}\cdot\bm{s}}
\Biggl[
\frac{\epsilon_{\rm NL, X}}{s_{1}}
\Biggl(
-1 + \mu^{2}_{k1} + {\rm i}f\frac{2}{ks_{1}}\mu_{k1}
+ {\rm i} b^{\rm E}_{\rm X} ks_{1} \mu_{k1}
\notag \\
& 
-2f\mu^{2}_{k1} + {\rm i} \frac{2}{ks_{1}}\mu_{k1} + {\rm i} f ks_{1} \mu^{3}_{k1}
\Biggr)
\Biggl(
b^{\rm E}_{\rm Y} + f\mu^{2}_{k2} - {\rm i} f \frac{2}{ks_{2}}\mu_{k2}
\Biggr)
\notag \\
&+
\frac{\epsilon_{\rm NL, Y}}{s_{2}}
\Biggl(
-1 + \mu^{2}_{k2} - {\rm i} f\frac{2}{ks_{2}}\mu_{k2}
- {\rm i} b^{\rm E}_{\rm Y} ks_{2} \mu_{k2}-2f\mu^{2}_{k2}
\notag \\
&
- {\rm i}\frac{2}{ks_{2}}\mu_{k2} - {\rm i} f ks_{2} \mu^{3}_{k2}
\Biggr)
\Biggl(
b^{\rm E}_{\rm X} + f\mu^{2}_{k1} + {\rm i} f \frac{2}{ks_{1}}\mu_{k1}
\Biggr)
\Biggr]
P_{\rm L}(k)
\label{eq: xi eps full appendix}
~,
\end{align}
where we define $\mu_{k1} = \hat{\bm{s}}_{1}\cdot\hat{\bm{k}}$ and $\mu_{k2} = \hat{\bm{s}}_{2}\cdot\hat{\bm{k}}$. The function $P_{\rm L}(k)$ stands for the linear power spectrum of the density field $\delta_{\rm L}$ given by
\begin{align}
\Braket{\delta_{\rm L}(\bm{k})\delta_{\rm L}(\bm{k}')} = (2\pi)^{3}\delta_{\rm D}(\bm{k}+\bm{k}')P_{\rm L}(k) ~. \label{eq: def PL(k)}
\end{align}

Eqs.~(\ref{eq: xi std full appendix})--(\ref{eq: xi eps full appendix}) involve the three-dimensional integrals over $\bm{k}$. Introducing the polar coordinate, the angular integral can be performed by using the following formulae:
\begin{align}
& \int{\frac{{\rm d}\Omega_{\bm{k}}}{4\pi}e^{i\bm{k}\cdot\bm{s}}} = j_{0}(ks) ~, \\
& \int{\frac{{\rm d}\Omega_{\bm{k}}}{4\pi}e^{i\bm{k}\cdot\bm{s}}} \left( i\hat{k}_{a} \right)
 = - j_{1}(ks)\hat{s}_{a} ~, \\
& \int{\frac{{\rm d}\Omega_{\bm{k}}}{4\pi}e^{i\bm{k}\cdot\bm{s}}} \left( \hat{k}_{a}\hat{k}_{b} \right)
 = - j_{2}(ks)\hat{s}_{a}\hat{s}_{b} + \frac{j_{1}(ks)}{ks}\delta_{ab} ~, \\
& \int{\frac{{\rm d}\Omega_{\bm{k}}}{4\pi}e^{i\bm{k}\cdot\bm{s}}} \left( i \hat{k}_{a}\hat{k}_{b}\hat{k}_{c} \right)
 = j_{3}(ks)\hat{s}_{a}\hat{s}_{b}\hat{s}_{c}
\notag \\
& \qquad\qquad\qquad
- \frac{j_{2}(ks)}{ks}(\hat{s}_{a}\delta_{bc} + \hat{s}_{b}\delta_{ca} + \hat{s}_{c}\delta_{ab}) ~, \\
& \int{\frac{{\rm d}\Omega_{\bm{k}}}{4\pi}e^{i\bm{k}\cdot\bm{s}}} \left( \hat{k}_{a}\hat{k}_{b}\hat{k}_{c}\hat{k}_{d} \right) 
 = j_{4}(ks)\hat{s}_{a}\hat{s}_{b}\hat{s}_{c}\hat{s}_{d}
\notag \\
& \qquad\qquad\qquad
- \frac{j_{3}(ks)}{ks}(\hat{s}_{a}\hat{s}_{b}\delta_{cd} + \hat{s}_{b}\hat{s}_{c}\delta_{ad} + 	\hat{s}_{b}\hat{s}_{d}\delta_{ac}
\notag \\
& \qquad\qquad\qquad\qquad\qquad
+ \hat{s}_{a}\hat{s}_{c}\delta_{bd} + \hat{s}_{a}\hat{s}_{d}\delta_{bc} + \hat{s}_{c}\hat{s}_{d}\delta_{ab})
\notag \\
& \qquad\qquad\qquad
 + \frac{j_{2}(ks)}{(ks)^{2}} (\delta_{ad}\delta_{bc} + \delta_{ac}\delta_{bd} + \delta_{ab}\delta_{cd}) ~, \\
& \int{\frac{{\rm d}\Omega_{\bm{k}}}{4\pi}e^{i\bm{k}\cdot\bm{s}}} \left( i\hat{k}_{a}\hat{k}_{b}\hat{k}_{c}\hat{k}_{d}\hat{k}_{e} \right) 
 = - j_{5}(ks)\hat{s}_{a}\hat{s}_{b}\hat{s}_{c}\hat{s}_{d}\hat{s}_{e}
\notag \\
& \qquad\qquad\qquad
+ \frac{j_{4(ks)}}{ks} (\hat{s}_{a}\hat{s}_{b}\hat{s}_{c}\delta_{de} + \mbox{9 perm.})
\notag \\
& \qquad\qquad\qquad
- \frac{j_{3(ks)}}{(ks)^{2}} (\hat{s}_{a}\delta_{bc}\delta_{de} + \mbox{14 perm.}) ~,
\end{align}
where $j_{\ell}$ stands for the spherical Bessel function.

As a result of the angular integration, the dependence of the correlation function on the vectors $\bm{s}_{1}$ and $\bm{s}_{2}$ in Eqs.~(\ref{eq: xi std full appendix})--(\ref{eq: xi eps full appendix}) is shown to be described by the following quantities:  $(\hat{\bm{s}}\cdot\hat{\bm{s}}_{1})$, $(\hat{\bm{s}}\cdot\hat{\bm{s}}_{2})$, $(\hat{\bm{s}}_{1}\cdot\hat{\bm{s}}_{2})$, $s_{1}$, and $s_{2}$. Note that these are re-expressed in terms of the three variables, i.e., separation $s = |\bm{s}_{2} - \bm{s}_{1}|$, the line-of-sight distance $d = |\bm{s}_{1} + \bm{s}_{2}|/2$, and directional cosine $\mu = \hat{\bm{s}} \cdot \hat{\bm{d}}$. Since we are interested in the cases with $s\ll d$, one can expand the quantities as   
\begin{align}
s_{1} &= d\left( 1 - \frac{s}{d}\mu + \frac{1}{4}\left( \frac{s}{d}\right)^{2}\right)^{1/2}
\simeq d\left( 1 - \frac{1}{2}\frac{s}{d} \mu\right) 
~, \\
s_{2} &= d\left( 1 + \frac{s}{d}\mu + \frac{1}{4}\left( \frac{s}{d}\right)^{2}\right)^{1/2}
\simeq d\left( 1 + \frac{1}{2}\frac{s}{d} \mu\right)
~, \\
(\hat{\bm{s}}\cdot\hat{\bm{s}}_{1}) &= \frac{\mu - \frac{1}{2}\frac{s}{d}}{\left( 1 - \frac{s}{d}\mu + \frac{1}{4}\left( \frac{s}{d} \right)^{2}\right)^{1/2}}
\simeq \mu - \frac{1}{2}(1-\mu^{2})\frac{s}{d}
~,\\
(\hat{\bm{s}}\cdot\hat{\bm{s}}_{2}) &= \frac{\mu + \frac{1}{2}\frac{s}{d}}{\left( 1 + \frac{s}{d}\mu + \frac{1}{4}\left( \frac{s}{d} \right)^{2}\right)^{1/2}}
\simeq \mu + \frac{1}{2}(1-\mu^{2})\frac{s}{d}
~,\\
(\hat{\bm{s}}_{1}\cdot\hat{\bm{s}}_{2}) &=
\frac{1 - \frac{1}{4}\left( \frac{s}{d} \right)^{2}}{\left( 1 - \frac{s}{d}\mu + \frac{1}{4}\left( \frac{s}{d} \right)^{2}\right)^{1/2}\left( 1 + \frac{s}{d}\mu + \frac{1}{4}\left( \frac{s}{d} \right)^{2}\right)^{1/2}}
\simeq 1
~,
\end{align}
where the last equalities in each equation is valid at $\mathcal{O}( s/d)$. Substituting these expressions into the cross-correlation function, the results are divided into the plane-parallel ($d\to\infty$) and leading-order wide-angle contributions ($O\left( s/d\right)$), in which the dependence of the directional cosine is factorized, and is expressed as a polynomial form of $\mu$. Thus, applying the multipole expansion, one easily derives the analytical expression for the multipole correlation functions, summarized in Appendix~\ref{app: other moments}.

\section{Comparison with approximate formula in \citet{2020MNRAS.498..981S}}
\label{app: comparison formula}

Employing the Zel'dovich approximation and combining the non-perturbative contribution from the halo potential, \citet{2020MNRAS.498..981S} have built a quasi-linear model of the dipole cross-correlation function, which successfully explains numerical simulations at both small and large scales. While a rigorous treatment of their model requires the time-consuming multi-dimensional integration, they also derived a simple approximate expression for the dipole moment, which resembles the analytical model presented in this paper. 
In this appendix, we clarify the similarity and difference between the approximate expression derived in Sec.~\ref{sec: model approx} and the one obtained from \citet{2020MNRAS.498..981S} (see their Eq.~(4.2) in Sec.~4.2)).

In \citet{2020MNRAS.498..981S}, 
the simplified expression of the dipole was derived based on a perturbative treatment of their rigorous quasi-linear model. Ignoring the non-perturbative halo potential, let us first denote the cross-correlation function of their model by $\xi_{\rm XY, \epsilon_{\rm NL}=0}(\bm{s}_1,\bm{s}_2)$. We then consider the gravitational redshift contributions arising from the non-perturbative halo potential, which gives a systematic offset of the redshift-space positions away from the observer (origin), i.e.,  $\bm{s}_{1,2}\to\bm{s}_{1,2}-\epsilon_{\rm NL, X/Y}\hat{\bm{s}}_{1,2}$. The resultant cross-correlation function taking the halo potential into account, $\xi_{\rm XY}$, is expressed as
\begin{align}
\xi_{\rm XY}(\bm{s}_{1},\bm{s}_{2})
&= \xi_{\rm XY, \epsilon_{\rm NL}=0} \left( \bm{s}_{1}-\epsilon_{\rm NL, X}\hat{\bm{s}}_{1},\, \bm{s}_{2}-\epsilon_{\rm NL, Y}\hat{\bm{s}}_{2} \right) 
\nonumber
\\
&\simeq\Bigl[1- \left\{ \epsilon_{\rm NL,X}\; \hat{\bm{s}}_{1}\cdot\bm{\nabla}_{s_{1}}
+ \epsilon_{\rm NL,Y}\; \hat{\bm{s}}_{2}\cdot\bm{\nabla}_{s_{2}}\right\}\Bigr] 
\nonumber
\\
&\qquad \times
\xi_{\rm XY,\epsilon_{\rm NL}=0}(\bm{s}_{1},\bm{s}_{2})
~.
\end{align}
Here, in the second equality, the systematic offset caused by the halo potential is treated as a small perturbation and is expanded at linear order, as similarly done by \citet{2020MNRAS.498..981S}.

Note that expanding the displacement field $\bm{\Psi}$ from the exponent and truncating it at linear order, the cross-correlation function $\xi_{\rm XY, \epsilon_{\rm NL}=0}$ is shown to be identical to the cross-correlation function $\xi^{({\rm std})}_{\rm XY}
+ \xi^{({\rm pot})}_{\rm XY}$ given in this paper  (see Eq.~(\ref{eq: xi std+rel+eps})). With this linearized treatment, the above expression is reduced to
\begin{align}
\xi_{\rm XY}(\bm{s}_{1},\bm{s}_{2}) &
\simeq \xi^{({\rm std})}_{\rm XY}(\bm{s}_{1},\bm{s}_{2})
+ \xi^{({\rm pot})}_{\rm XY}(\bm{s}_{1},\bm{s}_{2})
\notag \\
& \qquad 
- \left[ \epsilon_{\rm NL,X}\; \hat{\bm{s}}_{1}\cdot\bm{\nabla}_{s_{1}}
+ \epsilon_{\rm NL,Y}\; \hat{\bm{s}}_{2}\cdot\bm{\nabla}_{s_{2}}\right] \xi^{({\rm std})}_{\rm XY}(\bm{s}_{1},\bm{s}_{2})
\notag \\
& 
\equiv \xi^{({\rm std})}_{\rm XY}(\bm{s}_{1},\bm{s}_{2})
+ \xi^{({\rm pot})}_{\rm XY}(\bm{s}_{1},\bm{s}_{2})
+ \widetilde{\xi}^{({\epsilon_{\rm NL}})}_{\rm XY}(\bm{s}_{1},\bm{s}_{2})
~, \label{eq: expand xi}
\end{align}
where, in the first equality, we used the fact that the term $\xi^{({\rm pot})}_{\rm XY}$ only gives a sub-dominant contribution, and the contribution proportional to $\epsilon_{\rm NL,X/Y}\,\xi_{\rm XY}^{\rm(pot)}$ have been ignored from the second line. 
The function $\widetilde{\xi}^{({\epsilon_{\rm NL}})}_{\rm XY}(\bm{s}_{1},\bm{s}_{2})$ is explicitly given by
\begin{align}
\widetilde{\xi}^{({\epsilon_{\rm NL}})}_{\rm XY}(\bm{s}_{1},\bm{s}_{2})
&= \frac{\epsilon_{\rm NL, X}}{s_{1}}\int\frac{{\rm d}^{3}\bm{k}}{(2\pi)^{3}}\; {\rm e}^{{\rm i}\bm{k}\cdot \bm{s}}
\left( b_{\rm Y} + f\mu^{2}_{k2} - {\rm i} \frac{2f}{ks_{2}}\mu_{k2}\right)
\notag\\
&\times
\Biggl[
({\rm i} ks_{1}\mu_{k1})\left( b_{\rm X} + f\mu^{2}_{k1} + {\rm i} \frac{2f}{ks_{1}}\mu_{k1}\right)
+ {\rm i}\frac{2f}{k s_{1}}\mu_{k1}
\Biggr]
P_{\rm L}(k)
\notag \\
& 
+ \left( {\rm X}\leftrightarrow {\rm Y},\ s_{1}\leftrightarrow s_{2},\ \mu_{1}\leftrightarrow -\mu_{2}\right) ~. \label{eq: expand xi eps}
\end{align}

Thus, comparing Eq.~(\ref{eq: expand xi}) with the analytical model in Sec.~\ref{sec: model approx}, the difference essentially appears at the gravitational redshift contribution from the halo potential, i.e., $\widetilde{\xi}^{({\epsilon_{\rm NL}})}_{\rm XY}$ and $\xi^{({\epsilon_{\rm NL}})}_{\rm XY}$.  Taking their difference gives
\begin{align}
\xi^{({\epsilon_{\rm NL}})}_{\rm XY} - \widetilde{\xi}^{({\epsilon_{\rm NL}})}_{\rm XY}
& = 
\frac{\epsilon_{\rm NL, X}}{s_{1}}
\int\frac{{\rm d}^{3}\bm{k}}{(2\pi)^{3}}{\rm e}^{{\rm i}\bm{k}\cdot\bm{s}}
\Biggl( b_{\rm Y} + f\mu^{2}_{k2} - {\rm i} f \frac{2}{ks_{2}}\mu_{k2} \Biggr)
\notag \\
& \qquad \times 
\Biggl( -1 + \mu^{2}_{k1} + {\rm i}f\frac{2}{ks_{1}}\mu_{k1} \Biggr) P_{\rm L}(k)
\notag \\
& 
+ \left( {\rm X}\leftrightarrow {\rm Y},\ s_{1}\leftrightarrow s_{2},\ \mu_{1}\leftrightarrow -\mu_{2}\right) ~. 
\end{align}
As explicitly demonstrated in Fig.~\ref{fig: dipole}, this produces a  rather small difference, and the simple approximation presented in  \citet{2020MNRAS.498..981S} leads to the prediction of the dipole moment almost identical to the one from the present analytical model. 
\section{Multipole coefficients}
\label{app: other moments}

Here, we present the analytical expressions for the multipole moments of the cross-correlation functions. As we discussed in previous Appendix and Sec.~\ref{sec: correlation function}, the correlation function can be written as a function of the separation $s = |\bm{s}_{2}-\bm{s}_{1}|$, line-of-sight distance $d = |(\bm{s}_{1}+\bm{s}_{2})/2|$, and directional cosine between the line-of-sight and separation vectors, given by $\mu = \hat{\bm{s}}\cdot\hat{\bm{d}}$.
Based on the results in Appendix \ref{app: derivation}, the cross-correlation function can be expanded in powers of $(s/d)$. Further applying the multipole expansion, we obtain:
\begin{align}
\xi_{\rm XY}(s,d,\mu)
&= \sum_{\ell} \xi_{{\rm XY}, \ell}(s,d) \mathcal{L}_{\ell}(\mu) \\
&= \sum_{\ell} \Biggl[ \xi_{{\rm pp}, \ell}(s) + \left( \frac{s}{d}\right)\xi_{{\rm wa}, \ell}(s) + O\left( \left( \frac{s}{d} \right)^{2} \right) \Bigr] \mathcal{L}_{\ell}(\mu) ~, 
\end{align}
where the functions $\xi_{{\rm pp}, \ell}(s)$ and $\xi_{{\rm wa}, \ell}(s)$ respectively represent the contribution in the plane-parallel limit and wide-angle correction at leading order. These expressions involve only the one-dimensional integral given by
\begin{align}
\xi_{{\rm pp}, \ell}(s) &= (-{\rm i})^{\ell}\int \frac{k^{2}\, {\rm d}k}{2\pi^{2}}\,
P_{\rm pp, \ell}(k,z)j_{\ell}(ks) 
~, \\
\xi_{{\rm wa}, \ell}(s) &= (-{\rm i})^{\ell}\int \frac{k^{2}\, {\rm d}k}{2\pi^{2}}\, 
P_{\rm wa, \ell}(k,z) ~.
\end{align}
Below, we separately present the analytical expressions for the functions $P_{\rm pp,\ell}$ and $P_{\rm wa,\ell}$. While we focus on the dipole moment $(\ell=1)$ in the main text, we summarize all the non-vanishing moments valid at the order of $\mathcal{O}(s/d)$.

\subsection{Plane-parallel limit}
\label{app:plane-parallel_limit}

The non-vanishing multipoles in the plane-parallel limit are summarized as follows:
\begin{align}
P^{(\rm std)}_{\rm pp,0} & =
\left[ b_{\rm X}b_{\rm Y} + \frac{1}{3}(b_{\rm X}+b_{\rm Y})f + \frac{1}{5}f^{2} \right] P_{\rm L}(k)
~, \\
P^{(\rm std)}_{\rm pp,2} & = 
\left[ \frac{2}{3}f(b_{\rm X} + b_{\rm Y}) + \frac{4}{7}f^{2}\right] P_{\rm L}(k) ~, \\
P^{(\rm std)}_{\rm pp,4} & = \left[ \frac{8}{35}f^{2}\right] P_{\rm L}(k) ~,
\end{align}
for the standard Doppler contribution,
\begin{align}
P^{(\rm pot)}_{\rm pp,1} & =
\left[ - {\rm i} (b_{\rm X}-b_{\rm Y})\frac{\mathcal{M}}{k} \right] P_{\rm L}(k)~,
\end{align}
for the linear gravitational redshift contribution, and 
\begin{align}
P^{(\epsilon_{\rm NL})}_{\rm pp,1} & = \left[ - {\rm i} (\epsilon_{\rm NL,X}-\epsilon_{\rm NL,Y}) \left( b_{\rm X}b_{\rm Y} + \frac{3}{5}(b_{\rm X}+b_{\rm Y})f + \frac{3}{7}f^{2}\right) k \right] P_{\rm L}(k) ~, \\
P^{(\epsilon_{\rm NL})}_{\rm pp,3} & = \left[ - {\rm i} (\epsilon_{\rm NL,X}-\epsilon_{\rm NL,Y})\frac{2}{45}f\left( 9(b_{\rm X}+b_{\rm Y}) + 10 f\right)
k \right] P_{\rm L}(k)
~, \\
P^{(\epsilon_{\rm NL})}_{\rm pp,5} & = 
\left[ - {\rm i} \frac{8}{63} f^{2}(\epsilon_{\rm NL,X}-\epsilon_{\rm NL,Y}) k \right] P_{\rm L}(k) ~,
\end{align}
for the contribution arising from the non-perturbative halo potential.

\subsection{Wide-angle correction}

The non-vanishing multipoles of the wide-angle correction are summarized as follows:
\begin{align}
P^{(\rm std)}_{\rm wa,1} & ={\rm i} 2f(b_{\rm X}-b_{\rm Y}) \left[ - \frac{1}{5} j_{2}(ks) 
+ \frac{j_{1}(ks)}{ks}\right] P_{\rm L}(k) ~, \\
P^{(\rm std)}_{\rm wa,3} & = \left[ - {\rm i} \frac{2f}{5}(b_{\rm X}-b_{\rm Y}) j_{2}(ks) \right] P_{\rm L}(k) ~.
\end{align}
for the standard Doppler contribution,
\begin{align}
P^{(\rm pot)}_{\rm wa,0} & = \frac{\mathcal{M}}{k} \Biggl[ - \frac{1}{3} \left( b_{\rm X} + b_{\rm Y} -\frac{2f}{5} \right) j_{1}(ks)
\notag \\
& \qquad \qquad 
+ 2(b_{\rm X}+b_{\rm Y})\frac{j_{0}(ks)}{ks}
\Biggr] P_{\rm L}(k)~, \\
P^{(\rm pot)}_{\rm wa,2} & = \frac{\mathcal{M}}{k} \left[- \frac{1}{3} \left( b_{\rm X} + b_{\rm Y} - \frac{2}{5}f \right) j_{1}(ks) 
+ \frac{8f}{35} j_{3}(ks) \right] P_{\rm L}(k) ~, \\
P^{(\rm pot)}_{\rm wa,4} & = \left[ \frac{8}{35}\frac{\mathcal{M}f}{k} j_{3}(ks) \right] P_{\rm L}(k)~.
\end{align}
for the linear gravitational redshift contribution, and 
\begin{align}
P^{(\epsilon_{\rm NL})}_{\rm wa,0} = &
\Biggl[ \Biggl\{
-\frac{2f^{2}}{5}
-\frac{2}{3}b_{\rm Y}
-\frac{2f}{15}(1 + 5b_{\rm Y})
\Biggr\} ks\, j_{-1}(ks)
\notag \\
&+\Biggl\{
-\frac{13f^{2}}{35}
- \frac{f}{15} (2+b_{\rm X}+7b_{\rm Y})
+ \frac{1}{3} (b_{\rm X}-2)b_{\rm Y}
\Biggr\}
ks\, j_{1}(ks)
\notag \\
& + \frac{2f}{15}\Bigl( 5b_{\rm X} + 3f \Bigr) j_{0}(ks)
\Biggr]
\frac{ \epsilon_{\rm NL,X} }{s}\, P_{\rm L}(k)
\notag \\
& + ({\rm X} \leftrightarrow {\rm Y})
~, \\
P^{(\epsilon_{\rm NL})}_{\rm wa,2} = &
\Biggl[
- \frac{1}{5}\Biggl\{ f^{2}
+ \frac{f}{21}(2+7b_{\rm X}+7b_{\rm Y})
- \frac{1}{3}(2+5b_{\rm X})b_{\rm Y}
\Biggr\} ks j_{1}(ks)
\notag \\
& - \frac{1}{15}\Biggl\{
\frac{16f^{2}}{7} - 2b_{\rm Y}
+ \frac{2f}{7} (1 + 6b_{\rm X}- 4 b_{\rm Y})
\Biggr\}
ks j_{3}(ks)
\notag \\
& + \frac{4f}{21}( 7b_{\rm X} + 6f ) j_{2}(ks)
\Biggr]
\frac{ \epsilon_{\rm NL,X}}{s}P_{\rm L}(k)
\notag \\
& + ({\rm X} \leftrightarrow {\rm Y})
~, \\
P^{(\epsilon_{\rm NL})}_{\rm wa,4} = &
\Biggl[
\frac{4}{315}f (2-9b_{\rm X}+27b_{\rm Y}+2f)
ks j_{3}(ks)
\notag \\
&
+ \frac{8f}{3465}(11-7f) ks j_{5}(ks)
\Biggr]
\frac{\epsilon_{\rm NL,X} }{s}P_{\rm L}(k)
\notag \\
& + ({\rm X} \leftrightarrow {\rm Y})
~, \\
P^{(\epsilon_{\rm NL})}_{\rm wa,6} = &
\Biggl[
\frac{8}{231} f^{2} ks j_{5}(ks)
+ \frac{16f^{2}}{35} j_{4}(ks)
\Biggr]
\frac{\epsilon_{\rm NL,X}}{s}P_{\rm L}(k)
\notag \\
& + ({\rm X} \leftrightarrow {\rm Y})~.
\end{align}
for the non-perturbative contribution.

\section{On the impact of the magnification bias}
\label{app: magnification}

In this appendix, we discuss the impact of the magnification bias on the dipole signal. In general, flux limited galaxy samples inherently lead to the apparent density fluctuations through the fluctuation in luminosity distance, referred to as the magnification bias, which also induces the additional dipole signal beyond the plane-parallel limit~\citep{2011PhRvD..84f3505B,2017PhRvD..95d3530H}. The magnification bias mainly comes from two contributions: one is the lensing magnification and another is the Doppler magnification, among which the latter has been shown to produce a larger dipole signal~\citep{2017PhRvD..95d3530H}. 
At linear order, the Doppler magnification modulates the standard Doppler term. To be precise, the factor of $2/s$ in the last term at Eq.~(\ref{eq: def delta std}) is changed to  $2/s \to 5s_{\rm B}aH + (2-5s_{\rm B})/s$, where the quantity $s_{\rm B}$ is the slope of the luminosity function \citep[e.g.,][]{2011PhRvD..84f3505B,2017PhRvD..95d3530H}. Here, incorporating these contributions into our analytical model, we estimate the impact of the Doppler magnification on the dipole signal.

Coupling with other terms in the density field, the modulation due to the Doppler magnification mentioned above yields the following new contributions to the dipole cross correlation (see Eq.~(\ref{eq: xi std+rel+eps})):
\begin{align}
\Delta \xi^{(\rm std)}_{\rm XY,1} &= \left( \frac{s}{d} \right) (1- aHd) f
\notag \\
&\qquad \times
( 5b_{\rm Y}s_{\rm B,X}-5b_{\rm X}s_{\rm B,Y} + 3f(s_{\rm B,X}-s_{\rm B,Y}) ) \Xi^{(1)}_{1}~, \label{eq: mag xi1 std}\\
\Delta \xi^{(\rm pot)}_{\rm XY,1} &= -\left( \frac{s}{d}\right)10 aH f \mathcal{M}s^{2} (s_{\rm B,X}-s_{\rm B,Y})\left( \Xi^{(0)}_{0}+ \Xi^{(0)}_{2}\right)~, \label{eq: mag xi1 pot} \\
\Delta \xi^{(\epsilon_{\rm NL})}_{\rm XY,1} &=
-\left( \frac{s}{d}\right)\frac{2aHf}{7}(s_{\rm B,Y}\epsilon_{\rm NL,X}-s_{\rm B,X}\epsilon_{\rm NL,Y})
\notag \\
& \qquad \times
\left( 3f\Xi^{(0)}_{0} + (7+12f)\Xi^{(1)}_{1} \right) \label{eq: mag xi1 eps}
~.
\end{align}
In the above, all the corrections are found to be proportional to the factor $(s/d)$, thus implying that these corrections are insignificant at small separation or higher redshift.

\begin{figure}
\centering
\includegraphics[width=0.8\columnwidth]{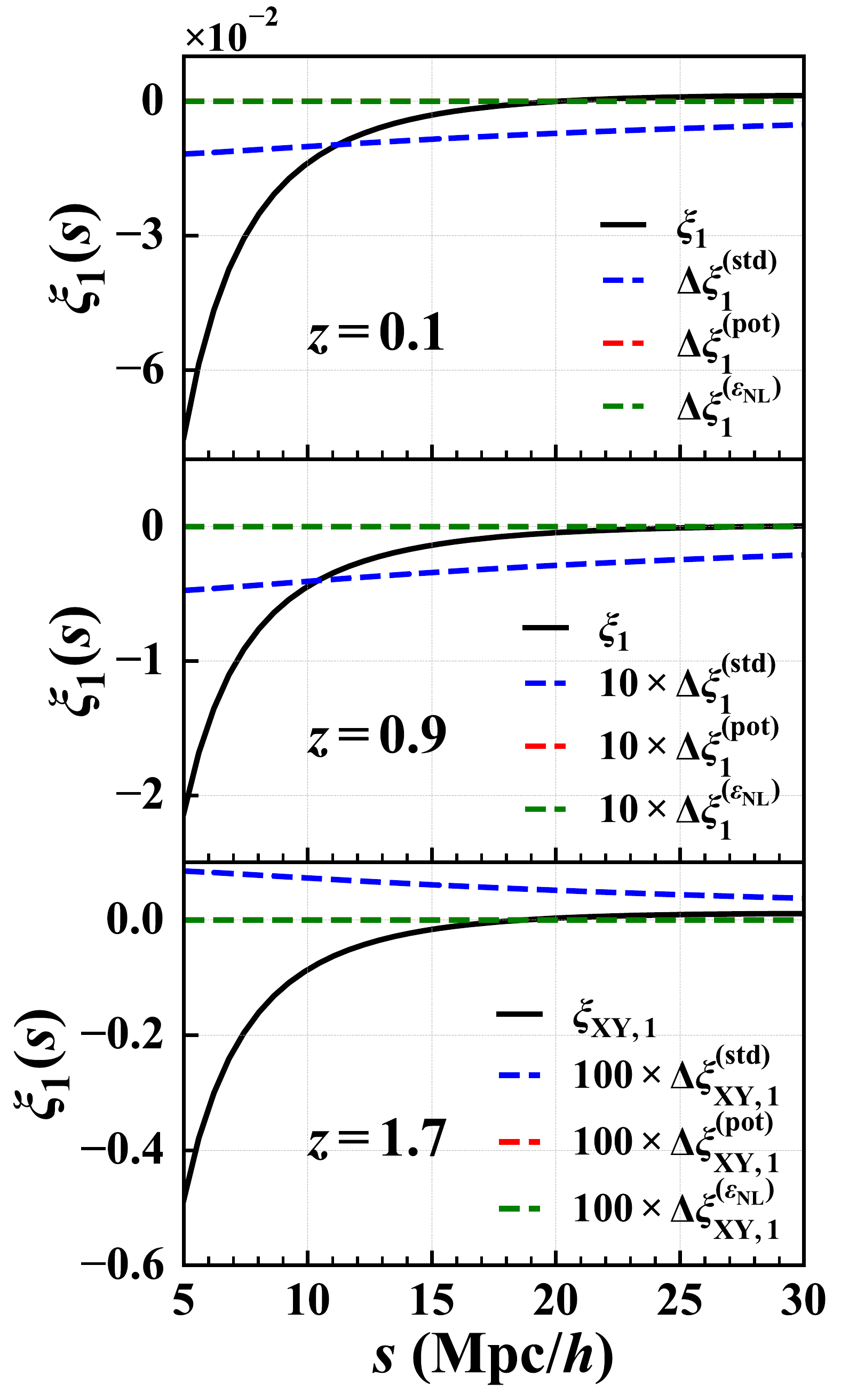}
\caption{Impact of the magnification bias on the dipole moment from $z=0.1$ ({\it top}) to $1.7$ ({\it bottom}), given by Eqs.~(\ref{eq: mag xi1 std})--(\ref{eq: mag xi1 eps}). The parameters including the bias are the same as Fig.~\ref{fig: xi eps NL}. We set the slope of the luminosity function as representative values of LRG and ELG for $s_{\rm B,X} = 1.2$ and $s_{\rm B,X} = 1.0$, respectively~\citep[][]{2017PhRvD..95d3530H}. As seen in these figures, the magnification bias has less contribution to the dipole, especially at high redshift.
}
\label{fig: xi1 magnification bias}
\end{figure}

Using the expressions at Eqs.~(\ref{eq: mag xi1 std})--(\ref{eq: mag xi1 eps}), we show in Fig.~\ref{fig: xi1 magnification bias} the impact of the Doppler magnification on the dipole signal, focusing particularly on small scales where the gravitational redshift effect becomes dominant. Here, we adopt the same parameter set as used in Fig.~\ref{fig: xi eps NL}, but for the slope of the luminosity function, we set $s_{\rm B,X} = 1.2$ and $s_{\rm B,X} = 1.0$ that are the typical values for the LRG and ELG samples~\citep[e.g.,][]{2017PhRvD..95d3530H}.
Fig.~\ref{fig: xi1 magnification bias} shows that the Doppler magnification can contribute about 10 percent to the dipole signal at low redshift, $z=0.1$. On the other hand, going to higher redshifts, the contribution from the magnification bias becomes negligibly smaller, as we expected. Thus, we conclude that the impact of the Doppler magnification on the dipole signal is neglected as long as we consider the high redshifts and small scales,  where the gravitational redshift effect dominates the dipole signal.

\section{Survey parameters and target samples}
\label{app: future surveys}

In Sec.~\ref{sec: future observations}, we examine the detectability for the dipole in future surveys: DESI, Euclid, Subaru-PFS, and SKA.
In this appendix, we summarize the survey parameters of each observation we used.

\subsection{Survey parameters and target samples}
\label{app: sum surveys}

When calculating the signal-to-noise ratio, we use the values of the central redshift, width of redshift bins, number density, bias, and the fractional sky coverage or survey volume, for each survey.
These survey parameters are summarized in Tables~\ref{table: DESI1} (DESI-BGS), \ref{table: DESI2} (DESI-LRG/ELG), \ref{table: Euclid} (Euclid), \ref{table: PFS} (Subaru-PFS), \ref{table: SKA1} (SKA1), and \ref{table: SKA2} (SKA2).
In these tables, we also include the ratio of the number densities $n_{\rm X}(M_{*})/n$ when the signal-to-noise ratio reaches its maximum (see Sec.~\ref{sec: future observations} in detail).
This will give us a guideline for future observations when we divide the sample into two subsamples.

Given the number density per unit redshift per square degree, ${\rm d}^{2}N/({\rm d}z\, {\rm d}{\rm deg}^{2})$, in order to obtain the number density per unit volume, $n$, we use the relation:
\begin{align}
n = \frac{{\rm d}^{2}N}{{\rm d}z\, {\rm d}{\rm deg}^{2}} \times \frac{ \Delta z \, f_{\rm sky}}{V} ~,
\end{align}
where the quantities $\Delta z$, $f_{\rm sky}$, and $V$ are the width of the redshift bin, the fractional sky coverage, and survey volume, respectively.

\begin{table}\centering
\caption{
DESI Bright Galaxy Survey (BGS) (taken from Table~2.5 of \citet{2016arXiv161100036D}).
The bias of BLG in \citet{2016arXiv161100036D} is assumed to be $b_{\rm BGS}(z) = 1.34/D_{+}(z)$.
The width of the redshift bin and fractional sky coverage are, respectively, $\Delta z = 0.1$ and $f_{\rm sky} = 0.339$.
}
\begin{tabular}{ccc}
$z$
& $n\, ({\rm Mpc}/h)^{-3}$
& $n_{\rm X}(M_{*})/n$
\\
\hline
 0.05 & $4.1\times 10^{-2}$ & $6.6\times10^{-3}$\\
 0.15 & $1.9\times 10^{-2}$ & $7.4\times10^{-3}$ \\
 0.25 & $4.6\times 10^{-3}$ & $8.3\times10^{-3}$ \\
 0.35 & $9.9\times 10^{-4}$ & $9.4\times10^{-3}$ \\
 0.45 & $1.1\times 10^{-4}$ & $1.1\times10^{-2}$ \\
\end{tabular}
\label{table: DESI1}
\end{table}

\begin{table}\centering
\caption{
DESI Luminous Red Galaxies (LRG) and Emission Line Galaxies (ELG) (taken from Table~2.3 of \citet{2016arXiv161100036D}).
The biases of LRG and ELG in \citet{2016arXiv161100036D} are assumed to be $b_{\rm LRG}(z) = 1.7/D_{+}(z)$ and $b_{\rm ELG}(z) = 0.84/D_{+}(z)$, respectively.
The width of the redshift bin and fractional sky coverage are, respectively, $\Delta z = 0.1$ and $f_{\rm sky} = 0.339$.
}
\begin{tabular}{ccccc}
\centering
& \multicolumn{2}{c}{ELG}
& \multicolumn{2}{c}{LRG}\\
$z$
& $n\, ({\rm Mpc}/h)^{-3}$
& $n_{\rm X}(M_{*})/n$
& $n\, ({\rm Mpc}/h)^{-3}$
& $n_{\rm X}(M_{*})/n$
\\
\hline
 0.65 & $1.6\times 10^{-4}$ & $1.8\times 10^{-3}$ & $4.4\times 10^{-4}$ & $2.3\times 10^{-2}$ \\
 0.75 & $1.0\times 10^{-3}$ & $2.3\times 10^{-3}$ & $4.2\times 10^{-4}$ & $2.6\times 10^{-2}$ \\
 0.85 & $7.4\times 10^{-4}$ & $2.8\times 10^{-3}$ & $2.5\times 10^{-4}$ & $2.2\times 10^{-2}$ \\
 0.95 & $7.2\times 10^{-4}$ & $2.3\times 10^{-3}$ & $9.3\times 10^{-5}$ & $2.6\times 10^{-2}$\\
 1.05 & $4.5\times 10^{-4}$ & $2.9\times 10^{-3}$ & $1.6\times 10^{-5}$ & $2.3\times 10^{-2}$ \\
 1.15 & $3.9\times 10^{-4}$ & $3.6\times 10^{-3}$ & $4.9\times 10^{-6}$ & $2.7\times 10^{-2}$ \\
 1.25 & $3.6\times 10^{-4}$ & $3.1\times 10^{-3}$ & - & -\\
 1.35 & $1.3\times 10^{-4}$ & $3.9\times 10^{-3}$ & - & -\\
 1.45 & $1.1\times 10^{-4}$ & $3.4\times 10^{-3}$ & - & -\\
 1.55 & $7.7\times 10^{-5}$ & $4.4\times 10^{-3}$ & - & -\\
 1.65 & $2.9\times 10^{-5}$ & $5.6\times 10^{-3}$ & - & -
\end{tabular}
\label{table: DESI2}
\end{table}

\begin{table}\centering
\caption{
Euclid with the fractional sky coverage $f_{\rm sky} = 0.364$, H$\alpha$ Emission Line Galaxies (taken from Table~3 of \citet{2019arXiv191009273E}).
}
\begin{tabular}{c c c c c}
$z$
& $\Delta z$
& $n\ ({\rm Mpc}/h)^{-3}$
& bias
& $n_{\rm X}(M_{*})/n$
\\
\hline
 1.0 & 0.2 & $6.86\times 10^{-4}$ & 1.46 & $4.5\times 10^{-3}$ \\
 1.2 & 0.2 & $5.58\times 10^{-4}$ & 1.61 & $4.8\times 10^{-3}$ \\
 1.4 & 0.2 & $4.21\times 10^{-4}$ & 1.75 & $7.4\times 10^{-3}$ \\
 1.65 & 0.3 & $2.61\times 10^{-4}$ & 1.90 & $7.8\times 10^{-3}$ \\
\end{tabular}
\label{table: Euclid}
\end{table}

\begin{table}\centering
\caption{
Subaru PFS with the fractional sky coverage $f_{\rm sky} = 0.0355$, $[{\rm OII}]$ Emission Line Galaxies (taken from Table~2 of \citet{2014PASJ...66R...1T}).
}
\begin{tabular}{c c c c c}
$z$
& $\Delta z$
& $n \, ({\rm Mpc}/h)^{-3}$
& bias
& $n_{\rm X}(M_{*})/n$
\\
\hline
 0.7 & 0.2 & $1.9\times 10^{-4}$ & 1.18 & $1.7\times 10^{-3}$\\
 0.9 & 0.2 & $6.0\times 10^{-4}$ & 1.26 & $2.5\times 10^{-3}$ \\
 1.1 & 0.2 & $5.8\times 10^{-4}$ & 1.34 & $2.7\times 10^{-3}$ \\
 1.3 & 0.2 & $7.8\times 10^{-4}$ & 1.42 & $2.9\times 10^{-3}$ \\
 1.5 & 0.2 & $5.5\times 10^{-4}$ & 1.50 & $3.2\times 10^{-3}$ \\
 1.8 & 0.4 & $3.1\times 10^{-4}$ & 1.62 & $3.3\times 10^{-3}$ \\ 
 2.2 & 0.4 & $2.7\times 10^{-4}$ & 1.78 & $3.1\times 10^{-3}$\\ 
\end{tabular}
\label{table: PFS}
\end{table}

\begin{table}\centering
\caption{
SKA1-MID with the fractional sky coverage $f_{\rm sky} = 0.121$ and the width of redshift bin $\Delta z = 0.1$, HI Galaxies (taken from Table~1 of \citet{2015aska.confE..24B}).
Only in the lowest redshift $z=0.05$, since the given bias parameter is too small, Eq.~(\ref{eq: b obs to Mmin}) does not have a solution $M_{\rm min}$. Therefore, we will fix $M_{\rm min} = 10^{8}\, M_{\odot}/h$ only for this case, based on \citet{2015MNRAS.450.2251Y}.
}
\begin{tabular}{c c c c}
$z$ & $n \ ({\rm Mpc}^{-3})$ & bias & $n_{\rm X}(M_{*})/n$ \\
\hline
 0.05 & $2.92\times 10^{-2}$ & 0.678 & $3.7\times 10^{-2}$\\
 0.15 & $6.74\times 10^{-3}$ & 0.727 & $8.2\times 10^{-6}$ \\
 0.25 & $1.71\times 10^{-3}$ & 0.802 & $8.1\times 10^{-5}$ \\
 0.35 & $4.64\times 10^{-4}$ & 0.886 & $3.5\times 10^{-4}$ \\
 0.45 & $1.36\times 10^{-4}$ & 0.975 & $7.6\times 10^{-4}$ \\
\end{tabular}
\label{table: SKA1}
\end{table}

\begin{table}\centering
\caption{
SKA2 with sky coverage with the fractional sky coverage $f_{\rm sky} = 0.727$ and the width of redshift bin $\Delta z = 0.1$, HI Galaxies (taken from Table~1 of \citet{2015aska.confE..24B}).
}
\begin{tabular}{c c c c}
$z$
& $n\ ({\rm Mpc}^{-3})$
& bias
& $n_{\rm X}(M_{*})/n$
\\
\hline
 0.23 & $4.43\times 10^{-2}$ & 0.713 & $2.0 \times 10^{-6}$\\
 0.33 & $2.73\times 10^{-2}$ & 0.772 & $4.6 \times 10^{-5}$ \\
 0.43 & $1.65\times 10^{-2}$ & 0.837 & $1.5 \times 10^{-4}$ \\
 0.53 & $9.89\times 10^{-3}$ & 0.907 & $3.6 \times 10^{-4}$ \\
 0.63 & $5.88\times 10^{-3}$ & 0.983 & $7.5 \times 10^{-4}$ \\
 0.73 & $3.48\times 10^{-3}$ & 1.066 & $1.0 \times 10^{-3}$ \\
 0.83 & $2.05\times 10^{-3}$ & 1.156 & $1.9 \times 10^{-3}$ \\
 0.93 & $1.21\times 10^{-3}$ & 1.254 & $2.4 \times 10^{-3}$ \\
 1.03 & $7.06\times 10^{-4}$ & 1.360 & $3.0 \times 10^{-3}$ \\
 1.13 & $4.11\times 10^{-4}$ & 1.475 & $3.7 \times 10^{-3}$ \\
 1.23 & $2.39\times 10^{-4}$ & 1.600 & $4.6 \times 10^{-3}$ \\
 1.33 & $1.39\times 10^{-4}$ & 1.735 & $5.6 \times 10^{-3}$ \\
 1.43 & $7.99\times 10^{-5}$ & 1.882 & $6.9 \times 10^{-3}$ \\
 1.53 & $4.60\times 10^{-5}$ & 2.041 & $8.5 \times 10^{-3}$ \\
 1.63 & $2.64\times 10^{-5}$ & 2.214 & $1.0 \times 10^{-3}$ \\
 1.73 & $1.51\times 10^{-5}$ & 2.402 & $1.3 \times 10^{-3}$ \\
 1.81 & $9.66\times 10^{-6}$ & 2.566 & $1.7 \times 10^{-3}$ \\
\end{tabular}
\label{table: SKA2}
\end{table}

\subsection{Cross-correlating two measurements with different redshift bins}
\label{app: cross sample}

Since the width of redshift bins is generally different for each observation, we perform the following procedure for different width of bins when cross-correlating in Sec.~\ref{sec: future observations}.

We have the survey parameters as summarized in Appendix~\ref{app: future surveys}: the mean redshift $z^{\rm X/Y}_{i}$, width of redshift bins $\Delta z^{\rm X/Y}_{i}$, number density $n^{\rm X/Y}_{i}$, and bias $b^{\rm X/Y}_{i}$ where the subscript $i$ stands for the $i$th redshift bin.
Then, we define the number density and bias for the survey Y as a function of redshift:
\begin{align}
n^{\rm Y}(z) &= n^{\rm Y}_{i} ~~~(z^{\rm Y}_{i} - \Delta z^{\rm Y}_{i}/2 \leq z \leq z^{\rm Y}_{i} + \Delta z^{\rm Y}_{i}/2) ~,\\
b^{\rm Y}(z) &= b^{\rm Y}_{i} ~~~(z^{\rm Y}_{i} - \Delta z^{\rm Y}_{i}/2 \leq z \leq z^{\rm Y}_{i} + \Delta z^{\rm Y}_{i}/2) ~,
\end{align}
where these functions correspond to the plots shown in Fig.~\ref{fig: observations}.
Then, we obtain the number density and bias for the survey Y in the mean redshift and redshift bin for the survey X by
\begin{align}
\tilde{n}^{\rm Y}_{i} & = \frac{1}{\Delta z^{\rm X}_{i}}\int^{z^{\rm X}_{i}+\Delta z^{\rm X}_{i}/2}_{z^{\rm X}_{i}-\Delta z^{\rm X}_{i}/2}\, n^{\rm Y}(z)\, {\rm d}z ~,\\
\tilde{b}^{\rm Y}_{i} & = \frac{1}{\int^{z^{\rm X}_{i}+\Delta z^{\rm X}_{i}/2}_{z^{\rm X}_{i}-\Delta z^{\rm X}_{i}/2}\, n^{\rm Y}(z)\, {\rm d}z }
\int^{z^{\rm X}_{i}+\Delta z^{\rm X}_{i}/2}_{z^{\rm X}_{i}-\Delta z^{\rm X}_{i}/2}\, b^{\rm Y}(z)n^{\rm Y}(z)\, {\rm d}z ~.
\end{align}
Thus, we obtain the survey parameters $(b^{\rm X}_{i}, \tilde{b}^{\rm Y}_{i}, n^{\rm X}_{i}, \tilde{n}^{\rm Y}_{i})$ in the common mean redshifts and redshift bins of the survey X.
In this definition, when the mean redshift and redshift bin for the survey X are the same as ones for the survey Y, we obtain $\tilde{b}^{\rm Y} = b^{\rm Y}$ and $\tilde{n}^{\rm Y}_{i} = n^{\rm Y}_{i}$.

\section{Signal-to-noise ratio in simulations: as a function of halo mass}
\label{sec: SN in sim}

\begin{figure*}
\centering
\includegraphics[width=0.8\textwidth]{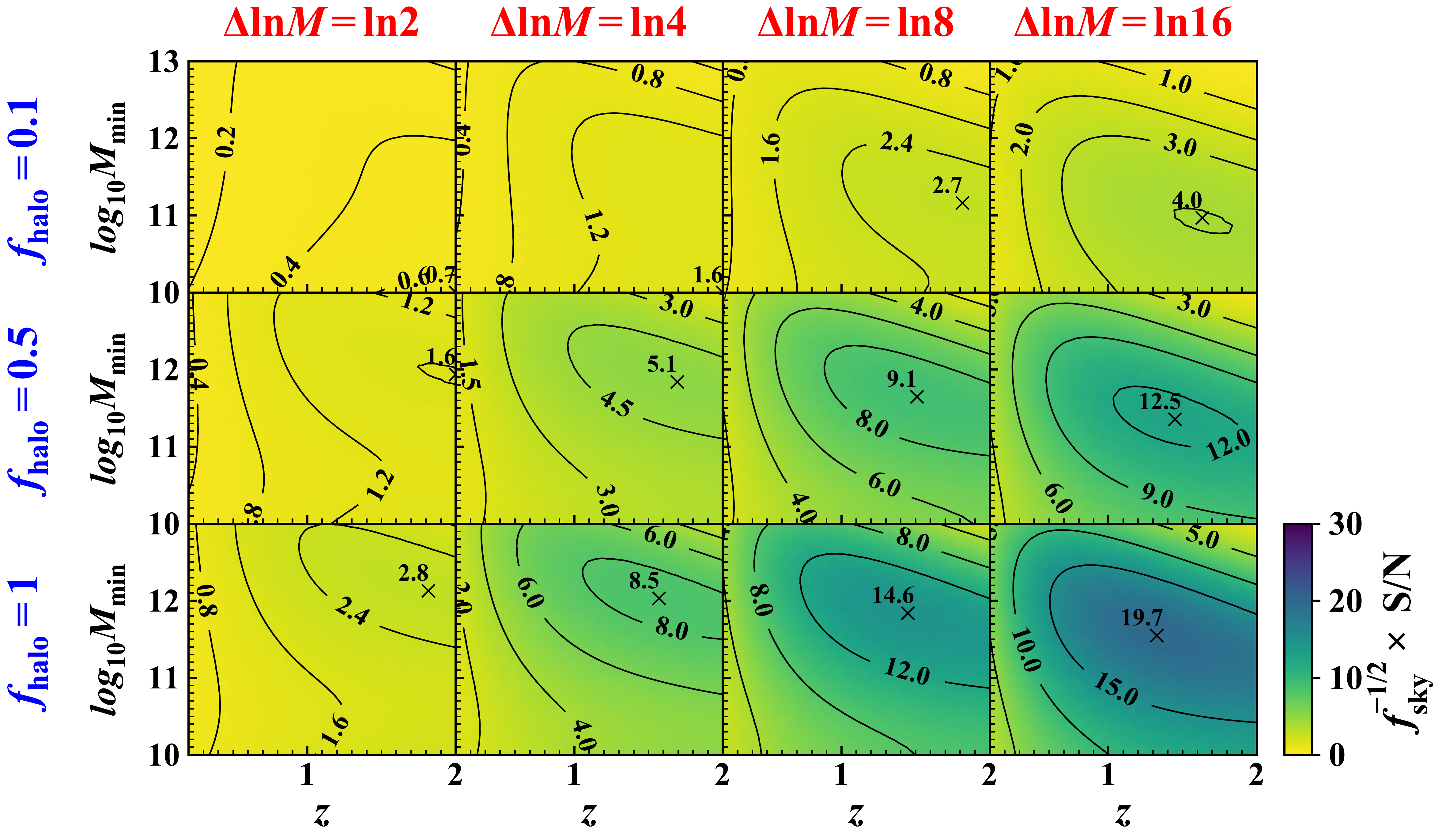}
\caption{
Signal-to-noise ratio as a function of the minimum halo mass $M_{\rm min}$ and mean redshift $z$.
From {\it left} to {\it right}, the logarithmic mass bin $\Delta \ln{M}$ is varied from $\ln{2}$ to $\ln{16}$, and from {\it top} to {\it bottom}, the parameter $f_{\rm halo}$ is varied from 0.1 to 1.
The cross symbols accompanied by a number indicate the parameters that give the maximum signal-to-noise ratio in the parameter space and the corresponding value of the signal-to-noise ratio.
The width of the redshift bins is fixed to $\Delta z = 0.1$.
}
\label{fig: SN 2D Mmin}
\end{figure*}
\begin{figure}
\centering
\includegraphics[width=\columnwidth]{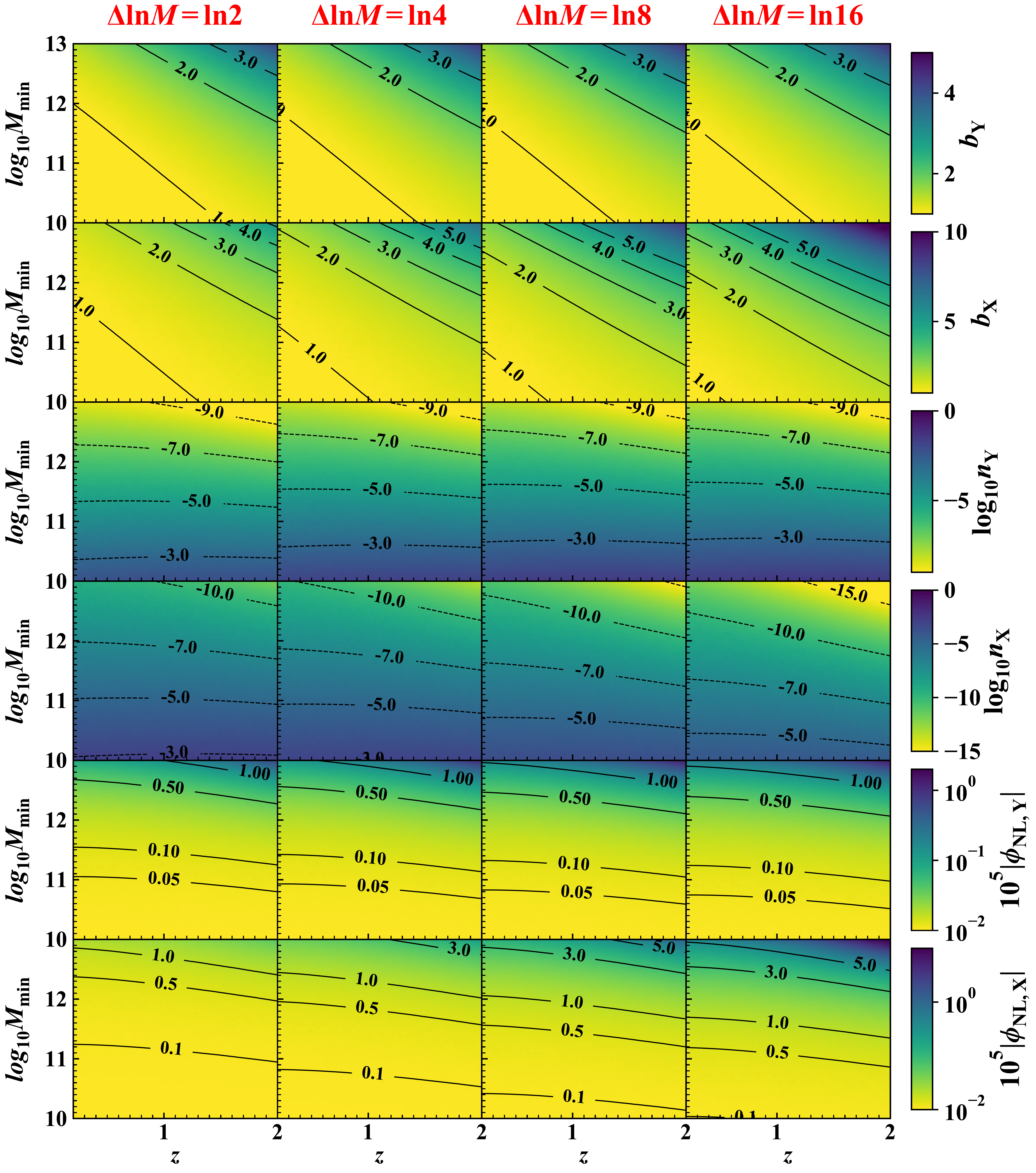}
\caption{
The relevant parameters to compute the signal-to-noise ratio in Fig.~\ref{fig: SN 2D Mmin}.
From {\it top} to {\it bottom}, we present the parameters as a function of the mean redshift and minimum mass, $b_{\rm Y}$, $b_{\rm X}$, $n_{\rm Y}$, $n_{\rm X}$, $\phi_{\rm NL\, Y}$, and $\phi_{\rm NL\, X}$, respectively.
From {\it left} to {\it right}, the logarithmic mass bin $\Delta \ln{M}$ is varied from $\ln{2}$ to $\ln{16}$.
}
\label{fig: 2D Mmin}
\end{figure}

When performing $N$-body simulations with a halo finder algorithm, we observe all haloes with their masses and number density.
In this appendix, assuming the minimum mass $M_{\rm min}$ and the width of logarithmic mass bins $\Delta\ln{M}$ in simulations, we ideally split two populations:
\begin{align}
(M_{1}, M_{2}, M_{3}) = (M_{\rm min}, M_{\rm min}e^{\Delta\ln{M}}, M_{\rm min}e^{2 \Delta\ln{M}})
\end{align}
and thereby we discuss the signal-to-noise ratio, as a function of $M_{\rm min}$ and $\Delta \ln{M}$.
This investigation provides us with an insight into the detectability in $N$-body simulations including special and general relativistic effects~\citep{2019MNRAS.483.2671B,2021MNRAS.501.2547G}.

Using two mass bins, the parameters to evaluate the dipole moment are given by
\begin{align}
n_{\rm Y} &= \int^{\ln{M_{2}}}_{\ln{M_{1}}} \frac{{\rm d}n}{{\rm d}\ln{M}}\, {\rm d}\ln{M}~, \\
\Braket{A_{\rm Y}} &=
\frac{1}{n_{\rm Y}}
\int^{\ln{M_{2}}}_{\ln M_{1}} \frac{{\rm d}n}{{\rm d}\ln{M}}\, A(M)\,{\rm d}\ln{M}
~, \\
n_{\rm X} &= \int^{\ln M_{3}}_{\ln M_{2}} \frac{{\rm d}n}{{\rm d}\ln{M}}\, {\rm d}\ln{M}~,\\
\Braket{A_{\rm X}} &=
\frac{1}{n_{\rm X}}
\int^{\ln M_{3}}_{\ln M_{2}} \frac{{\rm d}n}{{\rm d}\ln{M}}\, A(M)\,{\rm d}\ln{M}
~,
\end{align}
where we define $A = M$, $b_{\rm ST}(z,M)$, and $\phi_{\rm NFW,0}(z,M)$, and the function 
${\rm d}n/{\rm d}\ln{M}$ is the Sheth-Tormen mass function.

Since all the galaxies within haloes would not be detected in real observations, we introduce a suppression factor, the so-called halo occupation number $0<f_{\rm halo}\leq1$: the number of galaxies found in a virialized halo of a given mass, in the number density of haloes. Thus this factor can be regarded as a kind of halo occupation number.
If $f_{\rm halo} = 1$, all haloes in simulations are assumed to be detected.
In calculating the covariance matrix and signal-to-noise ratio, we multiply this factor by the number density of haloes.

In Fig.~\ref{fig: SN 2D Mmin}, we show the signal-to-noise ratio normalized by the fractional sky coverage $f_{\rm sky}$ as a function of the minimum halo mass $M_{\rm min}$ and mean redshift $z$.
This figure indicates that the signal-to-noise ratio becomes maximum at $z\approx 1.3$, slightly depending on the parameters $\Delta\ln{M}$ and $f_{\rm halo}$.
Note that the width of the redshift bins is fixed to  $\Delta z = 0.1$ in this figure.
This value of redshift at which the signal-to-noise ratio is maximum is different from Figs.~\ref{fig: SN 2D bY1} and \ref{fig: SN 2D bY2} because the number density is not constant in Fig.~\ref{fig: SN 2D Mmin}, but depends on the redshift following the Sheth-Tormen mass function.
In Fig.~\ref{fig: 2D Mmin}, from {\it top} to {\it bottom}, we have shown the parameters as a function of the mean redshift and minimum mass, $b_{\rm Y}$, $b_{\rm X}$, $n_{\rm Y}$, $n_{\rm X}$, $\phi_{\rm NL\, Y}$, and $\phi_{\rm NL\, X}$, respectively.

Fig.~\ref{fig: SN 2D Mmin} is useful to discuss the detectability for the dipole moment in simulations.
For example, comparing the amplitude of the signal with its error bars in Fig.~\ref{fig: dipole}, the signal-to-noise ratio is roughly given by ${(\rm S/N)} \approx 4$ in simulations with the following parameters: $\Delta \ln{M} \approx 2$, $\Delta z\approx 0.5$, $M_{\rm min} \approx 2\times 10^{12}\, M_{\rm sun}/h$, and $f_{\rm sky}=1$~\citep[see][]{2019MNRAS.483.2671B}, which lie at the region shown in the bottom-leftmost panel of Fig.~\ref{fig: SN 2D Mmin}.
Looking particularly at $z\approx 0.3$, we obtain the signal-to-noise ratio of ${\rm S/N} \approx 0.8$ for the width $\Delta z=0.1$. 
Accounting further for the width of the redshift bins, a simple multiplication by the factor $5$  results in ${\rm S/N} =4$, which reasonably agrees with the signal-to-noise ratio estimated from the measured dipole amplitudes and their error bars in  simulations~\citep[][]{2019MNRAS.483.2671B,2020MNRAS.498..981S}.

\bsp
\label{lastpage}
\end{document}